\documentclass[11pt,a4paper]{article}
\usepackage{jheppub}
\usepackage{amssymb}
\usepackage{amsmath}
\usepackage{amsfonts}
\usepackage[usenames,dvipsnames]{xcolor}
\usepackage{enumitem}
\usepackage{placeins}

\usepackage{mathpazo}  
\usepackage{makecell}
\usepackage{cancel}
\usepackage{mdframed}
\usepackage{comment}
\usepackage{multirow}

\usepackage{mathtools}

\usepackage{array,siunitx}

\usepackage{tikz}
\usetikzlibrary{patterns}
\usetikzlibrary{decorations.pathmorphing}
\usetikzlibrary{decorations.markings}
\usepgflibrary{arrows.meta}
\usetikzlibrary{intersections,through,hobby}
\usepackage{tkz-euclide}
\tikzset{
  WLBE/.style={double distance=1.1pt,postaction={decorate},
    decoration={markings,mark=at position .5 with {\arrow{Straight Barb[scale=0.5]}}}},  
  WLB/.style={double distance=1.1pt,postaction={decorate},
    decoration={markings,mark=at position .2 with {\arrow{Straight Barb[scale=0.5]}}}},
  WLBC/.style={double distance=1.1pt,postaction={decorate},
    decoration={markings,mark=at position .5 with {\arrow{Straight Barb[scale=0.5]}}}},
  WLBS/.style={double distance=1.1pt,postaction={decorate},
    decoration={markings,mark=at position .6 with {\arrow{Straight Barb[scale=0.5]}}}},
  GLUON/.style={decorate,
    decoration={coil, amplitude=2.5pt,segment length=3.25pt, aspect=0.65}},
  sglu/.style={
    draw=none,
    decoration={name=none},
    postaction={
      draw,
      line width = 0.6pt,
      decoration={
        coil,
        aspect=0.85,
        mirror,
        amplitude=2.0pt,
        segment length=4pt,
      },
      decorate=true},
    color=Gray},
  higgs/.style={line width=0.8pt,Gray, densely dashed},
  sghost/.style={line width=0.8pt,Gray, densely dotted,postaction={decorate},
    decoration={markings,mark=at position .4 with {\arrow{latex[scale=0.5]}}}},
  squark/.style={line width=0.8pt,Gray, postaction={decorate},
    decoration={markings,mark=at position .4 with {\arrow{latex[scale=0.5]}}}}
}

\definecolor{colA}{HTML}{c19277}
\definecolor{colB}{HTML}{e1bc91}
\definecolor{colD}{HTML}{62959c}
\hypersetup{
  citecolor=colA, 		%color of links to bibliography
  linkcolor=colA,    	%color of internal links
  urlcolor=colD,			%color of external links
}

\newcommand{\ep}{\ensuremath{\varepsilon}}
\newcommand{\dm}{\ensuremath{\textrm{d}}}

\newcommand{\dFtddA}{\ensuremath{\textrm{d}\Phi_{\theta\delta\delta}}}

\newcommand{\dFttdA}{\ensuremath{\textrm{d}\Phi_{\theta\theta\delta}}}
\newcommand{\dFtdtA}{\ensuremath{\textrm{d}\Phi_{\theta\delta\theta}}}
\newcommand{\dFdttA}{\ensuremath{\textrm{d}\Phi_{\delta\theta\theta}}}
\newcommand{\dFtttA}{\ensuremath{\textrm{d}\Phi_{\theta\theta\theta}}}
\newcommand{\dFfffA}{\ensuremath{\textrm{d}\Phi_{f_1f_2f_3}}}
\newcommand{\dFfffAnu}{\ensuremath{\textrm{d}\Phi_{f_1f_2f_3}^{\nu}}}
\newcommand{\dFttdB}{\ensuremath{\textrm{d}\Phi_{\theta\theta\delta}}}
\newcommand{\dFtdtB}{\ensuremath{\textrm{d}\Phi_{\theta\delta\bar{\theta}}}}
\newcommand{\dFdttB}{\ensuremath{\textrm{d}\Phi_{\delta\theta\bar{\theta}}}}
\newcommand{\dFtttB}{\ensuremath{\textrm{d}\Phi_{\theta\theta\bar{\theta}}}}
\newcommand{\dFfffB}{\ensuremath{\textrm{d}\Phi_{f_1f_2\bar{f_3}}}}
\newcommand{\dFfffBnu}{\ensuremath{\textrm{d}\Phi_{f_1f_2\bar{f_3}}^{\nu}}}
\newcommand{\dFrelabeledtbdt}{\ensuremath{\textrm{d}\Phi_{\bar{\theta}\delta\theta}}}
\newcommand{\dFrelabeledtbtt}{\ensuremath{\textrm{d}\Phi_{\bar{\theta}\theta\theta}}}

\newcommand{\normNepTrip}{\mathcal{N}_{\ep}^3}
\newcommand{\normNep}{\mathcal{N}_{\ep}}

\newcommand{\nf}{\ensuremath{n_f}}

\newcommand{\CF}{\ensuremath{C_F}}
\newcommand{\CA}{\ensuremath{C_A}}
\newcommand{\CR}{\ensuremath{C_R}}
\newcommand{\TF}{\ensuremath{T_F}}

\newcommand{\Gcusp}{\ensuremath{\Gamma}}
\newcommand{\Gsoft}{\ensuremath{\gamma^s}}
\newcommand{\Sbc}[1]{\ensuremath{S_{#1}}}
\newcommand{\boxS}[1]{\tikz[baseline]{\node[draw=MidnightBlue!25, anchor=base,
    line width=1pt]{$#1$};}}
\newcommand{\Ls}{\ensuremath{L_S}}
\newcommand{\as}{a_s}

\allowdisplaybreaks

\begin{document}

\title{Triple real-emission contribution to the zero-jettiness soft function at N3LO in QCD}

\preprint{
\begin{minipage}[c]{0.3\linewidth}
  \begin{flushright}
    MPP-2024-246\\
    P3H-24-097\\
    TTP24-045\\
    TUM-HEP-1544/24\\
    ZU-TH-64/24
  \end{flushright}
\end{minipage}}

\author[a]{Daniel~Baranowski,}
\author[b,d]{Maximilian~Delto,}
\author[c]{Kirill~Melnikov,}
\author[c]{Andrey~Pikelner}
\author[d]{and Chen-Yu~Wang}

\affiliation[a]{Physik Institut, Universität Zürich, Winterthurerstrasse 190, 8057 Zürich, Switzerland}
\affiliation[b]{Physics Department, Technical University of Munich, James-Franck-Strasse 1,  85748, Munich, Germany}
\affiliation[c]{Institute for Theoretical Particle Physics (TTP), Karlsruhe Institute of Technology, 76128, Karlsruhe, Germany}
\affiliation[d]{Max-Planck Institute for Physics, Boltzmannstr.\ 8, 85748 Garching, Germany}

\emailAdd{daniel.baranowski@physik.uzh.ch}
\emailAdd{maximilian.delto@tum.de}
\emailAdd{kirill.melnikov@kit.edu}
\emailAdd{andrey.pikelner@kit.edu}
\emailAdd{cywang@mpp.mpg.de}

\abstract{ Recently, we have presented the result for
the zero-jettiness soft function at next-to-next-to-next-to-leading order (N3LO)
in perturbative QCD~\cite{Baranowski:2024vxg}, without providing technical
details of the calculation. The goal of this paper is to describe the most
important element of that computation, the triple real-emission contribution. We
present a detailed discussion of the many technical aspects of the calculation,
for which a number of methodological innovations was required. Although some
elements of the calculation were discussed
earlier~\cite{Baranowski:2020xlp,Chen:2020dpk,Baranowski:2021gxe,Baranowski:2022khd,Baranowski:2024ene},
this paper is intended to provide a complete summary of the methods used in the
computation of the triple real-emission contribution to the soft function.
}

\maketitle

\newpage

\section{Introduction}
\label{sec:intro}

Experiments at the LHC and its high-luminosity upgrade will continue the
exploration of the Standard Model and searches for physics beyond it through
precise measurements. Eventually, interpretations of such measurements will be
dominated by theoretical systematics, and improvements in the description of
hard hadron collisions are therefore called for. Such improvements require
advancements in computations of two- and three-loop amplitudes for high
multiplicities, and in dealing with real-emission processes which are needed to
cancel infra-red divergences present in loop amplitudes~\cite{Bloch:1937pw,
Kinoshita:1962ur,Lee:1964is}.\footnote{Collinear divergencies associated with
the initial-state radiation are absorbed into the parton distribution
functions.}

The crucial difference between loop amplitudes and real-emission contributions
is that infra-red divergencies in virtual amplitudes appear explicitly after
integrating over the loop momenta, manifesting themselves as poles in the
dimensional regularization parameter $\epsilon=(4-d)/2$, with $d$ being the
dimensionality of space-time. On the contrary, real-emission contributions
involve complex observables and cannot be analytically integrated over the
entire phase space. Because of that, infra-red divergences need to be extracted
before the numerical integration.

It  is possible to do that because divergences only appear in phase-space regions where
additional radiation (with respect to Born processes) becomes soft and collinear
and, therefore, unresolved. Since infra-red safe observables are, by definition,
not sensitive to soft- and collinear radiation, and since matrix elements
exhibit universal factorization in these limits, it becomes possible to extract
infra-red divergences in a process-independent manner. Consequently, various
\emph{slicing and subtraction} schemes that isolate, extract, and cancel
infra-red poles in cross-section calculations, have been proposed and employed at
NLO~\cite{Ellis:1980wv,Fabricius:1981sx,Frixione:1995ms,Frixione:1997np,Catani:1996vz,Catani:2002hc,Nagy:2003qn,Nagy:2007ty,Chung:2010fx,Chung:2012rq,Bevilacqua:2013iha}, NNLO~\cite{Catani:2007vq,Bonciani:2015sha,Grazzini:2017mhc,Catani:2019iny,Catani:2019hip,
Kallweit:2020gcp,Catani:2020kkl,Catani:2021cbl,Boughezal:2015dva,Boughezal:2015aha,Gaunt:2015pea,
Boughezal:2016wmq,GehrmannDeRidder:2005cm,GehrmannDeRidder:2005aw,GehrmannDeRidder:2005hi,Daleo:2006xa,Gehrmann-DeRidder:2007nzq,Daleo:2009yj,Gehrmann:2011wi,Boughezal:2010mc,
GehrmannDeRidder:2012ja,Currie:2013vh,Currie:2016ytq,Currie:2017eqf,
Currie:2018xkj,Herzog:2018ily,Czakon:2010td,Czakon:2011ve,Czakon:2013goa,Czakon:2014oma,
Czakon:2019tmo,Chawdhry:2019bji,Magnea:2018hab,Magnea:2018ebr,Magnea:2020trj,
Somogyi:2005xz,Somogyi:2006cz,Somogyi:2006da,Somogyi:2006db,
Somogyi:2008fc,Aglietti:2008fe,Somogyi:2009ri,
Bolzoni:2009ye,Bolzoni:2010bt,DelDuca:2013kw,Somogyi:2013yk,
DelDuca:2016ily,Han:1992hr,Brucherseifer:2014ama,Cacciari:2015jma} and
N3LO~\cite{Chen:2021vtu,Neumann:2022lft,Chen:2022cgv,Chen:2022lwc,Campbell:2023lcy}.
Additionally, fully-numerical approaches exist which combine loop- and
real-emission integrands before the
integration~\cite{Soper:1999xk,Soper:2001hu,Kramer:2002cd,Kramer:2003jk,Soper:2003ya,Capatti:2020xjc,AH:2023kor}.
Further discussion of advanced computational methods in QCD and their
applications can be found in ref.~\cite{Heinrich:2020ybq}.

Modern slicing and subtraction schemes require integration of well-defined
functions encapsulating soft- and collinear dynamics, over phase spaces of
unresolved emissions. These integrals differ because of phase-space constraints
which are particular to a specific scheme. In this paper, we consider the
$N$-jettiness observable $\mathcal{T}_N$~\cite{Stewart:2009yx,Stewart:2010tn},
which can be used as a slicing variable for lepton- and hadron-collider
processes with final-state jets. The $N$-jettiness slicing scheme has already
been used in a number of high-profile NNLO QCD computations
~\cite{Boughezal:2015dva,Gaunt:2015pea,Boughezal:2015aha,
Boughezal:2016wmq,Neumann:2022lft,Campbell:2023lcy}.

In case of $N$-jettiness, integrals over unresolved phase-space regions are
further split into the so-called soft, beam and jet functions. Beam and jet
functions describe collinear emissions off incoming and outgoing hard partons.
These functions are process-independent and are currently known through N3LO in
perturbative QCD~\cite{Gaunt:2014cfa,Gaunt:2014xga,
Boughezal:2017tdd,Bruser:2018rad,Ebert:2020lxs,Ebert:2020unb,Behring:2019quf,Baranowski:2022vcn}.

The \emph{soft function} describes soft QCD radiation at large angles. For this
reason, it is sensitive to color charges in the full event. This fact makes the
soft function a much more complex quantity to calculate. The complexity of the
soft function is further exacerbated by the definition of the $N$-jettiness
variable which requires one to find the minima of certain combinations of scalar
products between soft and hard partons. The traditional approach to computing
the $N$-jettiness soft function, that goes back to the early papers on this
subject~\cite{Stewart:2009yx,Stewart:2010tn}, involves explicit resolution of
the $N$-jettiness constraint by an additional phase-space partitioning. Since
the complexity of such a partitioning strongly increases with the number
of external particles, the $N$-jettiness soft function was first calculated for
processes with two, three and four hard
partons~\cite{Kelley:2011ng,Monni:2011gb,
Boughezal:2015eha,Li:2016tvb,Campbell:2017hsw,Banerjee:2018ozf,Baranowski:2020xlp,Jin:2019dho}.

Recently, new computational approaches were developed which allowed the
calculation of the $N$-jettiness soft function for an \emph{arbitrary} number of
hard partons $N$~\cite{Bell:2023yso,Agarwal:2024gws}. This fact makes the
$N$-jettiness slicing scheme the first scheme in which all unresolved
real-emission ingredients are known analytically for arbitrary collider
processes (with massless partons) at NNLO QCD. We note in passing that the
computation of ref.~\cite{Agarwal:2024gws} represents a notable departure from
the original methods, because divergences are regulated at the level of the
entire soft function and the $N$-jettiness is treated as \emph{any other }
infra-red safe observable. Practically, one employs nested soft-collinear
subtraction scheme~\cite{Caola:2017dug} and the integrated double-soft
subtraction term computed in ref.~\cite{Caola:2018pxp}.

One may dream of replicating this success also at N3LO QCD but, given a rather
limited understanding of soft and collinear singularities at that perturbative
order, and a complex nature of the $N$-jettiness soft function, calculations at
N3LO have naturally focused on the $N=0$
case~\cite{Chen:2020dpk,Baranowski:2021gxe,Baranowski:2022khd,Baranowski:2024ene}.
The explicit partitioning of the radiative phase space into sectors where the
$N$-jettiness assumes a definite value, remains manageable in such a case.
However, instead of integrating the resulting expressions numerically using the
sector decomposition method~\cite{Binoth:2000ps}, as was done in NNLO QCD
computations for the $N=0,1,2$ soft functions~\cite{
Boughezal:2015eha,Li:2016tvb,Campbell:2017hsw,Banerjee:2018ozf,Jin:2019dho}, we
use techniques developed for multi-loop computations, such as the
integration-by-parts (IBP) relations~\cite{Tkachov:1981wb,Chetyrkin:1981qh} and
the method of differential
equations~\cite{Kotikov:1990kg,Kotikov:1991hm,Kotikov:1991pm,Remiddi:1997ny,Gehrmann:1999as},
and adapt them for our purposes.

In this paper, we discuss the triple-real contribution to the N3LO QCD
zero-jettiness soft function, accounting for both $ggg$ and $gq\bar{q}$ soft
final-state partons. We aim at explaining how to overcome the technical
challenges described in ref.~\cite{Baranowski:2024vxg}, where the result for the
zero-jettiness soft function at N3LO has been presented.

The outline of the paper is as follows. In section~\ref{sec:soft-lim}, we define
the zero-jettiness soft function, explain how the soft limits of tree-level
matrix elements are computed, and determine the integrand for the soft function
computation. In section~\ref{sec:unregDiv} we argue that some integrals,
required for the computation of the soft function, are not regulated
dimensionally. We also describe a constructive procedure called filtering, which
allows us to remove such integrals from the calculation. In
section~\ref{sec:IBP} we explain how the integration-by-parts method works for
phase-space integrals with Heaviside functions. In sections~\ref{sec:no123ints}
and \ref{sec:DEs}, we discuss computation of master integrals. We find it
necessary to divide them into two classes. The first class contains no integrals
with the propagator $1/k_{123}^2$, where $k_{123}$ is the sum of three
soft-parton momenta. The second class comprises integrals \emph{with} this
propagator. Integrals that belong to the first class are discussed in
section~\ref{sec:no123ints}; we compute them by integrating over the phase space
of three soft partons, subject to the zero-jettiness constraint. Integrals of
the second type are too complicated for a direct integration. In
section~\ref{sec:DEs}, we explain how we introduce an auxiliary parameter and
derive differential equations for such integrals. Calculation of the boundary
conditions is discussed in section~\ref{sec:boundaries}. We elaborate on the
numerical checks in section~\ref{sec:numchecks}, provide results for the soft
function in section~\ref{sec:results} and conclude in section~\ref{sec:concl}.
Some elements of the calculations and the various quantities used there are
presented in several appendices.

\section{Definition of the soft function and 
the soft limits of squared amplitudes}
\label{sec:soft-lim}

The zero-jettiness observable describes processes with exactly two hard partons.
It is defined as follows~\cite{Stewart:2009yx,Stewart:2010tn}
\begin{equation}
   {\cal T}_0(n) = \sum \limits_{i=1}^{n} {\rm min}
   \left [ \frac{2 p_a \cdot k_i}{P}, \frac{2 p_b \cdot k_i}{P} \right ].
\label{eq:zero-jettiness-def}
\end{equation}
In eq.~\eqref{eq:zero-jettiness-def}, $p_{a,b}$ are the four-momenta of hard
partons, either incoming or outgoing, $k_i$, $i \in \{1,..,n\}$ are the momenta
of additional (soft) partons, and $P$ is an arbitrary parameter of
mass-dimension one. The zero-jettiness soft function can be written as
\begin{equation}
S_\tau = \sum_{n=0}^{\infty} S_\tau^{(n)} \,,
\end{equation}
where $S_\tau^{(n)}$ describes a partonic process with $n$ soft particles in the final state. It reads
\begin{equation}
S_\tau^{(n)} = \frac1{\mathcal{N_{\text{sym}}}} \int \prod_{i=1}^{n} \; [\mathrm{d}k_i] \;   \delta(\tau-\mathcal{T}_0(n)) \;\text{Eik}(p_a,p_b,\{k_1,\dots,k_n\})\,.
\label{eq:soft-function-Sn-def}
\end{equation}
In eq.~\eqref{eq:soft-function-Sn-def}, $ \mathcal{N}_{\text{sym}}$ is a
symmetry factor to account for identical particles in the final state,
$[\mathrm{d}k]=\mathrm{d}^dk/(2\pi)^{d-1}\delta(k^2)\theta(k^0)$, and the
eikonal function is defined as follows
\begin{equation}
 {\rm Eik}  ( p_a,p_b,\{k_1,\dots,k_n\} )= \lim_{\lambda \to 0}^{} \lambda^{2n}
    \frac{|{\cal M}(p_a,p_b,\{\lambda k_1,\dots,\lambda k_n\})|^2}{|{\cal M}(p_a,p_b)|^2} \,,
    \label{eq:generic-eikonal-def}
\end{equation}
which corresponds to the leading soft approximation of the matrix element ${\cal
M}$.\footnote{In eq.~\eqref{eq:generic-eikonal-def} $|{\cal M}|^2$ refers to the
matrix element squared summed over colors and polarizations of all particles
involved.} Each eikonal function receives virtual corrections which also have to
be computed in the leading soft approximation; it follows that at leading order
$S_\tau^{(n)} \sim \alpha_s^{n}$.

\begin{figure*}[h]
\centering
\begin{align*}
 \includegraphics{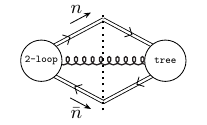}
 \includegraphics{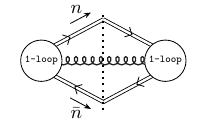}
 \includegraphics{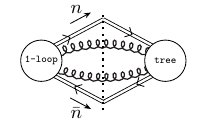}
 \includegraphics{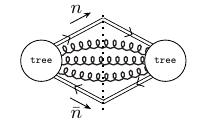}
\end{align*}
\caption{Different contributions to the zero-jettiness soft function at N3LO,
see text for details. Only contributions with final-state gluons are shown.
Diagrams to the right of the cut are complex-conjugated.}
\label{fig:diagsSF}
\end{figure*}

Hence, to obtain the N3LO contribution to the zero-jettiness soft function, one
requires ingredients that are familiar from computations of partonic cross
sections at that perturbative order. They include the radiation of three soft
partons, the one-loop corrections to the double-real emission, and the two-loop
corrections to the single-real emission. The notable difference to ordinary N3LO
computations is that the three-loop (purely virtual) corrections become
scaleless in the soft limit and vanish. For the case when soft partons are
gluons, the relevant contributions are shown in fig.~\ref{fig:diagsSF}. The two
non-vanishing virtual corrections (RRV and RVV) have been computed in
refs.~\cite{Chen:2020dpk,Baranowski:2022khd}. In the RRV case both phase-space
and loop integrations were treated on the same footing. In contrast, the RVV
contribution was computed directly by integrating
eq.~\eqref{eq:soft-function-Sn-def} for $n=1$. Such an integration is
straightforward thanks to the simple form of the single-soft current for two
hard partons at two loops~\cite{Duhr:2013msa,Li:2013lsa,Badger:2004uk}. We
discuss the calculation of the RVV contribution in
appendix~\ref{sec:RVVcorrections}.

In this paper, we elaborate on the computation of the triple real-emission
contribution, the most challenging part of the calculation of the zero-jettiness
soft function at N3LO. We begin with the explanation of how the four-momenta of
soft and hard partons are parametrized in this case. Since we aim at simplifying
the zero-jettiness measurement function and since this function depends on
projections of momenta $k_i$ on $p_{a,b}$, it is convenient to introduce
light-cone (Sudakov) coordinates. The two hard momenta $p_{a,b}$ define two
light-cone directions. We write
\begin{align}
p_a = \frac{\sqrt{s_{ab}}}{2} n,\;\;\;\;\; p_b = \frac{\sqrt{s_{ab}}}{2} \bar{n},
\end{align}
where $s_{ab}=2p_a\cdot p_b$. It follows that $n^2=\bar{n}^2=0$ and $n\cdot
\bar{n}=2$. The momenta of soft partons read
\begin{equation}
k_{i}^\mu = \frac{\alpha_i}{2} n^\mu + \frac{\beta_i}{2} \bar n^\mu + k_{\perp,i}^\mu,\;\;\;\; i = 1,2,3.
\label{eq:ki_LCD}
\end{equation}
We note that since $k_i^2=0$, $k_{\perp,i}^\mu=\sqrt{\alpha_i\beta_i} \,
e_{\perp,i}^\mu$, with $e_{\perp,i} \cdot n = e_{\perp,i} \cdot \bar n = 0$, and
$(e_{\perp,i})^2=-1$. It follows that $\alpha_i=k_i\cdot \bar{n}$ and
$\beta_i=k_i\cdot n$. We use the above definitions, take $P=\sqrt{s_{ab}}$, and
re-write eq.~\eqref{eq:zero-jettiness-def} as
\begin{equation}
   {\cal T}_0(n) = \sum \limits_{i=1}^{n} {\rm min}
   \left [ \alpha_i,\beta_i \right ].
\end{equation}
As can be seen from eq.~\eqref{eq:ki_LCD}, the minimum function differentiates
between soft partons emitted into the forward and the backward hemisphere,
defined with respect to the collision axis. In order to project the different
phase-space regions onto a unique value of ${\cal T}_0(n)$ without the minimum
function, we insert a partition of unity for each of the soft partons and write
\begin{equation}
1 = \prod \limits_{i=1}^{3} \big[ \theta(\alpha_i-\beta_i) + \theta(\beta_i-\alpha_i) \big].
\label{eq:partition-of-unity-theta}
\end{equation}
Upon expanding the product, eq.~\eqref{eq:partition-of-unity-theta} splits into
eight terms. These terms can be arranged in two groups. In the first group (two
terms) all soft partons are emitted into the same hemisphere whereas in the
second group two partons are emitted into the same hemisphere and one parton
into the opposite hemisphere (six terms). Assuming that the eikonal function is
symmetric under the permutation of soft partons,\footnote{This is the case for
three soft gluons, in case of soft gluon plus soft quark-antiquark pair emission
we symmetrize the integrand.} and under the exchange of forward and backward
directions ($n\leftrightarrow \bar{n}$), we only need to consider two of the
eight phase-space configurations, which we refer to as ``$nnn$'' and
``$nn\bar{n}$''. The phase-space measures for the two cases read\footnote{We
note that we have set $\tau=1$ in
eqs.~(\ref{eq:nnn-PS-TTT},\ref{eq:nnnbar-PS-TTT}) and in what follows. The
dependence on $\tau$ can be restored by means of simple dimensional analysis,
whenever required.}
\begin{align}
\dFtttA ={} & \frac1{\normNepTrip} \left( \prod \limits_{i=1}^{3} [\dm k_i] \right) \times \delta(1-\beta_{123}) \times \left( \prod \limits_{i=1}^{3} \theta(\alpha_i-\beta_i) \right)  \,, \label{eq:nnn-PS-TTT} \\
\dFtttB ={} & \frac1{\normNepTrip} \left(  \prod \limits_{i=1}^{3} [\dm k_i] \right) \times \delta(1-\beta_{12}-\alpha_3) \times \left( \prod \limits_{i=1}^{2} \theta(\alpha_i-\beta_i) \right) \times \theta(\beta_3-\alpha_3) \,. \label{eq:nnnbar-PS-TTT}
\end{align}
We assumed that in the $nn\bar n$ case, c.f.\ eq.~\eqref{eq:nnnbar-PS-TTT},
partons $1,2$ are emitted into the same hemisphere and parton $3$ into the
opposite one. Furthermore, we have introduced a normalization factor
\begin{align}
  \normNep
  = \frac{\Omega^{(d-2)}}{4(2\pi)^{d-1}}
  = \frac{(4\pi)^\ep}{16\pi^2\Gamma(1-\ep)} \,.
    \label{eqn:N-ep-def}
\end{align}
The N3LO triple-real emission contribution to
eq.~\eqref{eq:soft-function-Sn-def} is then written as
\begin{equation}
S_\tau^{(3)}\bigg\rvert_{\alpha_s^3} \equiv S_\tau^{RRR} = \normNepTrip \left[ 2\left(S_{nnn}^{ggg} +  S_{nnn}^{gq\bar{q}} \right) + 6\left(S_{nn\bar{n}}^{ggg} + S_{nn\bar{n}}^{gq\bar{q}} \right) \right] \,.
\label{eq:def-RRR-channel-config}
\end{equation}
This leaves us with four quantities to compute; they read
\begin{align}
  S_{nnn}^{ggg}
  & = \frac1{3!} \int \dFtttA
    \Big[ |J_{ggg}(n,\bar n, k_1,k_2,k_3)|^2
    + ( n \leftrightarrow \bar n )
    \Big], \label{eq:def-Snnn-ggg} \\
  S_{nn\bar{n}}^{ggg}
  & = \frac1{3!} \int \dFtttB
    \Big[ |J_{ggg}(n, \bar n, k_1,k_2,k_3)|^2
    + ( n \leftrightarrow \bar n )
    \Big], \label{eq:def-Snnnbar-ggg}
  \\
  S_{nnn}^{gq\bar{q}}
  & = \int \dFtttA
    \Big[ |J_{gq\bar{q}}(n,\bar n, k_1,k_2,k_3)|^2 + ( n \leftrightarrow \bar n ) \Big],
    \label{eq:def-Snnn-gqqbar} \\
  S_{n n \bar n}^{gq\bar{q}}
  & = \frac1{3} \int \dFtttB
    \Big[ \Big(|J_{gq\bar{q}}(n, \bar{n}, k_1,k_2,k_3)|^2+ (k_2\leftrightarrow k_3) +  (k_1\leftrightarrow k_3) \Big)
    + \Big( n \leftrightarrow \bar n \Big)
    \Big]. \label{eq:def-Snnnbar-gqqbar}
\end{align}
The eikonal functions $J_{ggg}(n,\bar{n},k_1,k_2,k_3)$ and
$J_{gq\bar{q}}(n,\bar{n},k_1,k_2,k_3)$ describe the soft limit $k_1\sim k_2\sim
k_3\to 0$ of the tree level matrix elements squared of the processes
$f_a(p_a)+f_b(p_b) \to X + g(k_1) + g(k_2) + g(k_3) $ and $f_a(p_a)+f_b(p_b) \to
X + g(k_1) + q(k_2) + \bar{q}(k_3) $, respectively, where $X$ is an arbitrary
color-neutral state. These quantities are known; the three-gluon eikonal
function was computed in ref.~\cite{Catani:2019nqv} and the eikonal function for
the $gq\bar{q}$ emission can be found in
refs.~\cite{DelDuca:2022noh,Catani:2022hkb}.

We have re-calculated both eikonal functions for the required case of two hard
partons. To do this, we generated all required diagrams for the above processes
using \texttt{DIANA} \cite{Tentyukov:1999is}, which internally calls
\texttt{QGRAF}~\cite{Nogueira:1991ex}, employed the ``soft-gluon
rules''~\cite{Catani:1999ss} and manipulated resulting expressions using
\texttt{FORM}~\cite{Vermaseren:2000nd,Kuipers:2012rf,Kuipers:2013pba,Ruijl:2017dtg}.
Complete agreement with the results in the
literature~\cite{Catani:2019nqv,DelDuca:2022noh,Catani:2022hkb} was found.

The eikonal functions that need to be integrated over zero-jettiness phase space
contain the following inverse propagators
\begin{align}
1/D_i \in \big\{ q \cdot k_{i} , q \cdot k_{ij} , q \cdot k_{123} , k_{ij}^2~, k_{123}^2 \big\} ,~~ q=n,\bar{n} ,~~ i \neq j = 1,2,3 \,,
\label{eq:eikonal-props}
\end{align}
where $k_{ij} = k_i + k_j$ and $k_{123} = k_1+k_2 +k_3$. In
section~\ref{sec:IBP} we explain how to use the integration-by-parts
technology~\cite{Tkachov:1981wb} to express integrals in
eqs.~(\ref{eq:def-Snnn-ggg}-\ref{eq:def-Snnnbar-gqqbar}) through a smaller set
of master integrals. However, before doing this, we will discuss a peculiar
issue that we encountered while working on the computation of the soft function,
namely the existence of integrals that are not regulated dimensionally.

\FloatBarrier
\section{Integrals  not regulated dimensionally}
\label{sec:unregDiv}

In refs.~\cite{Baranowski:2021gxe,Baranowski:2022khd} we have pointed out that
some master integrals required for the triple-real contribution to the N3LO
zero-jettiness soft function \emph{are not regulated
dimensionally}.\footnote{Note that this problem did not arise for the
real-emission contributions with \emph{two} real partons both at NNLO and
N3LO~\cite{Baranowski:2020xlp,Baranowski:2024ene}.} Our calculation is not the
first one to face this problem in QCD perturbation
theory~\cite{Collins:1992tv,Li:2016axz,Chen:2024mbk}. The standard way to treat
this problem is to introduce an analytic regulator into the integration measure
for all integrals. Unfortunately, the new regulator -- which appears alongside
with the dimensional one -- makes the required computations significantly more
complex, so that finding ways to avoid ill-defined integrals becomes important.

Our experience with computing the N3LO real-emission master integrals for the
soft function indicates that several conditions need to be satisfied for such
integrals to become unregulated. In particular, it is necessary that one of the
Sudakov variables is small, and the other one is large, that their product is
${\cal O}(1)$ and that it appears in one of the propagators. This can be
understood from the fact that the integration measure $\dm \Phi$ depends on the
dimensional regularization parameter through a factor $\prod \limits_{i=1}^{3}
(\alpha_i \beta_i)^{-\ep}$. Obviously, if $\alpha_i \sim 1/\alpha_j$, $\beta_{i}
\sim 1 / \beta_{j}$ or $\alpha_i \sim 1/\beta_j$, the dimensional regulator
becomes ineffective.

To understand if such scalings lead to integrals with a non-vanishing support,
we can assume, without loss of generality, that the smallest variable is
$\beta_1 \sim \lambda \to 0$ and that the largest variable is either $\alpha_1
\sim \lambda^{-1}$ or $\alpha_2 \sim \lambda^{-1}$, or $\beta_2 \sim
\lambda^{-1}$. We are interested in finding integrals where the $\lambda \to 0$
limit is non-trivial, and leads to non-vanishing integrals that are not
regulated dimensionally.

Consider the case $\beta_1\sim\lambda$ and $\alpha_1\sim 1/\lambda$, with
$\lambda \to 0$. Inverse propagators with non-trivial dependence on $\alpha_1$
and $\beta_1$ become
\begin{equation}
\alpha_1 + x \cdots \to \alpha_1,\;\;
\beta_1 + x \to x,\;\;
  k_{12}^2\to \alpha_1 \beta_2,
  \quad
  k_{13}^2 \to \alpha_1 \beta_3,
  \quad
  k_{123}^2 \to \alpha_1( \beta_2 + \beta_3),
\end{equation}
where $x$ denotes a generic ${\cal O}(1)$ quantity composed of Sudakov
parameters of the soft partons. We also drop $\beta_1$ from the
$\delta$-function constraining the zero-jettiness. It follows that integrations
over $\alpha_1$ and $\beta_1$ factorize, which implies that the region with
$\beta_1 \sim \lambda, \alpha_1 \sim \lambda^{-1}$ has no support.

We continue with the region $\beta_1\sim \lambda, \beta_2\sim \lambda^{-1}$,
where the following relations for inverse propagators hold
\begin{equation}
\beta_2 + x \cdots \to \beta_2,
\;\; 
\beta_1 + x \to x,
\;\;
k_{12}^2\to \beta_2 \alpha_1,
  \quad
  k_{23}^2 \to \beta_2 \alpha_3,
  \quad
  k_{123}^2 \to \beta_2( \alpha_1 + \alpha_3).
\end{equation}
Similar to the previous case, the factorization of $\beta_1$ and $\beta_2$
integrations implies that this region has no support.

For the remaining option $\beta_1\sim \lambda, \alpha_2\sim 1/\lambda$, the main
difference compared to the previous cases is that the propagator
\begin{equation}
k_{12}^2
= \alpha_1 \beta_2 + \alpha_2 \beta_1 - 2 \sqrt{\alpha_1 \beta_1 \alpha_2 \beta_2} \cos \phi_{12},
\label{eq3.3}
\end{equation}
does not simplify, because all terms on the right-hand side of eq.~\eqref{eq3.3}
are of the same order. Since we are interested in the region $\beta_1 \ll
\alpha_1$ and $\beta_2 \ll \alpha_2$, partons $1$ and $2$ are emitted into the
same hemisphere. Incorporating the zero-jettiness constraints for the two
partons, $\theta(\alpha_1 - \beta_1)\theta(\alpha_2-\beta_2)$, we change the
integration variables $\beta_{1} \to r_{1} \alpha_{1}, \alpha_{2} \to
\beta_{2}/r_{2} $, with $r_{1,2} \in [0,1]$.

As we already mentioned, the critical new element in this region is the inverse
propagator $k_{12}^2$ which does not simplify. This propagator depends on the
relative angle between directions of the transverse components of the
four-vectors $k_{1,2}$. To perform the integration over $k_{1,2}$, we write
\begin{equation}
  \label{eq:k21toAng}
  k_{12}^2 = \alpha_1 \beta_2 \left( 1 + \frac{r_1}{r_2} \right) \left( \rho \cdot \rho_{12} \right),
\end{equation}
where we have introduced two $(d-1)$-dimensional vectors $\rho$ and $\rho_{12}$
such that $\rho^2=0$ and $\rho_{12}^2 = 1- 4r_1 r_2 / \left( r_1+r_2\right)^2$.
The required integral over angles is then easily computed using
eq.~\eqref{eq:In1def}. We find
\begin{equation}
  \label{eq:angIntk1k2}
X_n  =  \int
  \frac{\dm \Omega_1^{(d-2)}}{\Omega_{(d-2)}}
  \frac{\dm \Omega_2^{(d-2)}}{\Omega_{(d-2)}}
  \frac{1}{\left( k_{12}^2 \right)^n} = \frac{I_{d-2;n}^{(1)}\left(\rho_{12}^2 \right)}{\alpha_1^n  \beta_2 ^n \left( 1 + \frac{r_1}{r_2} \right)^n}, 
\end{equation}
where $I_{d-2;n}^{(1)}\left( \rho^2 \right)$ can be extracted from
eq.~\eqref{eq:angInt-1-m}. Using eq.~\eqref{eq:f21tr69quad}, the result can be
simplified, and we obtain
\begin{equation}
  \begin{split}
\label{eq:xi1xi2AngIntForNu}
  X_n & = 
  \frac{1}{(\alpha_1 \beta_2)^n} \Big [
 \theta\left(r_2 - r_1 \right)
    {}_2F_1\left( n,n+\ep;1-\ep; \frac{r_1}{r_2}\right)
    \\
    & \phantom{= {}} +
   \left( \frac{r_2}{r_1} \right)^n \theta\left(r_1 - r_2 \right)
    {}_2F_1\left( n,n+\ep;1-\ep; \frac{r_2}{r_1}\right)
    \Big ].
    \end{split}
\end{equation}
Next, we need to integrate over $r_1,r_2$. The generic integral reads
\begin{equation}
\begin{split}
  \label{eq:Jxi1xi2}
  \Xi_{a,b}^{(n)}  & = \int\limits_0^1 \dm r_2 \int\limits_0^{r_2} \dm r_1 \, r_1^{a -\ep} r_2^{b+\ep}
                     {}_2F_1\left( n,n+\ep;1-\ep; \frac{r_1}{r_2}\right)
                     \\
                   & \phantom{= {}} + \int\limits_0^1 \dm r_1 \int\limits_0^{r_1} \dm r_2 \, r_1^{a -\ep} r_2^{b+\ep}
                     \left(\frac{r_2}{r_1}\right)^n  
                       {}_2F_1\left( n,n+\ep;1-\ep; \frac{r_2}{r_1}\right)
                      ,
\end{split}
\end{equation}
where $a$ and $b$ are integers, and powers of $\ep$ arise from the integration
measure after the variable transformation. We change the integration variables
as $r_1 = z y, r_2 = y$, and $r_2 = z y$, $r_1 = y$ in the first and the second
integral, respectively. We find
\begin{equation}
\Xi^{(n)}_{a, b} 
= \int \limits_{0}^{1} \frac{\dm y}{y} \; y^{a+b+2} 
\int \limits_{0}^{1} \dm z 
\left ( z^{a-\ep} + z^{b+\ep + n} 
\right ) 
{}_2F_1(n,n+\ep;1-\ep, z).
\label{eq3.8}
\end{equation}
The integral over $z$ is regulated dimensionally, but the integral over $y$ is
not. In fact, the integral exists for $a+b+2 > 0$ but not for other values.

To obtain a \emph{fully-regulated integral}, we need an additional regulator. We
do that by introducing a factor $\beta_1^\nu \beta_2^\nu \beta_3^\nu$ into the
measure for $nnn$ integrals and $\beta_1^\nu \beta_2^\nu \alpha_3^\nu$ for $nn
\bar n$ integrals. Of course, in the above discussion only the $\beta_1^\nu$
factor is relevant, but since we aim at modifying the measure in such a way that
\emph{all} integrals are regulated, momenta components of all soft partons must
appear. Since $\beta_1^\nu = \left( r_1\alpha_1\right)^\nu$, the computation
proceeds unaffected, except that in eqs.~(\ref{eq:Jxi1xi2},\ref{eq3.8}) $a$
becomes $a + \nu$ and the potential divergences for $2 + a + b \le 0$ are
regulated. The integral in eq.~\eqref{eq3.8} with $a \to a+\nu$ evaluates to
\begin{equation}
\begin{split}
  \label{eq:IntXi1Xi2res}
  \Xi^{(n)}_{a+\nu,b}
  & =
  \frac{{}_3F_2\left( n,n+\ep,1+a+\nu -\ep;1-\ep,2+a+\nu-\ep;1 \right)}{(1 + a+\nu-\ep) (2 + a + \nu + b)}
 \\
  & \phantom{= {}} +
  \frac{{}_3F_2\left( n,n+\ep,1+b+n+\ep;1-\ep,2+b+n+\ep;1 \right)}{(1 + b+\ep+n) (2 + a+\nu + b)}.
\end{split}
\end{equation}
The $1 / \nu$ pole arises for $2+a+b = 0$.

We note in passing that even if the equation $a+b =-2$ is satisfied, this does
not immediately imply that a particular integral has a $1/\nu$ divergence. The
reason for this is that such a divergence may be multiplied by an unconstrained
(scaleless) integral over another Sudakov parameter, or that the residue of
$\Xi_{a+\nu,b}^{(n)}$ at $\nu = 0$ vanishes.

 For the particular choice of small and large parameters that we discussed
above, the first option may occur because of the integration over $\alpha_1$.
Hence, unregulated integrals in this case \emph{must} involve denominators of
the form $\alpha_1+x$. The master formula for such integrals reads
\begin{equation}
  \label{eq:al1InfInt}
  \int\limits_0^\infty \frac{\dm \alpha_1}{\alpha_1} \frac{\alpha_1^{n_1}}{(\alpha_1 + x)^{n_2}} = \frac{\Gamma\left( n_1 \right)\Gamma\left( n_2-n_1 \right)}{\Gamma\left( n_2 \right)}
  x^{n_1-n_2},
\end{equation}
where $x$ stands for other ${\cal O}(1)$ Sudakov parameters. Note that in our
example, $x$ in eq.~\eqref{eq:al1InfInt} can only be $\alpha_3$, which shows
that one needs at least three partons for the unregulated term to occur.

\begin{table}[ht]
    \centering
    \begin{tabular}{|c||c|c|}
        \hline
        \multirow{2}*{$D_{i}$} & \multicolumn{2}{c|}{$\beta_{1} \sim \lambda$, $\alpha_{2} \sim \lambda^{-1}$}
        \\
        \cline{2-3} & integrand & scaling
        \\
        \hline \hline
        $k_{1 2}^{2}$ & $k_{1 2}^{2}$ & $1$
        \\
        $k_{1 3}^{2}$ & $\alpha_{1} \beta_{3}$ & $1$
        \\
        $k_{2 3}^{2}$ & $\alpha_{2} \beta_{3}$ & $\lambda^{-1}$
        \\
        $k_{1 2 3}^{2}$ & $\alpha_{2} \beta_{3}$ & $\lambda^{-1}$
        \\
        \hline
        $\alpha_{1} + \alpha_{2}$ & $\alpha_{2}$ & $\lambda^{-1}$
        \\
        $\alpha_{1} + \alpha_{3}$ & $\alpha_{1} + \alpha_{3}$ & $1$
        \\
        $\alpha_{2} + \alpha_{3}$ & $\alpha_{2}$ & $\lambda^{-1}$
        \\
        $\alpha_{1} + \alpha_{2} + \alpha_{3}$ & $\alpha_{2}$ & $\lambda^{-1}$
        \\
        \hline
        $\beta_{1} + \beta_{2}$ & $\beta_{2}$ & $1$
        \\
        $\beta_{1} + \beta_{3}$ & $\beta_{3}$ & $1$
        \\
        $\beta_{2} + \beta_{3}$ & $\beta_{2} + \beta_{3}$ & $1$
        \\
        $\beta_{1} + \beta_{2} + \beta_{3}$ & $\beta_{2} + \beta_{3}$ & $1$
        \\
        \hline
        $\alpha_{1}$ & $\alpha_{1}$ & $1$
        \\
        $\alpha_{2}$ & $\alpha_{2}$ & $\lambda^{-1}$
        \\
        $\alpha_{3}$ & $\alpha_{3}$ & $1$
        \\
        \hline
        $\beta_{1}$ & $\beta_{1}$ & $\lambda$
        \\
        $\beta_{2}$ & $\beta_{2}$ & $1$
        \\
        $\beta_{3}$ & $\beta_{3}$ & $1$
        \\
        \hline
    \end{tabular}
    \caption{Scalings of all possible propagators that appear in the soft
      function. All other configurations can be obtained through permutations of soft
      momenta $k_{i}$. To classify an integral, one can simply replace all inverse
      propagators with their $\lambda$-scalings and read off the overall factor
      $\lambda^{p}$. Integrals with $p > 0$ are well-defined without the regulator,
      while those with $p \le 0$ are ill-defined. Integrals with $p = 0$ are
      potentially $1 / \nu$-divergent which can be checked directly without much
      effort.}
    \label{tab:divergent-integral-scaling}
\end{table}

The above analysis can be used to identify integrals that are not regularized
dimensionally, and to remove all IBP relations that involve them from the system
of equations that one employs to find master integrals. We refer to this
procedure as \emph{filtering}. To determine affected integrals, we find all
integrals that contain at least one propagator $1/k_{i j}^{2}$, make sure that
$\theta$-functions' constraints are such that partons $i$ and $j$ are emitted
into the same hemisphere, consider the two cases $\beta_i \sim 1/\alpha_j \to 0$
and $\beta_j \sim 1/\alpha_i \to 0 $, expand remaining propagators assuming that
all other Sudakov parameters are ${\cal O}(1)$ and determine parameters $n, a$
and $b$ that appear in the function $\Xi^{(n)}_{a, b}$. If for a particular
integral $a+b +2 \le 0$, it is declared to be ill-defined without the regulator
and all equations that contain such an integral are removed the system of IBP
equations. If, however, the condition $a+b +2 >0$ holds, all equations that
contain such integrals are retained and used in the course of the reduction to
master integrals. This procedure can be conveniently summarized as a set of
power-counting rules for the denominators listed in
table~\ref{tab:divergent-integral-scaling}. We emphasize that, for this
analysis, it is important to derive the IBP relations keeping the analytic
regulator non-vanishing, $\nu \ne 0$, since an ill-defined integral may have a
coefficient that is proportional to $\nu$. Finally, we note that the redundancy
of IBP equations and a very specific set of conditions that integrals must
fulfill to be ill-defined, ensures that only a small set of $\nu$-dependent
master integrals\footnote{Furthermore, only divergent parts of those integrals
are needed.} is needed at the end of the calculation. Such master integrals are
explicitly computed in section~\ref{sec:divergent-integral}.

\section{Integral reduction in the presence of theta functions and additional regulators}
\label{sec:IBP}

It is well-known~\cite{Anastasiou:2002yz} that one can simplify the calculation
of phase-space integrals by mapping them onto loop integrals and treating them
using conventional multi-loop methods. In this section, we explain how to do
this for the integrals of the eikonal functions in
eqs.~(\ref{eq:def-Snnn-ggg}-\ref{eq:def-Snnnbar-gqqbar}). We would like to
express integrals that are needed for the computation of the soft function
through a smaller number of independent ``master integrals''. Computational
methods that we use to achieve that have been discussed in
ref.~\cite{Baranowski:2021gxe}; we repeat them here for completeness.

Reduction to master integrals simplifies the computation in multiple ways. On
the one hand, it minimizes the number of integrals that we need to calculate; on
the other hand, it also decreases the complexity of the calculation because
master integrals are often simpler than the original ones. Furthermore, the
reduction establishes algebraic relations between integrals offering the
flexibility in choosing which integrals to compute and which to obtain from such
relations. Also, availability of the reduction enables useful crosschecks by
comparing explicit computation of integrals with their expressions through
master integrals. Finally, a working reduction is the prerequisite for deriving
differential equations satisfied by master integrals, see section~\ref{sec:DEs}.

The reduction to master integrals follows Laporta's
algorithm~\cite{Laporta:2001dd} and proceeds in the following way. One derives a
large-enough system of linear relations among the integrals relevant for the
problem at hand using integration-by-parts (IBP)
identities~\cite{Tkachov:1981wb} and symmetries of the integrals. Then,
introducing a measure to order integrals in complexity, one solves the
homogeneous linear system using the Gauss' elimination method, expressing many
complex integrals, in terms of a few simpler integrals. These remaining
integrals are called ``master integrals''.

We note that the general approach described above is well-known and broadly used
for perturbative calculations in quantum field theory. However, it proved to be
very challenging to realize it for integrals that are needed to compute the soft
function. Below we describe the challenges and explain how they are overcome.

\paragraph{IBP relations}  
Linear relations between integrals are obtained using the IBP
technology~\cite{Tkachov:1981wb}, based on the observation that for properly
regularized integrals the following formula is valid
\begin{equation}
\label{eq4.1}
0 = \int  \prod_i \mathrm{d}^d k_i 
 \; \frac{\partial}{\partial k_j^\mu} \,  \left ( v^\mu \, f(\{k\},\{p\})  \right ) \, .
\end{equation}
In the above equation, $v$ is either one of the loop momenta $k_i$ or one of the
external momenta $p_i$, and the integrand $f$ is a product of propagators (see
eq.~\eqref{eq:eikonal-props}) raised to arbitrary powers, $\delta$-functions
that ensure that partons are on-shell and that the zero-jettiness has a definite
value, and $\theta$-functions that allow us to resolve the zero-jettiness
constraint.\footnote{ As we already mentioned, not all integrals that are needed
for computing the soft function are regularized dimensionally. However, once the
analytic regulator is employed, IBP relations are applicable. } In principle,
computing derivatives under the integral sign in eq.~\eqref{eq4.1} is
straightforward and, once this is done, that equation provides algebraic
relations between different integrals that we seek to exploit. In practice,
there is a problem of differentiating $\delta$- and $\theta$-functions that we
now discuss.
 
We deal with all $\delta$-functions by employing the reverse unitarity
idea~\cite{Anastasiou:2002yz,Anastasiou:2003ds}, which amounts to writing
\begin{align}
\delta(g(x)) = \lim_{\sigma\to 0} \, \frac{i}{2\pi} \left[ \frac{1}{g(x) + i \sigma} - \frac{1}{g(x) - i \sigma} \right] \equiv \left[ \frac1{g(x)} \right]_{\text{c}}.
\end{align}
The ``cut propagators'' appearing on the right-hand side of the above equation
can be easily accommodated into the IBP technology.

It is useful to classify appearing integrals in terms of the integral families,
defined by complete and linearly-independent sets of generalized propagators
with respect to algebraic relations that involve scalar products of all
four-vectors in the problem. Such a classification allows one to re-write scalar
products that appear in the numerators of phase-space integrals uniquely through
the denominators, and to represent integrals in a compact way using denominators
raised to positive or negative powers. We will explain shortly how the integral
families for the computation of the soft function are constructed.

However, before doing that, we discuss how the $\theta$-functions, that
determine a hemisphere into which a particular soft parton is emitted, are
incorporated into the IBP formalism. Schematically, modified IBP relations can
be written as follows
\begin{equation}
\begin{split}
0 ={} & \int \left( \prod_i \mathrm{d}^d k_i \right) ~ \frac{\partial}{\partial k_j^\mu} \,  v^\mu \, f \, \theta\left[ g \right]  \\
= {} &\int \left( \prod_i \mathrm{d}^d k_i \right) \bigg\{ d \;\delta_{v,k_j} \; f \; \theta[g]  + \theta[g] \; v^\mu \frac{\partial}{\partial k_j^\mu} f + f \;\delta\!\left[ g  \right]  v^\mu  \frac{\partial}{\partial k_j^\mu} g \bigg\} \,.
\label{eq:mIBP}
\end{split}
\end{equation}
The three terms that appear on the right-hand side of the above equation are
quite different. In the first two terms the original $\theta$-function is
unaffected, which means that they would not change even if the $\theta$-function
constraint was removed from the integrand. The last term in eq.~\eqref{eq:mIBP},
where the $\theta$-function has turned into a $\delta$-function with the same
argument, is the contribution of the boundary that now occurs at finite values
of partons' momenta.

Making use of this observation, we derive IBP relations that connect integrals
with a certain number of $\theta$- and $\delta$-functions to integrals with the
same number of $\theta$- and $\delta$-functions, \emph{as well as} integrals
where the number of $\theta$-functions is decreased and the number of
$\delta$-functions is increased by one. Repeated application of IBP relations to
boundary-type integrals decreases the number of $\theta$-functions further,
until we arrive at integrals where all $\theta$-functions are replaced by
$\delta$-functions. Integrals with $\delta$-functions and no $\theta$-functions
close under the integration-by-parts identities thanks to reverse unitarity, so
once this stage is achieved, no new types of integrals appear. To incorporate
both $\theta$- and $\delta$-functions with arguments $\pm(\alpha_i-\beta_i)$, we
generalize the notation for the phase-space in
eqs.~(\ref{eq:nnn-PS-TTT},\ref{eq:nnnbar-PS-TTT}) and write
\begin{align}
\dFfffA ={} & \frac1{\normNepTrip} \left( \prod \limits_{i=1}^{3} [\dm k_i] \right) \times \delta(1-\beta_{123}) \times \left( \prod \limits_{i=1}^{3} f_i(\alpha_i-\beta_i) \right)  \,, \label{eq:nnn-PS-fff} \\
\dFfffB ={} & \frac1{\normNepTrip} \left(  \prod \limits_{i=1}^{3} [\dm k_i] \right) \times \delta(1-\beta_{12}-\alpha_3) \times \left( \prod \limits_{i=1}^{2} f_i(\alpha_i-\beta_i) \right) \times f_3(\beta_3-\alpha_3) \,. \label{eq:nnnbar-PS-fff}
\end{align}
The functions $f_{1,2,3}$ represent $\theta$- and $\delta$-function constraints. 

As an example, consider the IBP relation\footnote{We note that the derivative on
the left-hand side does not act on the volume differential $\dm^d k_2$.
Furthermore, recall that $\alpha_i=k_i \cdot \bar{n}$ and $\beta_i=k_i \cdot
n$.}
\begin{align}
0 ={} & \int \frac{\partial}{\partial k_2^\mu} \,  \frac{ \bar{n}^\mu   \dFtttB}{(k_1 \cdot k_2)(k_{12}\cdot\bar{n})^2(k_2\cdot n)(k_1\cdot \bar{n})} =  \int \frac{\dFtttB ~ \bar{n}_\mu}{(k_1\cdot k_2)(k_{12}\cdot\bar{n})^2(k_2\cdot n)(k_1\cdot \bar{n})} \nonumber \\
& ~~ \times \left[ - \frac{2k_2^\mu}{[k_2^2]_{\text{c}}} - \frac{n^\mu}{[1-k_{12}\cdot n-k_3\cdot \bar{n}]_{\text{c}}} - \frac{k_1^\mu}{(k_1\cdot k_2)} - \frac{\bar{n}^\mu}{(k_{12}\cdot \bar{n})} - \frac{n^\mu}{(k_2\cdot n)} \right] 
\label{eqn:IBP-example}
\\
& - (n\cdot\bar{n}) \int 
\prod \limits_{i=1}^{3} [\dm k_i] \; \frac{ \delta(1-\beta_{12}-\alpha_3) \theta(\alpha_1-\beta_1) \theta(\beta_3-\alpha_3)}{(k_1\cdot k_2)(k_{12}\cdot\bar{n})^2(k_2\cdot n)(k_1\cdot \bar{n})} \delta(k_2\cdot \bar{n} - k_2\cdot n ) \,.
\nonumber 
\end{align}
The first five terms on the right-hand side of eq.~\eqref{eqn:IBP-example} arise
when the derivative acts on either an eikonal or a cut propagator from the phase
space; we will refer to such terms as ``homogeneous''. The last --
``inhomogeneous'' -- term in eq.~\eqref{eqn:IBP-example} describes the non-zero
boundary contribution. Writing the $\delta$-function which originated from the
derivative of the $\theta$-function as a cut propagator, it is easy to see that
the resulting propagators cease being linearly independent. Although the
presence of linearly-dependent propagators is in itself not fatal for the IBP
technology, it does create multiple inconveniences because of hidden linear
dependencies between integrals which may significantly increase the complexity
of intermediate expressions.

\begin{figure}[t]
\centering
  \includegraphics[width=\textwidth]{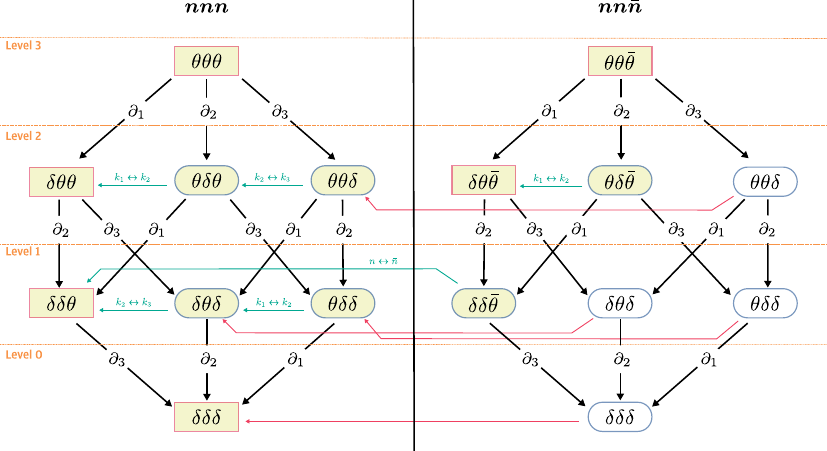}
  \caption{
   Relations between integrals.  IBP
identities relate integrals to those where one of the $\theta$-functions is replaced
by a $\delta$-function, as illustrated by  black arrows. 
Rectangular boxes represent  sets of  integrals 
which are processed using integration by parts, 
and no mapping  onto other integral families is performed. 
Oval-shaped  empty boxes are 
mapped  into the same-hemisphere configuration as indicated by red arrows.
After momenta re-naming, shown above  the
green arrows, filled oval-shaped  boxes are mapped onto unique topologies.
  }
\label{fig:ibps}
\end{figure}

To get rid of these problems, we perform a partial fraction decomposition by
multiplying the last term in eq.~\eqref{eqn:IBP-example} by
$1=\left[k_1\cdot\bar{n}+(k_2\cdot\bar{n}-k_2\cdot n) + k_2\cdot n \right] /
\left[ k_{12}\cdot\bar{n} \right]$. We arrive at\footnote{We note that the
second term in the last line of eq.~\eqref{eqn:IBP-example-PF} has the form
$\sim x\delta(x)$ and integrates to zero.}
\begin{equation}
\begin{split}
0 ={} &  \int \frac{\dFtttB }{(k_1\cdot k_2)(k_{12}\cdot\bar{n})^2(k_2\cdot n)(k_1\cdot \bar{n})}  \\
& ~~ \times \left[ - \frac{2(k_{12}\cdot\bar{n}-k_1\cdot\bar{n})}{[k_2^2]_{\text{c}}} - \frac{n\cdot\bar{n}}{[1-k_{12}\cdot n-k_3\cdot \bar{n}]_{\text{c}}} - \frac{k_1\cdot\bar{n}}{(k_1 \cdot k_2)}  - \frac{n\cdot\bar{n}}{(k_2\cdot n)} \right]  \\
& - (n \cdot \bar{n}) \int \frac{ \dFtdtB}{(k_1\cdot k_2)(k_{12}\cdot\bar{n})^3}  \left[ \frac1{(k_2\cdot n)} + \underbrace{\frac{(k_2\cdot \bar{n}-k_2 \cdot n)}{(k_2\cdot n)(k_1\cdot \bar{n})}}_{=0} + \frac1{(k_1\cdot \bar{n})} \right] \,.
\label{eqn:IBP-example-PF}
\end{split}
\end{equation}
After this step, all propagators in the two integrals in the above equations are
linearly independent, and we proceed with defining the integral families. The
primary distinction between families is the number of $\theta$- and
$\delta$-functions, and the types of propagators they contain. We show the IBP
relations and the way they connect the various types of integral families in
fig.~\ref{fig:ibps}.

\paragraph{Analytic regulator}
In section~\ref{sec:unregDiv}, we have argued that some integrals, needed to
compute the zero-jettiness soft function, are not regularized dimensionally. We
have also explained that one can regularize such integrals by introducing the
analytic regulator $\nu$. The modified measures read
\begin{align}
\dFfffAnu ={} & \frac1{\normNepTrip} \left( \prod \limits_{i=1}^{3} [\dm k_i] \right) \times \delta(1-\beta_{123}) \times \left( \prod \limits_{i=1}^{3} f_i(\alpha_i-\beta_i) \beta_i^\nu \right)  \,, \label{eq:nnn-PS-fffnu} \\
\dFfffBnu ={} & \frac1{\normNepTrip} \left(  \prod \limits_{i=1}^{3} [\dm k_i] \right) \times \delta(1-t_{123}) \times \left( \prod \limits_{i=1}^{2} f_i(\alpha_i-\beta_i) \beta_i^\nu \right) \times f_3(\beta_3-\alpha_3) \alpha_3^\nu \,, \label{eq:nnnbar-PS-fffnu}
\end{align}
where $t_{123} = \beta_{12} + \alpha_3$. A significant drawback in using the
analytic regulator is that it changes the IBP relations. To illustrate this,
consider the same IBP relation as before, but with the analytic regulator. We
find
\begin{equation}
\begin{split}
0 ={} & \int \left( \prod_i \mathrm{d}^d k_i \right) ~ \frac{\partial}{\partial k_j^\mu} \,  v^\mu \, f \, \theta\left[ g \right] (k_j\cdot q)^\nu  \\
= {} &\int \left( \prod_i \mathrm{d}^d k_i \right) \bigg\{ d \delta_{v,k_j} f \theta  + \theta v^\mu \frac{\partial}{\partial k_j^\mu} f + f \delta\!\left[ g \! \right]  v^\mu  \frac{\partial}{\partial k_j^\mu} f + \frac{\nu f \theta\left[ g \right] (v\cdot q) (k_j\cdot q)^\nu}{(k_j\cdot q)}  \bigg\} \,,
\label{eq:mIBPreg}
\end{split}
\end{equation}
where $q=n,\bar{n}$ depending on the the choice of $k_j$ and the configuration
of the integral. The last term on the r.h.s.\ of eq.~\eqref{eq:mIBPreg} is
caused by the regulator and, generally, also requires an additional partial
fraction decomposition.

\paragraph{Symmetry relations}
In addition to linear relations provided by the IBP equations, there are also
symmetry relations between integrals. In the $nnn$ case, the phase space is
symmetric under the re-labeling of partons' momenta. In the $nn\bar{n}$ case,
the phase space is symmetric under the relabeling $k_1 \leftrightarrow k_2 $.

In the $nn\bar{n}$ case, integrals with $\delta$-functions can be simplified
further. For example, interchanging $k_2\leftrightarrow k_3$ and
$n\leftrightarrow \bar{n}$ leaves the $\delta\theta\bar{\theta}$ phase space
unchanged. Furthermore, there are two types of $nn\bar{n}$ integrals that can be
mapped onto the $nnn$ configuration \emph{entirely}. First, for integrals with
$f_3=\delta$, we can write
\begin{align}
\delta(1-\beta_{12}-\alpha_3)\delta(\alpha_3-\beta_3)=\delta(1-\beta_{123})\delta(\beta_3-\alpha_3) \,.
\end{align}
Second, for $f_1=f_2=\delta$, we find
\begin{equation}
\begin{split}
& \delta(1-\beta_{12}-\alpha_3)\delta(\alpha_1-\beta_1)\delta(\alpha_2-\beta_2)f_3(\beta_3-\alpha_3) \\
\stackrel{n\leftrightarrow \bar{n}}{=} {} & \delta(1-\alpha_{12}-\beta_3)\delta(\beta_1-\alpha_1)\delta(\beta_2-\alpha_2)f_3(\alpha_3-\beta_3) \\
\stackrel{\hphantom{n\leftrightarrow \bar{n}}}{=} {} & \delta(1-\beta_{123})\delta(\alpha_1-\beta_1)\delta(\alpha_2-\beta_2)f_3(\alpha_3-\beta_3).
\end{split}
\end{equation}
Hence, we only need to consider integrals of the $\theta\theta\bar{\theta}$-,
$\delta\theta\bar{\theta}$- and $\theta\delta\bar{\theta}$-type for the
configuration $nn\bar{n}$. The symmetry relations are also illustrated in
fig.~\ref{fig:ibps}; they are heavily used to simplify the reduction.

\paragraph{Details of the technical implementation}
Following the sequence of steps described below, we express the phase-space
integrals in eqs.~(\ref{eq:def-Snnn-ggg}-\ref{eq:def-Snnnbar-gqqbar}) through
fewer and less complex master integrals.
\begin{itemize}
  
\item Using the partial fraction decomposition to resolve dependencies between
  the zero-jettiness constraining $\delta$-function and the eikonal propagators,
  we map all $\theta\theta\theta$ integrals onto a set of $\mathcal{O}(100)$
  families. They are constructed in such a way that they close under IBP relations
  with the analytic regulator $\nu$.
  
\item Starting from these families, we determine all lower-level families which
  arise when $\theta$-functions turn into $\delta$-functions. Again, these
  families are constructed in such a way, that they close when the term with the
  analytic regulator in the measure is differentiated. We arrive at additional
  $\mathcal{O}(100)$ and $\mathcal{O}(200)$ integral families for the $nnn$ and
  $nn\bar{n}$ configuration, respectively. The partial fraction decompositions we
  use in this step are not unique; to ensure that these definitions suffice to
  uniquely identify integrals in terms of families in the entire IBP setup, we had
  to choose a global ordering of propagators that we keep through the entire
  calculation.

\item It is easy to derive the \emph{homogeneous} parts of the IBP relations for
  generic powers of propagators for each integral family. To obtain a complete IBP
  relation, we need to add terms that arise from the derivatives of Heaviside
  functions and measure factors raised to power $\nu$. We add these terms
  on-the-fly, while generating relations for a specific set of indices (i.e.\ for
  a particular seed integral) which makes this step rather slow. We do so, because
  inhomogeneous terms require partial fraction decomposition, using the chosen
  ordering, as well as the integral-family identification.
  
  The issue of generating sufficiently many linear relations, to have a reduction
  to the minimal set of master integrals, did play a crucial role in the
  calculation. Choosing a large-enough seed-list for complete reduction to occur,
  yet the seed-list that we could still handle with available resources, proved
  non-trivial and required a significant amount of trial and error.
  
\item We use \texttt{Feynson}~\cite{Maheria:2022dsq} to derive symmetry
  relations between integrals.
  
\item We use
  \texttt{Kira}$\oplus$\texttt{FireFly}~\cite{Maierhofer:2017gsa,Maierhofer:2018gpa,Klappert:2020nbg}
  to solve the resulting linear system of equations as a function of $d$ and
  $\nu$. Since we expect that the soft function is regular in the limit $ \nu \to
  0$, it is useful to choose master integrals in such a way, that they remain
  independent in the $\nu \to 0$ limit. Good candidates for such basis can be
  found by studying the system at $\nu = 0$ and requiring that the $\nu=0$ master
  integrals remain master integrals also in the $\nu \ne 0$ case. With this
  informed choice of basis, the reduction requires a runtime of about $10$ days on
  $32$ cores, compared to about $70$ days without this optimization.
\end{itemize}

Proceeding along the lines described above, we arrive at the following result
for the soft function\footnote{We note that not all integrals contribute to all
channels/configurations, so some of the $a_i,b_i,c_i$ are zero.}
\begin{align}
S^{ggg,gq\bar{q}}_{nnn,nn\bar{n}}
=
\sum_{i=1}^{123} a_i I^{\bcancel{k_{123}^2}}_i
+
\sum_{i=1}^{139} b_i I^{k_{123}^2}_i
+
\nu \sum_{i=1}^{4} c_i I^{1/\nu}_i + \mathcal{O}(\nu)\,,
\label{eqn:SRRR-reduced}
\end{align}
where we have taken the $\nu \to 0$ limit wherever possible, such that the
reduction coefficients $a_i$, $b_i$, $c_i$ only depend on $\ep$. Using the power
counting rules from section~\ref{sec:unregDiv}, we find four integrals
$\{I^{1/\nu}_i\}$ that contain a $1/\nu$ singularity but, thanks to the choice
of basis, they appear in the soft function with a prefactor $\nu$. Furthermore,
in eq.~\eqref{eqn:SRRR-reduced}, we have separated $\nu$-regular integrals
according to whether or not they contain the $1/k_{123}^2$ propagator.

The reason for this separation is that these three classes of integrals are
computed differently. Indeed, in the following section, we will show that direct
integration allows us to compute master integrals without the $1/k_{123}^2$
propagator $\{I^{\bcancel{k_{123}^2}}_i \}$, as well as the $1/\nu$ parts of
master integrals $\{ I^{1/\nu}_i \}$. After that, we consider the most
complicated set of integrals $\{I^{k_{123}^2}_i \}$, which we compute by
modifying these master integrals through the introduction of an auxiliary
parameter, followed by constructing and solving differential equations with
respect to this parameter that the modified master integrals satisfy.

\section{Integrals without $1/k_{123}^2$ propagator 
and $1/\nu$-divergent integrals}
\label{sec:no123ints}

In this section we discuss the calculation of master integrals without the
$1/k_{123}^2$ propagator, and the $1/\nu$-divergent integrals. These integrals
are computed by a direct integration over the light-cone coordinates of soft
partons $\alpha$ and $\beta$. All in all, 127 integrals are calculated in this
way.

\subsection{Integrals that do not need an analytic regulator}
\label{sec:no123intsNoNu}

\begin{table}[t]
  % \centering
  \renewcommand{\arraystretch}{1.5}
  \begin{tabular}{|c|ccc|c|c|}\hline
&    $f_1$ & $f_2$ & $f_3$ & Variable change &  Splitting \\\hline\hline
$\dFdttA$ &    $\delta\left( \alpha_1 - \beta_1 \right)$ & $\theta\left( \alpha_2 - \beta_2 \right)$ & $\theta\left( \alpha_3 - \beta_3 \right)$  & $\alpha_1 \to \beta_1, \alpha_2 \to \frac{\beta_2}{r_2}, \alpha_3 \to \frac{\beta_3}{r_3}$ & \textemdash\\
$\dFdttB$ &    $\delta\left( \alpha_1 - \beta_1 \right)$ & $\theta\left( \alpha_2 - \beta_2 \right)$ & $\theta\left( \beta_3  - \alpha_3 \right)$ &  $\alpha_1 \to \beta_1, \alpha_2 \to \frac{\beta_2}{r_2}, \beta_3 \to \frac{\alpha_3}{r_3}$  &  \textemdash \\\hline
$\dFtddA$ &    $\theta\left( \alpha_1 - \beta_1 \right)$ & $\delta\left( \alpha_2 - \beta_2 \right)$ & $\delta\left( \alpha_3 - \beta_3 \right)$  & $\alpha_1 \to \frac{\beta_1}{r_1}, \alpha_2 \to \beta_2,  \alpha_3 \to \beta_3$  & \textemdash \\
$\dFtdtA$ &    $\theta\left( \alpha_1 - \beta_1 \right)$ & $\delta\left( \alpha_2 - \beta_2 \right)$ & $\theta\left( \alpha_3 - \beta_3 \right)$  & $\alpha_1 \to \frac{\beta_1}{r_1}, \alpha_2 \to \beta_2, \alpha_3 \to \frac{\beta_3}{r_3}$ & $r_1 | r_3$
    \\
$\dFtdtB$&    $\theta\left( \alpha_1 - \beta_1 \right)$ & $\delta\left( \alpha_2 - \beta_2 \right)$ & $\theta\left( \beta_3  - \alpha_3 \right)$  & $\alpha_1 \to \frac{\beta_1}{r_1}, \alpha_2 \to \beta_2, \beta_3 \to \frac{\alpha_3}{r_3}$  & \textemdash \\
$\dFtttB$ &    $\theta\left( \alpha_1 - \beta_1 \right)$ & $\theta\left( \alpha_2 - \beta_2 \right)$ & $\theta\left( \beta_3  - \alpha_3 \right)$  & $\alpha_1 \to \frac{\beta_1}{r_1}, \alpha_2 \to \frac{\beta_2}{r_2}, \beta_3 \to \frac{\alpha_3}{r_3}$ & $r_1 | r_2$
    \\\hline
$\dFrelabeledtbdt$ &    $\theta\left( \beta_1 - \alpha_1  \right)$ & $\delta\left( \alpha_2 - \beta_2 \right)$ & $\theta\left( \alpha_3 - \beta_3 \right)$  & $\beta_1 \to \frac{\alpha_1}{r_1}, \alpha_2 \to \beta_2, \beta_3 \to \frac{\alpha_3}{r_3}$ & \textemdash \\
$\dFrelabeledtbtt$ &    $\theta\left( \beta_1 - \alpha_1 \right)$ & $\theta\left( \alpha_2 - \beta_2 \right)$ & $\theta\left( \alpha_3 - \beta_3 \right)$  & $\beta_1 \to \frac{\alpha_1}{r_1}, \alpha_2 \to \frac{\beta_2}{r_2}, \beta_3 \to \frac{\alpha_3}{r_3}$ & \textemdash \\\hline  
  \end{tabular}
  \caption{Summary of applied variable changes and integral splittings according
  to $\delta-$ and $\theta$-function constraints $f_i$. Splitting of 
  the integration domain into 
 $r_i < r_j$ and $r_j < r_i$ sectors 
is  required only if the  integral contains the propagator  $1/k_i\cdot k_j$.}
  \label{tab:sp2intsReg0}
\end{table}

An important property of the eikonal function is that all integrals without the
$1/k_{123}^2$ propagator, needed for computing the N3LO contribution to the soft
function, can have up to two scalar products between momenta of different soft
partons. Hence, owing to the possibility of relabelling soft momenta, we write
such integrals as
\begin{equation}
  \label{eq:intk1k2k3reg0ep}
  I_{nm} = \int \frac{\dm \Phi_{f_1 f_2 f_3}
  \; R(\{\alpha\}, \{\beta\})}{\left[ 2 k_1 \cdot k_2\right]^n \left[ 2 k_1 \cdot k_3\right]^m }
  =
  \int \prod\limits_{i=1}^3 \left(\frac{\dm \alpha_i \dm \beta_i}{\left( \alpha_i\beta_i \right)^\ep} 
    f_i(\alpha_i,\beta_i) \right) R(\{\alpha\}, \{\beta\}) \; \Omega_{nm}.
\end{equation}
In the above equation, the phase-space measure $\dm \Phi_{f_1 f_2 f_3}$ is given
in eqs.~(\ref{eq:nnn-PS-fff},\ref{eq:nnnbar-PS-fff}), and we used the Sudakov
decomposition in eq.~\eqref{eq:ki_LCD} to express all denominators other than $2
k_1 k_2$ and $2 k_1 k_3$ through $\{ \alpha \}$ and $\{\beta \}$, collecting
them into the rational function $R$. The non-trivial dependence on the relative
azimuthal angles between soft-parton momenta is encapsulated in the angular
integral $\Omega_{nm}$. It reads
\begin{equation}
  \label{eq:om123def}
  \Omega_{nm} = \frac{1}{\left(\Omega^{(d-2)}  \right)^3} \int \frac{ \dm \Omega^{(d-2)}_{k_1} \dm \Omega^{(d-2)}_{k_2} \dm \Omega^{(d-2)}_{k_3}} {\left[ 2 k_1 \cdot k_2\right]^n \left[ 2 k_1 \cdot k_3\right]^m}.
\end{equation}
To compute this integral we use the fact that we can integrate over angles of
partons $2$ and $3$ independently. Furthermore,
\begin{equation}
  \frac{1}{\Omega^{(d-2)}} \int \frac{ \dm \Omega^{(d-2)}_{k_i} }{\left[ 2 k_i \cdot k_j\right]^n} =  \frac{I_{d-2;n}^{(1)}(\rho^2_{ij})}{
    \left ( 
      \alpha_i \beta_j + \alpha_j \beta_i
    \right )^n},
\end{equation}
where $I_{d,n}^{(1)}$ can be found in 
eq.~\eqref{eq:angInt-1-m} and 
$\rho_{ij}^2 = \left( \alpha_i\beta_j - \alpha_j \beta_i
\right)^2 /\left( \alpha_i\beta_j + \alpha_j \beta_i
\right)^2$.  
It follows that 
\begin{equation}
\Omega_{nm} = \frac{I_{d-2;n}^{(1)}(\rho_{12}^2)}{\left( \alpha_1 \beta_2 + \alpha_2 \beta_1 \right)^n} \frac{I_{d-2;m}^{(1)}(\rho_{13}^2)}{\left( \alpha_1 \beta_3 + \alpha_3 \beta_1 \right)^m}.
\end{equation}

For each specific set of the jettiness constraints represented by functions
$f_i$ in eq.~\eqref{eq:intk1k2k3reg0ep}, we change integration variables
following table~\ref{tab:sp2intsReg0}. We then simplify the hypergeometric
functions appearing in functions $I^{(1)}$ by applying the transformation in
eq.~\eqref{eq:f21tr69quad}. For configurations where two soft partons are
emitted into the same hemisphere we split the integration region and introduce
new variables $r_i$, c.f.\ table~\ref{tab:sp2intsReg0}, to map the integration
regions onto the intervals $[0,1]$.

We note that the transformation eq.~\eqref{eq:f21tr69quad} that we apply to
simplify angular integrals involves the \emph{absolute value} of $\rho_{ij}$. In
principle, $\rho_{ij}$ does not need to be positive-definite, but there are
cases when it can be. Indeed, this happens if the two constraints $f_i$ and
$f_j$ are $f_if_j \in \{\delta\theta,\theta\delta,\theta\bar{\theta} \}$. Then,
using variables defined in table~\ref{tab:sp2intsReg0}, we easily find
\begin{equation}
  \left|\rho_{ij}^{\delta\theta}  \right| = \frac{1-r_j}{1+r_j},
  \quad
  \left|\rho_{ij}^{\theta\delta}  \right| = \frac{1-r_i}{1+r_i},
  \quad
  \left|\rho_{ij}^{\theta\bar{\theta}}  \right| = \frac{1-r_i r_j}{1+ r_i r_j},
\end{equation}
where $0 \le r_{i, j} \le 1$. Superscripts in the above equations are introduced
to indicate constraints on partons $i$ and $j$. However, if $f_i f_j = \theta
\theta $, so that the two partons are emitted into the same hemisphere, we find
$|\rho_{ij}| = |r_i-r_j|/(r_i+r_j)$. In this case, we have to write
\begin{equation}
  \lvert \rho_{ij} \rvert =
  \theta(r_i-r_j) \; \frac{r_i-r_j}{r_i + r_j} 
  + 
  \theta(r_j-r_i) \; \frac{r_j-r_i}{r_j+r_i},
\end{equation}
to get rid of the absolute value.

Hence, after integrating over azimuthal angles in integrals $I_{nm}$ in
eq.~\eqref{eq:intk1k2k3reg0ep}, and changing the integration variables as in
table~\ref{tab:sp2intsReg0}, we obtain integrals over variables
$\alpha_i,\beta_i,r_i$ which are at most six-dimensional. Furthermore, at least
one of the integrations can be performed by removing the $\delta$ function that
fixes the zero-jettiness value.

Calculation of several representative $I_{nm}$ integrals for the $nnn$ case were
discussed in ref.~\cite{Baranowski:2021gxe}. Below, we will consider examples of
new master integrals with three $\theta$ functions, which appear for the first
time in the $nn \bar n$ configuration.
%
% 
% INT["Tpppb119tttarm", 6, 3087, 6, 0, {1, 1, 1, 1, 0, 0, 0, 0, 0, 0, 1, 1}]
% 
%
We begin by considering an integral without  scalar products between momenta $k_i$ and $k_j$. The integral reads
\begin{align}
  \label{eq:exampleIntNNNB-ttt-noang}
  J = \int \frac{\dFtttB}{\left( k_{123} \cdot n\right) \left( k_{123} \cdot \bar{n} \right)}.
\end{align}
We choose the parametrization according to table~\ref{tab:sp2intsReg0} and note
that for this integral the splitting of integration variables $r_{1,2}$ into
$r_1 < r_2$ and $r_2 < r_1$ is not necessary. We then integrate over $\alpha_3$
to remove the zero-jettiness constraint, and change the integration variables,
$\beta_1 \to x \ y$ and $\beta_2 \to x \ (1-y)$. We obtain
\begin{equation}
  J = \int\limits_0^1 \dm x \dm y \dm r_1 \dm r_2 \dm r_3 
  \frac{(1-x)^{1-2\ep}x^{3-4\ep}r_1^{-1+\ep}r_2^{-1+\ep}r_3^{-1+\ep}(1-y)^{1-2\ep}y^{1-2\ep}}{\left(x r_2 y+ r_1\left(r_2(1-x) +x(1-y)\right)\right) \left(1-x+r_3 x\right)}.
\end{equation}
The integrations over $r_1$ and $r_3$ can be performed in terms of the
hypergeometric functions. We find
\begin{equation}
\begin{split}
\label{eq:exampleIntNNNB-ttt-nok12pt2}
  J & = \frac{1}{\ep^2} \int \limits_{0}^{1}  \dm x \dm y  \dm r_2
 (1-x)^{-2\ep}x^{2-4\ep}r_2^{-2+\ep}(1-y)^{1-2\ep}y^{-2\ep} \\
  & \phantom{= {}} \times {}_{2}F_{1} \left(1,\ep;1+\ep;-\tfrac{x}{1-x}\right) {}_{2}F_{1} \left(1,\ep;1+\ep; - \tfrac{r_2(1-x) + x(1-y)}{x r_2 y}\right).
\end{split}
\end{equation}
We note that arguments of the hypergeometric functions in the above equation
look complicated, suggesting an increased difficulty compared to integrals
needed for the $nnn$ case discussed in ref.~\cite{Baranowski:2021gxe}. However,
upon closer inspection, the singularity structure of the integrand in
eq.~\eqref{eq:exampleIntNNNB-ttt-nok12pt2} turns out to be quite simple, as
could be expected on general grounds. Indeed, since the hypergeometric functions
in eq.~\eqref{eq:exampleIntNNNB-ttt-nok12pt2} diverge logarithmically at the
integration boundaries, the singularity structure is entirely determined by the
double pole at $r_2=0$. To isolate it, we apply the transformation shown in
eq.~\eqref{eq:f21trInv1} to both hypergeometric functions. We then find
\begin{equation}
\begin{split}
\label{eq:exampleIntNNNB-ttt-nok12pt3a}
  J & =
  \frac{1}{\ep^2} \int\limits_0^1 \dm x \;  \dm y  \; \dm r_2 \; 
 (1-x)^{-\ep} x^{2-3\ep}r_2^{-2+2\ep} \left(r_2+x(1-r_2)(1-y)\right)^{-\ep} \\
  & \phantom{= {}} \times (1-y)^{1-2\ep}y^{-\ep} {}_{2}F_{1} \left(\ep,\ep;1+\ep;x\right) {}_{2}F_{1} \left(\ep,\ep;1+\ep;\tfrac{r_2+x \left(1-r_2-y\right)}{r_2+x(1-r_2)(1-y)}\right).
\end{split}
\end{equation}
To subtract the double-pole singularity at $r_2 = 0$, we need to extract the
different $r_2 \to 0$ branches that are contained in the last hypergeometric
function in eq.~\eqref{eq:exampleIntNNNB-ttt-nok12pt3a}. This can be easily done
using eq.~\eqref{eq:f21tr2zb}, and we find
\begin{equation}
J = J^{(a)}+ J^{(b)},
\end{equation}
where 
\begin{equation}
\begin{split}
\label{eq:exampleIntNNNB-ttt-nok12pt3}
  J^{(a)} & =
  \frac{ \Gamma(1-\ep)}{\Gamma(1+\ep)} \int\limits_0^1 \dm x \dm y  \dm r_2 \; 
 (1-x)^{-\ep} x^{2-3\ep}r_2^{-2+2\ep}(1-y)^{1-2\ep}y^{-\ep} \\
  & \phantom{= {}} \times \left(r_2+x(1-r_2-y)\right)^{-\ep}  {}_{2}F_{1} \left(\ep,\ep;1+\ep;x\right),
  \end{split} 
  \end{equation}
  \begin{equation}
  \begin{split} 
    J^{(b)} & =
  \frac{1}{\ep(\ep-1)} \int\limits_0^1 \dm x \dm y  \dm r_2 \; 
 (1-x)^{-\ep} x^{3-3\ep}r_2^{-1+2\ep}(1-y)^{1-2\ep}y^{1-2\ep} \\
  & \phantom{= {}} \times \frac{{}_{2}F_{1} \left(\ep,\ep;1+\ep;x\right)}{\left(r_2+x(1-r_2)(1-y)\right)}  {}_{2}F_{1} \left(1,1;2-\ep;\tfrac{x\ y \ r_2 }{\left(r_2+x(1-r_2)(1-y)\right)}\right).
  \end{split}
\end{equation}
Integral $J^{(a)}$ can be integrated over  $r_2$ using eq.~\eqref{eq:hfIntReprF21}; the result reads
\begin{equation}  
  \begin{split}
    J^{(a)}  & =
               \frac{ \Gamma (1-\ep) \Gamma (\ep) }{\ep (2 \ep-1)}
               \int\limits_0^1 \dm x \dm y (1-x)^{-\ep} x^{2-4 \ep} (1-y)^{1-3 \ep} y^{-\ep}   \\
             & \phantom{= {}}
               \times {}_2F_1(\ep,\ep;1+\ep;x)  {}_2F_1\left(\ep,-1 + 2 \ep;2 \ep;\frac{x-1}{x (1-y)}\right).
  \end{split}
  \label{eq5.14}
\end{equation}
The integrand of $J^{(b)}$ has a simple pole at $r_2=0$ which can be easily
subtracted. Finally, we expand the integrand for $J^{(a)}$ in eq.~\eqref{eq5.14}
and the subtracted integrand for $J^{(b)}$ in $\ep$ using the package
\texttt{HypExp}~\cite{Huber:2005yg,Huber:2007dx}, and integrate order by order
in $\ep$ with the help of \texttt{HyperInt}~\cite{Panzer:2014caa}. The result
for the full integral reads
\begin{equation}
  \label{eq:exampleIntNNNB-ttt-nok12-res}
  J =
 - \frac{5}{12\ep^2} - \frac{215}{72\ep}
  + \left( \frac{11}{72} \pi^2-\frac{343}{24} \right) +\ep \left(
 -\frac{1999}{36} + \frac{7}{8} \pi^2+\frac{14}{3}  \zeta_3
  \right)  + \mathcal{O}\left( \ep^2 \right).
\end{equation}
%
% 
% INT["Tpppb29tttarm", 8, 3135, 8, 0, {1, 1, 1, 1, 1, 1, 0, 0, 0, 0, 1, 1}]
% 
%

We continue with the  discussion of an integral involving  two
scalar products of different   soft-parton momenta
\begin{align}
  \label{eq:exampleIntNNNB-ttt-k13k23}
  I = \int \frac{\dFtttB}{\left( k_{13} \cdot \bar{n} \right) \left( k_{2} \cdot \bar{n} \right)\left( k_{1} \cdot k_3 \right) \left( k_{2} \cdot k_3 \right)}.
\end{align}
Proceeding as discussed earlier, we integrate over the azimuthal angles of the
three partons, change the integration variables following
table~\ref{tab:sp2intsReg0}, and integrate over $\alpha_3$ removing the
zero-jettines $\delta$ function. We find
\begin{equation}
\begin{split}
  \label{eq:exampleIntNNNB-ttt-k13k23-cube-delta}
  I & = 4 \int\limits_0^1 \dm \beta_1\dm \beta_2 \; \theta(1-\beta_{12})
    \int\limits_0^1 \dm r_1 \dm r_2 \dm r_3 \frac{\beta_1 \left( r_1 r_2 r_3 \right)^\ep}{\left( \beta_1 \beta_2 \bar \beta_{12} \right)^{1+2\ep}\left( \beta_1 + r_1 \bar \beta_{12} \right)} \\
  & \phantom{= {}} \times {}_{2}F_{1} \left(1,1+\ep;1-\ep;r_1 r_3\right) {}_{2}F_{1} \left(1,1+\ep;1-\ep;r_2 r_3\right),
\end{split}
\end{equation}
where $\bar \beta_{12} = 1- \beta_{12}$. We change the integration variable
$\beta_1 = t(1-\beta_2)$, integrate over $r_2$ using the definition of the
generalized hypergeometric function in eq.~\eqref{eq:hfIntReprFPQ}, and obtain
\begin{equation}
\begin{split}
  \label{eq:exampleIntNNNB-ttt-k13k23-cube}
 I  & =  \frac{4}{(1+\ep)} \int\limits_0^1 \dm \beta_2 \dm t \dm r_1 \dm r_3\frac{\left( 1-\beta_2 \right)^{-1-4\ep}\beta_2^{-1-2\ep}\left( 1-t \right)^{-1-2\ep} t^{-2\ep}}{r_1 + t - r_1 t} \\
 & \phantom{= {}} \times {}_{2}F_{1} \left(1,1+\ep;1-\ep;r_1 r_3\right) \; {}_{3}F_{2} \left(1,1+\ep,1+\ep;1-\ep,2+\ep; r_3\right).
\end{split}
\end{equation}
 This expression can be further integrated over $\beta_2$ and $t$. Integration
over $\beta_2$ results in $\Gamma$-functions, and integration over $t$ leads to
an ${}_2F_1$ function with the argument $1-1/r_1$, which can be simplified using
eq.~\eqref{eq:f21trInv1}. Proceeding along these lines, we obtain
\begin{equation}
\begin{split}
  \label{eq:exampleIntNNNB-ttt-k13k23-2d}
  I & = \frac{ 2 \Gamma^3(-2\ep)}{(1+\ep)\Gamma(-6\ep)}
  \int\limits_0^1\dm x \dm y \; 
  \left( \frac{y}{x}  \right)^\ep  {}_{2}F_{1} \left(1,1+\ep;1-\ep; x y\right)\\
  & \phantom{= {}} \times {}_{2}F_{1} \left(1-2\ep,-4\ep;1-4\ep; 1-x\right)
  {}_{3}F_{2} \left(1,1+\ep,1+\ep;1-\ep,2+\ep; y\right).
\end{split}
\end{equation}
Since this integral is finite, we compute it by expanding the integrand in $\ep$
and integrating the resulting expression. To do this, we use the packages
\texttt{HypExp}~\cite{Huber:2005yg,Huber:2007dx} and
\texttt{HyperInt}~\cite{Panzer:2014caa}. The final result reads
\begin{equation}
\begin{split}
  \label{eq:exampleIntNNNB-ttt-k13k23-res}
  I & =
  \frac{\pi^2}{2\ep^2}
  - \left( \frac{3}{2}\pi^2 - 18\zeta_3 \right)\frac{1}{\ep}
  + \left( \frac{9}{2}\pi^2 - 54\zeta_3 + \frac{1}{120}\pi^4 \right) \\
  & \phantom{= {}} - \ep \left(
  \frac{27}{2}\pi^2 - 162 \zeta_3 + \frac{1}{40}\pi^4 + \frac{125}{2}\pi^2\zeta_3 - \frac{837}{2}\zeta_5
  \right)  + \mathcal{O}\left( \ep^2 \right).
\end{split}
\end{equation}

The above examples demonstrate how the calculation of integrals that are
regulated dimensionally and do not contain the $1/k_{123}^2$ propagator is
performed. Although these computations are not easy and, quite often, integral
representations for $nn \bar n$ integrals look quite complex, in comparison with
their $nnn$ counterparts they have a simpler structure of singularities and
require a smaller number of subtractions before the expansion of integrands in
$\ep$ can be performed.

\subsection{$1 / \nu$-divergent integrals}
\label{sec:divergent-integral}

There are 4 master integrals that become divergent if the analytic regulator is sent to zero. They are 
\begin{align}
  I^{1 / \nu}_1
    & =
      \int \frac{\dFttdA \left( \beta_1\beta_2\beta_3 \right)^\nu}{k_{1 2 3}^{2} (k_{1} \cdot k_{2}) (\alpha_{1} + \alpha_{3}) \beta_{1}}
    ,
    \\
  I^{1 / \nu}_2
    & =
      \int \frac{\dFttdA \left( \beta_1\beta_2\beta_3 \right)^\nu}{(k_{1} \cdot k_{2}) (\alpha_{1} + \alpha_{3}) \alpha_{2} \beta_{1}}
    ,
    \\
  I^{1 / \nu}_3
    & =
      \int \frac{\dFtttB \left( \beta_1\beta_2\alpha_3 \right)^\nu}{k_{1 2 3}^{2} (k_{1} \cdot k_{2}) (\alpha_{1} + \alpha_{3}) (\beta_{1} + \beta_{2} + \beta_{3}) \beta_{1}}
    ,
    \\
  I^{1 / \nu}_4
    & =
      \int \frac{\dFtttB \left( \beta_1\beta_2\alpha_3 \right)^\nu}{(k_{1} \cdot k_{2}) (\alpha_{1} + \alpha_{3}) \alpha_{2} (\beta_{1} + \beta_{2} + \beta_{3}) \beta_{1}}.
\end{align}
The integrals $I^{1 / \nu}_{1, 2}$ have been computed previously in
ref.~\cite{Baranowski:2021gxe}; they read
\begin{align}
  I^{1 / \nu}_1
  & =
    \nu^{-1}
    \frac{\Gamma^2(-2 \ep) \Gamma(-4 \ep - 1) \Gamma(1 + 2 \ep ) }{\Gamma(-6 \ep - 1)}
    C(\ep)
    +
    \mathcal{O}(\nu^{0})
    ,
    \label{eq5.25}
  \\
  I^{1 / \nu}_2
  & =
    \nu^{-1}
    \frac{\Gamma^2(-2 \ep) \Gamma(-4 \ep) \Gamma(1 + 2 \ep)}{\Gamma(-6 \ep)}
    C(\ep)
    +
    \mathcal{O}(\nu^{0})
    \label{eq5.26}
    ,
\end{align}
where
\begin{equation}
\begin{split}
    C(\ep)
    & =
    \lim \limits_{\nu \to 0}  \; \left [ \nu \Xi_{\nu - 1,  -1}^{(1)} \right ]
    \\
    & =
    \frac{{}_{3}F_{2}(1, 1 + \ep, 1 + \ep; 1 - \ep, 2 + \ep; 1)}{1 + \ep}
    -
    \frac{{}_{3}F_{2}(1, 1 + \ep, - \ep ; 1 - \ep, 1 - \ep; 1)}{ \ep }
    .
\end{split}
    \label{eq3.25}
\end{equation}

We have discussed the origin of the $1/\nu$ singularities in
section~\ref{sec:unregDiv}, and described steps required to extract them.
Although those steps are sufficient for obtaining the $1/\nu$ poles of $I^{1 /
\nu}_{3,4}$, we find it instructive to discuss the calculation of the $1/\nu$
poles of these integrals in detail.

We begin with $I^{1 / \nu}_4$. The $1/\nu$ singularity in this case originates
from the integration region $\beta_1 \sim \alpha_2^{-1} \to 0$. We integrate
over the relative azimuthal angle between $k_{1}$ and $k_2$, and obtain
\begin{align}
  I^{1 / \nu}_4
    & =
    \int \limits_{0}^{\infty} \mathrm{d} \beta_{1} \mathrm{d} \beta_{2} \mathrm{d} \alpha_{3}
    \beta_{1}^{-2 \ep + \nu}
    \beta_{2}^{-2 \ep + \nu}
    \alpha_{3}^{-2 \ep + \nu}
    \int \limits_{0}^{1} \mathrm{d} r_{1} \mathrm{d} r_{2} \;
    r_{1}^{\ep - 2}
    r_{2}^{\ep - 2}
    \int \limits_{\alpha_{3}}^{\infty} \mathrm{d} \beta_{3}
    \beta_{3}^{- \ep}
    \nonumber
    \\
    & \phantom{= {}}
    \times
    \frac{\delta(1 - \beta_{1} - \beta_{2} - \alpha_{3})}{(\beta_{1} / r_{1} + \alpha_{3}) (\beta_{2} / r_{2}) (\beta_{1} + \beta_{2} + \beta_{3}) \beta_{1}}
    \;  2
    \Big [ \; 
        r_{2} \;
        \theta(r_{1} - r_{2})
        \\ 
        & \phantom{= {}}
        \times 
        {}_{2}F_{1} \left( 1, 1 + \ep; 1 - \ep; \frac{r_{2}}{r_{1}} \right)
        +
        r_{1}
        \theta(r_{2} - r_{1})
        {}_{2}F_{1} \left( 1, 1 + \ep; 1 - \ep; \frac{r_{1}}{r_{2}} \right)
    \Big ]
    \nonumber 
    ,
\end{align}
where $r_{i} = \beta_{i} / \alpha_{i}$ are the new integration variables.
Because of the scaling relation between $\beta_1$ and $\alpha_2$, and since
other Sudakov variables are ${\cal O}(1)$, we find $r_1 \sim \beta_1 \to 0$ and
$r_2 \sim \alpha_2^{-1} \to 0$. Approximating the integrand, we find
\begin{align}
  I^{1 / \nu}_4
    & \sim
    \int \limits_{0}^{\infty} \mathrm{d} \beta_{1} \mathrm{d} \beta_{2} \mathrm{d} \alpha_{3}
    \beta_{1}^{-2 \ep + \nu}
    \beta_{2}^{-2 \ep + \nu}
    \alpha_{3}^{- \ep + \nu}
    \int \limits_{0}^{1} \mathrm{d} r_{1} \mathrm{d} r_{2} \;
    r_{1}^{\ep - 2}
    r_{2}^{\ep - 2}
    \int \limits_{\alpha_{3}}^{\infty} \mathrm{d} \beta_{3}
    \beta_{3}^{- \ep}
    \nonumber
    \\
    & \phantom{= {}}
    \times
    \frac{\delta(1 - \beta_{2} - \alpha_{3})}{(\beta_{1} / r_{1} + \alpha_{3}) (\beta_{2} / r_{2}) (\beta_{2} + \beta_{3}) \beta_{1}}
       \; 2
    \Big[  r_{2} \theta(r_{1} - r_{2}) 
    \\
    & \phantom{= {}}
                  \times {}_{2}F_{1} \left( 1, 1 + \ep; 1 - \ep; \frac{r_{2}}{r_{1}} \right)
        +
        r_{1} \theta(r_{2} - r_{1})
        {}_{2}F_{1} \left( 1, 1 + \ep; 1 - \ep; \frac{r_{1}}{r_{2}} \right)
    \Big ]
    \nonumber 
    .
\end{align}
The integral over $\beta_1$ is easily computed using eq.~\eqref{eq:al1InfInt}.
Keeping the $1/\nu$ pole, we write
\begin{equation}
\begin{split}
    \label{eq:divergent-integral-lb2a}
  I^{1 / \nu}_4
    & = 
    2
    \Gamma(-2 \ep )
    \Gamma(1 + 2 \ep) \; 
\frac{C(\ep)}{\nu}
    \\
    & \phantom{= {}}
    \times
    \left[
        \int \limits_{0}^{1} \mathrm{d} \beta_{2} \mathrm{d} \alpha_{3}
        \beta_{2}^{-2 \ep }
        \alpha_{3}^{-3 \ep - 1}
        \delta(1 - \beta_{2} - \alpha_{3})
        \int \limits_{\alpha_{3}}^{\infty} \mathrm{d} \beta_{3}
        \beta_{3}^{- \ep}
        \frac{1}{\beta_{2} + \beta_{3}}
        \frac{1}{\beta_{2}}
    \right]
    +
    \mathcal{O}(\nu^{0})
    ,
\end{split}
\end{equation}
where we have used eq.~\eqref{eq3.25} to express $\Xi_{\nu - 1, -1}^{(1)}$ in
terms of $C(\ep)$. The last integral in eq.~\eqref{eq:divergent-integral-lb2a}
can be easily calculated. We obtain
\begin{equation}
\begin{split}
    & \phantom{= {}}
    \int \limits_{0}^{1} \mathrm{d} \beta_{2} \mathrm{d} \alpha_{3}
    \beta_{2}^{-2 \ep }
    \alpha_{3}^{-3 \ep - 1}
    \delta(1 - \beta_{2} - \alpha_{3})
    \int \limits_{\alpha_{3}}^{\infty} \mathrm{d} \beta_{3}
    \beta_{3}^{- \ep}
    \frac{1}{\beta_{2} + \beta_{3}}
    \frac{1}{\beta_{2}}
    \\
    & =
      \frac{ \Gamma(-2 \ep ) \Gamma(1 - 4 \ep )}{\ep \Gamma(1 - 6 \ep )} \;
      {}_3F_2\left(1, 1, -2 \ep ; 1 + \ep,1 - 6 \ep ; 1 \right).
\end{split}
\end{equation}
Finally, combining the various contributions and extracting the $1/\nu$ pole, we
find
\begin{align}
  I^{1 / \nu}_4
    & =
    \frac{
      \Gamma^2(-2 \ep) \Gamma(1 + 2 \ep)
       \Gamma(1 - 4 \ep )}{
      \nu \ep \Gamma(1 - 6 \ep)}
      \; C(\ep) \;
      {}_3F_2\left(1, 1, -2 \ep; 1 + \ep,1 - 6 \ep; 1 \right)
    +
    \mathcal{O}(\nu^{0}),
\end{align}
where the function $C(\ep)$ is given in eq.~\eqref{eq3.25}.

The computation of integral $I^{1 / \nu}_3$ proceeds analogously. The only
difference in comparison to $I^{1 / \nu}_4$ is the presence of the propagator
$1/k_{1 2 3}^{2}$, which simplifies to $1/(\alpha_{2} \beta_{3})$ in the region
responsible for producing the $1/\nu$ singularity. We find
\begin{align}
  I^{1 / \nu}_3
    & =
      \frac{
       \Gamma(1 + 2 \ep)
      \Gamma(-4 \ep) \Gamma^2(-2 \ep)} { \nu \Gamma(-6 \ep) (1 + \ep)} \;  C(\ep) \;
      {}_3F_2\left(1, 1, -2 \ep ; -6 \ep, 2 + \ep ; 1 \right)
    +
    \mathcal{O}(\nu^{0})
    .
\end{align}

The four integrals discussed above are the only $1/\nu$-divergent integrals that
are required for computing the N3LO QCD contribution to the soft function.

\section{Computing  integrals with  $1/k_{123}^2$ propagators using   differential equations}
\label{sec:DEs}

It remains to compute $139$ integrals that contain the $1/k_{123}^2$ propagator.
Since we did not find a way to calculate them by direct integration, we follow
the approach described in ref.~\cite{Baranowski:2021gxe} and modify this
propagator by introducing an auxiliary parameter $m^2$,
\begin{equation}
    \frac{1}{k_{1 2 3}^{2}}
    \to
    \frac{1}{k_{1 2 3}^{2} + m^{2}}
    .
    \label{eq6.1}
\end{equation}
This step, applied to an original master integral $I(\ep)$ transforms it to an
$m^2$-dependent integral $J(\ep, m^{2})$.

There are two reasons for introducing $m^2$ in this way. First, it allows us to
derive differential equations for the integrals $\boldsymbol{J}(\ep,m^2)$, to
solve them with high numerical precision and to extrapolate solutions to the
point $m^2 =0 $. Second, inclusion of $m^2$ into the propagator $1/k_{123}^2$
enables computation of boundary conditions for the differential equations at the
point $m^2 = \infty$, where significant simplifications occur. Although these
simplifications are not as radical as may be naively expected, they are
sufficient for an analytic computation of the required boundary constants, as we
explain in section~\ref{sec:boundaries}.

\subsection{Constructing  the differential equations}

The differential equations are constructed following the standard procedure.

\begin{enumerate}
\item After the reduction of all integrals needed to compute the soft function,
  we select a set of dimensionally-regulated master integrals that contain the
  $1/k_{123}^2$ propagator. We will refer to this set as $\{ I^{k_{1 2 3}^{2}}(\ep) \}$.
  
\item  For integrals from this set, we modify the $1/k_{123}^2$ propagator
  as in eq.~\eqref{eq6.1}. We will refer to this new set as $\{ J^{k_{1 2 3}^{2}}(\ep, m^{2}) \}$.
  
\item For $\{J^{k_{123}^2}(\ep,m^2) \}$ integrals, we generate a system of
  linear equations using the\\integration-by-parts method, following the
  discussion in section~\ref{sec:IBP}. We need to extend the list of integrals
  that we consider and include additional integrals with $1/(k_{123}^2 + m^2)$ and
  also integrals without this propagator, to ensure that integration-by-parts
  identities close. We will refer to the new list of integrals as
  $\{J(\ep,m^2)\}$. We note that the presence of the parameter $m^2$ does not
  affect the classification of integral families.
  
\item Computing derivatives of integrals from $\{ J^{k_{123}^2}(\ep,m^2) \}$ with respect to
  $m^2$, and expressing them through master integrals, we obtain a linear system
  of first-order differential equations
  \begin{equation}
    \frac{\partial}{\partial m^2} \boldsymbol{J}(\ep,m^2) = 
    \boldsymbol{M}(\ep,m^2) \boldsymbol{J}(\ep,m^2).
  \end{equation}
  
\end{enumerate}

In principle, the above procedure should be performed for integrals defined with
the analytic regulator. We have attempted to do that and found that the
construction of differential equations becomes extremely complicated, as the
reduction to master integrals involves three parameters $d,\nu$ and $m^2$ making
it very slow and inefficient. However, it is important to realize that it is
\emph{unnecessary} to do that. Indeed, since the set of integrals
$I^{k_{123}^{2}}(\ep)$ includes integrals that \emph{are} regularized
dimensionally, and since introduction of $m^2$ cannot affect this property,
derivatives of $J(\ep,m^2)$ integrals cannot depend on $\nu$ as well. Hence, it
should be possible to construct a $\nu$-independent system of differential
equations that these integrals satisfy. Our experience shows that the
construction of such a system is possible but highly non-trivial. To achieve
this, we relied heavily on the idea of \emph{filtering} described at the end of
section~\ref{sec:unregDiv}, which allows us to remove ill-defined
integration-by-parts identities from the sets of linear equations and set $\nu =
0$ everywhere \emph{before} attempting to solve it. We note that we have
performed extensive checks of the filtering process by comparing exact
$\nu$-dependent reductions with reductions performed using a filtered system of
IBP relations.

Remarkably, the filtered reduction actually achieves a more ``complete''
reduction than the $\nu$-dependent reduction. In a test of the filtered
reduction, we reduce those well-defined integrals needed for the computation of
the soft function and compare the result with the $\nu$-dependent
reduction.\footnote{Note that in our setup, the soft function is expressed as a
linear combination of both well-defined integrals and ill-defined integrals. The
filtered IBP system can only work with well-defined integrals.} Some of the
integrals are reduced to well-defined master integrals, which we reproduce
exactly, while other integrals are reduced to the four $1 / \nu$-divergent
integrals $I^{1 / \nu}$ discussed in section~\ref{sec:divergent-integral} with
$\mathcal{O}(\nu^{1})$ reduction coefficients. We find that the filtered IBP
system reduces all such integrals, equivalently, to \emph{two} well-defined
integrals with $\mathcal{O}(\nu^{0})$ reduction coefficients instead. It turns
out that there are linear relations among the $1 / \nu$ poles of the four
integrals $I^{1 / \nu}$
\begin{align}
    I^{1 / \nu}_{1}
    & =
    \frac{1 + 6 \ep}{1 + 4 \ep}
    I^{1 / \nu}_{2} + {\cal O}(\nu^0)
    ,
    \\
    I^{1 / \nu}_{3}
    & =
    - \frac{1 + 6 \ep}{1 + 3 \ep}
    I^{1 / \nu}_{4}
    +
    \frac{1 + 7 \ep}{\ep (1 + 3 \ep)}
    I^{1 / \nu}_{2} + {\cal O}(\nu^0)
    .
\end{align}
In the $\nu$-dependent reduction, the $I^{1 / \nu}$ integrals are independent,
which is understandable since higher order terms in the expansion in $\nu$ of
these integrals are unrelated to each other. This fact shows that the filtered
reduction correctly captures the contributions that are relevant for the soft
function and avoids the redundancies introduced by the analytic regulator $\nu$.

\subsection{Solving the  differential equation and 
constructing solutions at $m^2=0$}

In the previous section we described the construction of the system of
differential equations for integrals that contain the $1/k_{123}^2$ propagator.
Any system of differential equations requires boundary conditions. We find it
convenient to compute them at $m^2 \to \infty$; we discuss the details of their
computation in section~\ref{sec:boundaries}. In this section we assume that the
boundary conditions are known, explain how to solve the system of differential
equations numerically, and recover the $m^2 = 0$ master integrals $I(\ep)$ that
we actually require.

 The system of differential equations reads
\begin{equation}
    \partial_{m^{2}}
    \boldsymbol{J}(\ep, m^{2})
    =
    \boldsymbol{M}(\ep, m^{2})
    \boldsymbol{J}(\ep, m^{2})
    .
    \label{eq6.3}
\end{equation}
It contains 630 integrals. Our goal is to solve it numerically, starting from
$m^{2} = \infty$, and obtaining the desired $m^2=0$ integrals as
\begin{equation}
    \boldsymbol{I}(\ep)
    =
    \lim_{m^{2} \to 0^{+}}
    \boldsymbol{J}(\ep, m^{2})
    .
\end{equation}

In deriving the system of differential equations it is critical to choose a
basis of master integrals that keeps the differential equations simple. In
particular, it is important to ensure that the matrix $\boldsymbol{M}(\ep,
m^{2})$ contains no denominators which mix $\epsilon$ and $m^{2}$. We were able
to achieve this for the system of equations that we have to solve. Even for good
bases, elements of the matrix $\boldsymbol{M}$ are rational functions of $m^2$,
comprised of high-degree polynomials with many poles in the complex $m^2$-plane.

Before describing how the system of differential equations can be solved, it is
useful to make a few remarks about singularities of the integrals
$\boldsymbol{J}(\ep,m^2)$. Since these are phase-space integrals with $k_{123}^2
> 0$, the mass dependent propagator $1/(k_{1 2 3}^{2} + m^{2})$ cannot develop
any non-analyticity for real positive values of $m^2$, except for $m^2 = 0$ and
$m^2 = \infty$. Hence, in principle, as long as we stay away from the negative
real axis in the complex $m^2$-plane, we can reach the point $m^2 = 0$ without
having to worry about crossing singular surfaces, where values of integrals can
branch.

However, it is to be noted that the matrix $\boldsymbol{M}(\ep, m^{2})$, that
appears in the differential equation, does have singularities also in the
half-plane where ${\rm Re}( m^2) > 0$. Altogether, there are 38 poles at various
values of $m^{2}$ in the matrix $\boldsymbol{M}(\ep, m^{2})$, coming from 29
different polynomials in the denominator. These polynomials read
\begin{equation}
\begin{split}
  &
    m^{2},
    \pm 1 + m^{2},
    \pm 1 + 2 m^{2},
    \pm 1 + 4 m^{2},
    \pm 1 + 8 m^{2},
    \pm 9 + 16 m^{2},
    4 + m^{2},
    -16 + m^{2},
  \\
  & 
    1 + 16 m^{2},
    1 + 3 m^{2},
    -9 + 4 m^{2},
    -3 + 4 m^{2},
    1 + 5 m^{2},
    -3 + 8 m^{2},
    4 + 9 m^{2},
    1 + 64 m^{2},
  \\
  &
    % m^4 part
    1 + 4 m^{4},
    1 + 4 m^{2} + 16 m^{4},
    4 \pm 13 m^{2} + 32 m^{4},
    -27 + 64 m^{4},
    -7 - 36 m^{2} + 96 m^{4},
  \\
  &
    16 + 87 m^{2} + 1024 m^{4},
    1 + 108 m^{2} - 304 m^{4} + 64 m^{6}.
\end{split}
\end{equation}

\begin{figure}
  \centering
  \includegraphics[width=\textwidth]{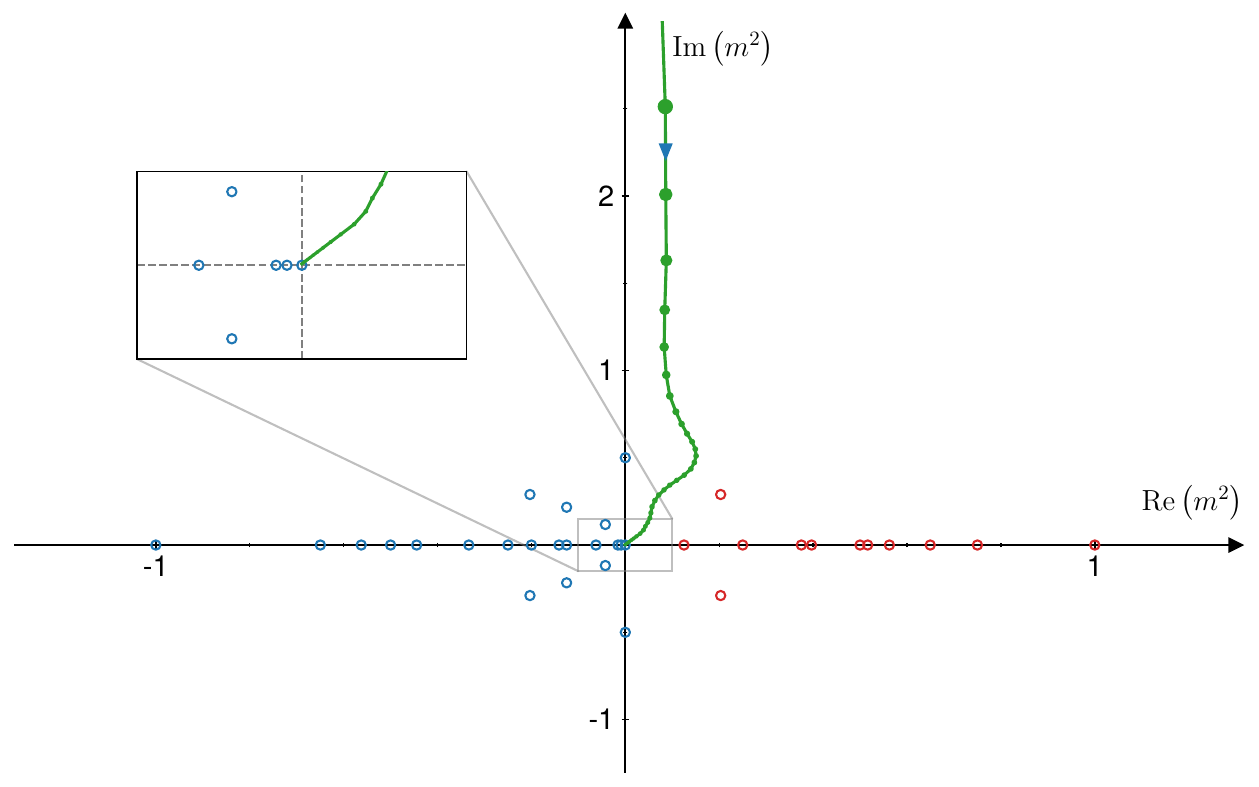}
  \caption{Singularities of $\boldsymbol{M}(\ep, m^{2})$ in the vicinity of
    $m^{2} = 0$. The singularities located in the right half-plane of $m^{2}$ are
    colored in red, while other singularities are colored in blue. The path that we
    follow to move from $m^2 = \infty$ to $m^2 = 0$ is shown in green.}
  \label{fig:singularties}
\end{figure}
Among the zeros of these polynomials, there is a branch point at the origin,
which is our target point. There are 22 singularities located at the half-plane
with ${\rm Re}(m^2) < 0$ or on the imaginary axis of $m^{2}$
\begin{equation}
  \label{eq6.8}
  \begin{split}
    m^2 \approx \{
    &
      {
      -4, \quad
      -1,\quad
      -0.649511, \quad
      -0.5625, \quad
      -0.5, \quad
      -0.444, \quad
      }
    \\
    &
      {
      -0.33, \quad
      -0.25, \quad
      -0.203125 \pm 0.289379 i, \quad
      -0.2, \quad
      }
    \\
    &
      {
      -0.1412, \quad
      -0.125, \quad
      -0.125 \pm 0.21651 i, \quad
      -0.0625, \quad
      }
    \\
    &
      {
      -0.042480 \pm 0.11756 i, \quad
      -0.015625, \quad
      -0.00903, \quad
      \pm 0.5 i
      }
    \}.
  \end{split}
\end{equation}

Finally, there are 15 poles in the half-plane where ${\rm Re}(m^2) >0$,
\begin{equation}
  \label{eq6.9}
  \begin{split}
    m^2 \approx \{
    & 
      {
      0.125, \quad
      0.25, \quad
      0.203125 \pm 0.289379 i, \quad
      0.375, \quad 
      }
      \\
    &
      {
      0.3966835638, \quad    
      0.5, \quad
      0.5162444550, \quad
      0.5625, \quad
      }
      \\
    &
      {
      0.6495190528, \quad
      0.75, \quad
      1, \quad
      2.25, \quad
      4.362345770, \quad
      16
      }
    \}.
  \end{split}
\end{equation}

These different singularities of the matrix $\boldsymbol{M}$ are illustrated in
fig.~\ref{fig:singularties}, and, as we already mentioned, all
$\boldsymbol{J}(\ep,m^2)$ integrals have to remain regular in the half-plane to
the right of the imaginary axis.

Furthermore, when $m^{2}$ is real and positive, the phase-space integrals should
also be real. While this sounds completely obvious, it provides a useful
consistency check for the solutions of the differential equations, especially if
one starts at complex infinity and moves towards a positive real axis.

The solution at $m^{2} = \infty$ takes the following form
\begin{equation}
\begin{split}
    \label{eq:equation-boundary}
    \boldsymbol{J}_{\infty}(\ep, m^{2})
    & =
    \sum_{n l k}
    \left( m^{2} \right)^{-n - l \ep} \log^{k} m^{2} \boldsymbol{B}_{n l k}(\ep)
     =
    \sum_{n}
    \left( m^{2} \right)^{-n} \boldsymbol{B}^{0}_{n}(\ep)
\\
   & \phantom{= {}} +
    \left( m^{2} \right)^{- \ep}
    \Bigg [
        \sum_{n}
        \left( m^{2} \right)^{-n}\boldsymbol{B}^{- \ep}_{n}(\ep)
                 +
        \sum_{n}
        \left( m^{2} \right)^{-n} \log m^{2} \boldsymbol{B}^{- \ep, \log}_{n}(\ep)
    \Bigg ]
 \\
    & \phantom{= {}}
    +
    \left( m^{2} \right)^{-2 \ep}
    \Bigg [
        \sum_{n}
        \left( m^{2} \right)^{-n}\boldsymbol{B}^{-2 \ep}_{n}(\ep)
        +
        \sum_{n}
        \left( m^{2} \right)^{-n} \log m^{2} \boldsymbol{B}^{-2 \ep, \log}_{n}(\ep)
        \\
         & \phantom{= {}} +
        \sum_{n}
        \left( m^{2} \right)^{-n} \log^{2} m^{2} \boldsymbol{B}^{-2 \ep, \log^{2}}_{n}(\ep)
    \Bigg ]
    ,
\end{split}
\end{equation}
where in the second step we write the relevant terms of the expansion
\emph{actually} allowed by the differential equations.

We find that we require 49 constants of the type $\boldsymbol{B}_{i}^{0}$, 28
constants of the type $\boldsymbol{B}_{i}^{-\ep}$, 26 constants of the type
$\boldsymbol{B}_{i}^{-2\ep}$, and 1 constant of the type
$\boldsymbol{B}_{i}^{-\ep, \log} $ to fully specify the solution. Integrals,
from which these constants are determined, can be chosen arbitrarily, but
simpler integrals are preferred.

Having determined a sufficient number of the expansion coefficients in
eq.~\eqref{eq:equation-boundary}, we can evaluate
$\boldsymbol{J}_{\infty}(\ep,m^2)$ at a point $m^2=m_i^2$ that is different from
infinity. To choose this point, we need to determine the radius of convergence
of the expansion in eq.~\eqref{eq:equation-boundary}, which is controlled by the
closest singularity to $m^2 = \infty$ in the complex $m^{2}$-plane, or the
farthest singularity away from the origin. For our equation $m^{2} = 16$
dictates the radius of convergence of the solution at the boundary
eq.~\eqref{eq:equation-boundary}, any $m^{2}$ that satisfies $|m^{2}| > 16$ is a
valid choice of $m_{i}^{2}$. In practice $m_{i}^{2}$ is taken to be $64 i$.
Having chosen the first evaluation point $m_i^2$, within the radius of
convergence, we obtain $\boldsymbol{J}(\ep, m_{i}^{2}) =
\boldsymbol{J}_{\infty}(\ep, m_{i}^{2})$. To move forward, we represent
solutions of the differential equation as Taylor series
\begin{equation}
    \label{eq:equation-regular}
    \boldsymbol{J}_{m_i^2}(\ep, m^2)
    =
    \sum_{n=0}^{}
    \left( m^{2} - m_{i}^{2} \right)^{n} \boldsymbol{c}_{n}(\ep),
\end{equation}
which is possible since $m_i^2$ is a regular point of the differential equation.
It should be apparent that $\boldsymbol{c}_{0}(\ep) = \boldsymbol{J}(\ep,
m_{i}^{2})$ and that other coefficients in the above equation are
\begin{equation}
c_n(\ep) = \frac{1}{n!}\frac{\dm^n\; \boldsymbol{J}_{\infty}(\ep,m^2) }{\dm m^{2n} }
\Big |_{m^2 = m_i^2}.
\end{equation}
We find it convenient to compute these derivative by utilizing the differential
equation. The $n$-th derivative satisfies
\begin{equation}
  \frac{\dm^n\; \boldsymbol{J}(\ep,m^2) }{\dm m^{2n} }
  = \sum \limits_{k=0}^{n-1} \frac{(n-1)!}{(n-1-k)! k!}
  \left [ \frac{\dm^k \boldsymbol{M}}{\dm m^{2k}}
    \;\;
    \frac{\dm^{(n-1-k)}\; \boldsymbol{J}(\ep,m^2) }{\dm m^{2(n-1-k)} }
  \right ].
\label{eq6.13}
\end{equation}
Using these equations recursively, we easily obtain the coefficients
$c_n(\ep)$.\footnote{In practice, $c_{n}(\ep)$ can be evaluated efficiently by
casting the differential equation system into a system of linear recurrence
relations and solving it order by order.} We then use
eq.~\eqref{eq:equation-regular} to move away from the point $m_i^2$, staying
within its radius of convergence, and then repeat the above procedure at another
regular point.

Continuing this process, after about 50 steps we reach a point within the radius
of convergence of the solution at $m^{2}=0$. We will refer to this last regular
point as $m_f^2$. The point $m^{2} = 0$ is a singular point of the differential
equation, so the expansion around this point has a power-logarithmic form
\begin{equation}
    \label{eq:equation-origin}
    \boldsymbol{J}_{0}(\ep, m^{2})
    =
    \sum_{l n  k}
    \left( m^{2} \right)^{n + l \ep} \log^{k} m^{2} \boldsymbol{c}_{ n l k}(\ep)
    .
\end{equation}
We compute coefficients of the expansion in eq.~\eqref{eq:equation-origin} by
comparing it with the value of the Taylor-expanded integrals
$\boldsymbol{J}(\ep, m_{f}^{2})$ at $m_f^2$. We emphasize that in the formal
solution shown in eq.~\eqref{eq:equation-origin} we include all branches that
are consistent with the behavior of the differential equation eq.~\eqref{eq6.3}
at $m^2 =0$.

Finally, we take the limit $m^{2} \to 0$ \emph{at fixed} $\ep$, recovering
original massless integrals
\begin{equation}
    \boldsymbol{I}(\ep)
    =
    \lim_{\ep \to 0}
    \lim_{m^{2} \to 0^{+}}
    \boldsymbol{J}_{0}(\ep, m^{2})
    =
    \boldsymbol{c}_{0 0 0}(\ep)
    ,
\end{equation}
where the limit $\ep \to 0$ is understood as an expansion through 
the required  order in $\ep$.

It is interesting to note that there are more independent integrals in the list
$\boldsymbol{J}(\ep,m^2)$ than in $\boldsymbol{I}(\ep)$. This means that the
massless integrals that we obtain as the limit of massive integrals are not
independent and relations between them can be found using the IBP relations for
massless integrals. This provides an opportunity for a highly non-trivial and
powerful check, and we find that our solutions do pass it.

 To conclude this section, we note that in comparison to our previous
work~\cite{Baranowski:2021gxe,Baranowski:2022khd}, where both parameters $d$ and
$\nu$ were retained in the differential equations, the current approach is
simpler and more transparent. The reason for this is that it is difficult to
construct the $\nu$-dependent differential equations and to analyze the behavior
of the solutions around singular points $m^2 = 0$ and $\infty$. For example, the
solution $\boldsymbol{J}_{0}(\ep, \nu, m^{2})$ of the $\nu$-dependent system of
differential equations at $m^{2} = 0$ depends on both regulators $\ep$ and
$\nu$, and the original massless integrals are recovered in the following
way\footnote{The $\ep \to 0$ and $\nu \to 0$ limits should be understood as the
expansion of integrals through an appropriate order in these variables.}
\begin{equation}
    \boldsymbol{I}(\ep, \nu)
    =
    \lim_{\ep \to 0}
    \lim_{\nu \to 0}
    \lim_{m^{2} \to 0^{+}}
    \boldsymbol{J}_{0}(\ep, \nu, m^{2})
    .
    \label{eq6.16}
\end{equation}
 The need to compute these limits forces us to develop full understanding of the
$\nu$ dependence of master integrals at the singular points, which is highly
non-trivial. Furthermore, evolving solutions of the differential equations from
$m^2 = \infty$ to $m^2 = 0$ is much more challenging in the presence of the
$\nu$ regulator.

Previously~\cite{Baranowski:2021gxe,Baranowski:2022khd}, we observed that the
$\nu$-dependent solutions $(m^{2})^{n + l \ep + l' \nu} \log^{k} m^{2}$ take a
relatively simple form and, in the limit $\nu \to 0$, can be accounted for
correctly without the need of explicitly introducing the regulator $\nu$ to the
differential equation. This allows us to take the $\nu \to 0$ limit before
starting to solve the differential equations, making it similar to cases where
dimensional regularization is sufficient. However, due to an increased
complexity of master integrals, this was hardly a viable option for the $nn \bar
n$ case. The use of a filtered reduction was crucial for solving the problem, as
all integrals $\boldsymbol{I}(\ep)$ that appear in that case are guaranteed to
be well-defined without the need of an additional regulator.

\section{Boundary conditions}
\label{sec:boundaries}

In the previous section, we have assumed that all the boundary conditions at
$m^2=\infty$ are known. In this section we discuss how they are computed. By
analyzing the matrix $\boldsymbol{M}(\ep,m^2)$ at $m^2 \to \infty $, it is
possible to identify the possible branches for various integrals and deduce the
minimal set of constants for the boundary conditions. Many of these branches
provide vanishing contributions, and only three branches $\left(m^2\right)^{-n
\ep}$ with $n=0,1,2$, which have already appeared in the $nnn$
case~\cite{Baranowski:2021gxe,Baranowski:2022khd}, contribute.

\subsection{The Taylor branch}

Since we discuss phase-space integrals and consider the limit $m^2 \to \infty$,
the most natural contribution arises from the Taylor expansion of integrands in
$1/m^2$. This implies
\begin{equation}
\frac{1}{k_{123}^2+m^2} \to \frac{1}{m^2} + {\cal O}(k_{123}^2/m^2),
\end{equation}
and the complicated denominator $k_{1 2 3}^{2}$ disappears from the computation
entirely. Integrals that appear in the calculation of the Taylor branch can be
computed by direct integration in the same way as the master integrals discussed
in section~\ref{sec:no123ints}.

\subsection{Region $\left(m^2\right)^{-\ep}$}
\label{sec:regmm1}

A peculiar feature of \emph{phase-space} integrals required for the computation
of the soft function is their \emph{ultra-violet} sensitivity, manifesting
itself in the appearance of non-trivial branches in the $m^2 \to \infty$ limit.
We start with the discussion of the $(m^2)^{-\ep}$ branch. This branch occurs
when a larger light-cone component of one of the partons' momenta is ${\cal
O}(m^2) $ and all other momenta components and momenta of other partons are
${\cal O}(1)$. This can only happen if at least one $\theta$-function is present
in the integrand.

In table~\ref{tab:densRescale1} we provide original and simplified expressions
for all inverse propagators, for different choices of the large light-cone
coordinate.
  \begin{table}[t]
    \centering
    \renewcommand{\arraystretch}{1.2}
    \begin{tabular}{|l||c|c|c|c|}
      \hline
      $D_i$ & $\alpha_1 \sim m^2 $ & $\alpha_2 \sim m^2 $ & $\alpha_3 \sim m^2 $ & $\beta_3 \sim m^2 $ \\
      \hline\hline
      $k_{12}\cdot n$ & $\beta_{12}$ & $\beta_{12}$ & $\beta_{12}$ & $\beta_{12}$ \\
      $k_{23}\cdot n$ & $\beta_{23}$ & $\beta_{23}$ & $\beta_{23}$ & $\boldsymbol{\beta_{3}}$\\
      $k_{13}\cdot n$ & $\beta_{13}$ & $\beta_{13}$ & $\beta_{13}$ & $\boldsymbol{\beta_{3}}$ \\
      $k_{123}\cdot n$ & $\beta_{123}$ & $\beta_{123}$ & $\beta_{123}$ & $\boldsymbol{\beta_{3}}$ \\
      \hline
      $k_{12}\cdot \bar{n}$ & $\boldsymbol{\alpha_{1}}$ & $\boldsymbol{\alpha_{2}}$ & $\alpha_{12}$ & $\alpha_{12}$ \\
      $k_{23}\cdot \bar{n}$ & $\alpha_{23}$ & $\boldsymbol{\alpha_{2}}$ & $\boldsymbol{\alpha_{3}}$ & $\alpha_{23}$ \\
      $k_{13}\cdot \bar{n}$ & $\boldsymbol{\alpha_{1}}$ & $\alpha_{13}$ & $\boldsymbol{\alpha_{3}}$ & $\alpha_{13}$ \\
      $k_{123}\cdot \bar{n}$ & $\boldsymbol{\alpha_{1}}$ & $\boldsymbol{\alpha_{2}}$ & $\boldsymbol{\alpha_{3}}$ & $\alpha_{123}$ \\
      \hline
      $k_{12}^2$ & $\boldsymbol{\alpha_1 \beta_2}$ & $\boldsymbol{\alpha_2 \beta_1}$ & $k_{12}^2$ & $k_{12}^2$ \\
      $k_{23}^2$ & $k_{23}^2$ & $\boldsymbol{\alpha_2 \beta_3}$ & $\boldsymbol{\alpha_3 \beta_2}$ & $\boldsymbol{\beta_3 \alpha_2}$ \\
      $k_{13}^2$ & $\boldsymbol{\alpha_1 \beta_3}$ & $k_{13}^2$ & $\boldsymbol{\alpha_3 \beta_1}$  & $\boldsymbol{\beta_3 \alpha_1}$\\
      \hline
      $k_{123}^2+ m^2$ & $\boldsymbol{\alpha_{1}\beta_{23}} + m^2$ & $\boldsymbol{\alpha_{2}\beta_{13}} + m^2$ & $\boldsymbol{\alpha_{3}\beta_{12}} + m^2$ & $\boldsymbol{\beta_{3}\alpha_{12}} + m^2$ \\
      \hline
    \end{tabular}
    \caption{Scalings in the region $(m^2)^{-\ep}$.}
    \label{tab:densRescale1}
  \end{table}
It follows from that table, that there is one (and only one) case-specific
scalar product of two four-momenta that cannot be simplified. These scalar
products depend on the relative angle between momenta of two partons, and it is
this dependence that makes the computation of boundary conditions for this
branch challenging.

We denote two partons, whose momenta scale as ${\cal O}(1)$, as $i$ and $j$, and
the parton with one ${\cal O}(m^2)$ Sudakov component as $h$. We also assume
that the larger momentum component is $\tilde \gamma_h$ and the smaller one is
$\gamma_h$. It follows from table~\ref{tab:densRescale1} that any integral that
provides the $(m^2)^{-\ep}$ branch can be written as
\begin{equation}
  J \sim \frac{1}{\normNep^2}\int \limits_{0}^{\infty} \dm \gamma_h \; \gamma_h^{-\ep} \;
  \int [\dm k_i][\dm k_j]
  \frac{f_i f_j R_0(\alpha_i,\beta_i,\alpha_j,\beta_j,\gamma_h)}{[k_{ij}^2]^n}
  \int \limits_{0}^{\infty}  
  \frac{\dm \tilde \gamma_h \; \tilde{\gamma}_h^{-\ep} }{ {\tilde \gamma_h}^{n_1} \; ( \tilde \gamma_h \; x_{ij} + m^2)^{n_2} },
\end{equation}
where the rational function $R_0(\alpha_i,\beta_i,\alpha_j,\beta_j, \gamma_h)$
contains the zero-jettiness $\delta$-function, and depends on the light-cone
components of all partons. The polynomial $x_{ij}$ that depends on the
light-cone components of momenta $k_{i,j}$ can be read off from
table~\ref{tab:densRescale1}. The integral over $\tilde \gamma_h$ gives
\begin{equation}
  \label{eq:intgammahreg1ep}
  \begin{split}
    & \int \limits_{0}^{\infty}
      \frac{\dm \tilde \gamma_h \; \tilde{\gamma}_h^{-\ep} }{ {\tilde \gamma_h}^{n_1} \; ( \tilde \gamma_h \; x_{ij} + m^2)^{n_2} } = (m^2)^{1-\ep-n_1 - n_2} \;
      \frac{\Gamma(n_1+n_2-1+\ep) \Gamma(1-n_1 - \ep) }{\Gamma(n_2)} 
      \; x_{ij}^{n_1-1+\ep}.
  \end{split}
\end{equation}

As the next step, we discuss integration  over momenta 
of partons $i$ and $j$  and consider the following integral
\begin{equation}
  \label{eq:intk1k2reg1ep}
  J_{ij} = \frac{1}{\normNep^2}\int [\dm k_i][\dm k_j] \frac{1} { [k_{ij}^2]^n } \; f_i \; f_j \;
  R(\alpha_i, \beta_i, \alpha_j,\beta_j, \gamma_h).
\end{equation}
In eq.~\eqref{eq:intk1k2reg1ep} we have introduced a new function
$R(\alpha_i,\beta_i,\alpha_j,\beta_j,\gamma_h) = x_{ij}^{n_1-1+\ep}
R_0(\alpha_i,\beta_i,\alpha_j,\beta_j,\gamma_h)$. The inverse propagator
$k_{ij}^2$ reads
\begin{equation}
  \label{eq:kikjToAng}
  k_{ij}^2 = 2 k_i\cdot k_j
  = \left( \alpha_i \beta_j + \alpha_j \beta_i \right) \left( 1 - 2 \frac{\sqrt{\alpha_i\beta_j\alpha_j\beta_i}}{\alpha_i \beta_j + \alpha_j \beta_i} \vec{e}_{\perp, i} \cdot \vec{e}_{\perp, j} \right)
  = \left( \alpha_i \beta_j + \alpha_j \beta_i \right) \left( \rho \cdot \rho_m \right),
\end{equation}
where $\rho$ and $\rho_m$ are two $(d-1)$-dimensional \emph{Minkowski} vectors.
The vector $\rho$ is light-like, $\rho^2 = 0$, and the vector $\rho_m$ is
time-like with
\begin{equation}
  \rho_m^2 =  \frac{\left( \alpha_i
      \beta_j - \alpha_j \beta_i \right)^2}{\left( \alpha_i
      \beta_j + \alpha_j \beta_i \right)^2}.
\end{equation}
The angular integration in the transverse space is easily performed with the
help of eq.~\eqref{eq:In1def}, and $J_{ij}$ becomes
\begin{equation}
  \begin{split}
    \label{eq:intk1k2reg1epSud}
    J_{ij}
    & = \frac{1}{\left(\Omega^{(d-2)}  \right)^2} \int \frac{\dm \alpha_i \dm \beta_i \dm \alpha_j \dm \beta_j}{\left( \alpha_i \beta_i \alpha_j \beta_j \right)^\ep} \; f_i \; f_j \; R(\alpha_i, \beta_i, \alpha_j,\beta_j,\gamma_h) \int \frac{\dm\Omega^{(d-2)}_i  \dm\Omega^{(d-2)}_j}{[ k_{ij}^2]^n}  \\
    & =  \int \frac{\dm \alpha_i \dm \beta_i \dm \alpha_j \dm \beta_j}{\left( \alpha_i \beta_i \alpha_j \beta_j \right)^\ep} \; f_i \; f_j \; \frac{R(\alpha_i, \beta_i, \alpha_j,\beta_j, \gamma_h )}{\left( \alpha_i \beta_j + \alpha_j \beta_i \right)^n} \; I_{d-2;n}^{(1)}(\rho_m^2).
  \end{split}
\end{equation}
The function $I_{d,n}^{(1)}$ is defined in eq.~\eqref{eq:angInt-1-m}.

Further integrations  require us to specify the constraints 
$f_i$ and $f_j$. There are five  cases to be considered.
\begin{enumerate}[label=(\Alph*)]
  % dd
\item\label{item:dd} 
If $f_i = \delta(\alpha_i - \beta_i)$ and  $f_j =
  \delta(\alpha_j - \beta_j)$,  we integrate over $\alpha_i$ and $\alpha_j$ to obtain 
  \begin{equation}
    J_{ij} = 
    \frac{\Gamma\left(1-\ep \right)\Gamma\left( 1-2\ep -2 n\right)}{\Gamma\left(1-\ep - n  \right)\Gamma\left( 1-2\ep -n \right)}
          \int\limits_0^1 \dm \beta_i \dm \beta_j \left(  \beta_i  \beta_j \right)^{-n-2\ep} R(\beta_i, \beta_i, \beta_j,\beta_j,\gamma_h).
  \end{equation}

  % dt
\item\label{item:dt}  if $f_i = \delta(\alpha_i - \beta_i)$ and $ f_j =
  \theta(\alpha_j - \beta_j)$,  we integrate over  $\alpha_i$ and 
  perform a variable transformation $\alpha_j
  \to \beta_j/r_j$. We find 
  \begin{equation}
    \label{eq:reg1integrandDT}
    J_{ij} =  
    \int\limits_0^1 \dm \beta_i \dm \beta_j \int\limits_0^1 \dm r_j
    \frac{R\left(\beta_i, \beta_i, \frac{\beta_j}{r_j},\beta_j,\gamma_h\right)}
    {\beta_i^{n+2\ep}\beta_j^{n-1+2\ep}r_j^{2-n-\ep}} \; 
    {}_2F_1\left( n,n + \ep;1-\ep; r_j \right).
  \end{equation}

  % dtb
\item\label{item:dtb} if $f_i = \delta(\alpha_i - \beta_i)$ and $ f_j =
  \theta(\beta_j - \alpha_j)$ we integrate over  $\alpha_i$, and replace $\beta_j = \alpha_j/r_j$

  \begin{equation}
  \label{eq:reg1integrandDTB}
    J_{ij} =  
    \int\limits_0^1 \dm \beta_i \dm \alpha_j \int\limits_0^1 \dm r_j
    \frac{R\left(\beta_i, \beta_i, \alpha_j,\frac{\alpha_j}{r_j},\gamma_h\right)}
    {\beta_i^{n+2\ep}\alpha_j^{n-1+2\ep}r_j^{2-n-\ep} } \;
    {}_2F_1\left( n,n+\ep;1-\ep; r_j \right).
  \end{equation}

  % tt
\item\label{item:tt} if $f_i = \theta(\alpha_i - \beta_i)$ and $ f_j =
  \theta(\alpha_j - \beta_j)$, we replace $\alpha_i = \beta_i/r_i,\;
  \alpha_j\to \beta_j/r_j$,  apply the transformation
  shown in eq.~\eqref{eq:f21tr69quad} to the angular integral, where we also need to split 
  the original integration region into two regions  $r_i < r_j$ and $r_j < r_i$. 
  Writing $r_i = r_j \xi$ 
  and $r_j = \xi r_i$ 
  in the first and in the second region, respectively,
  we obtain the following representation for the integral
  \begin{equation}
  \begin{split}
    \label{eq:reg1integrandTT}
    J_{ij} &
             = 
             \int\limits_0^1 \dm \beta_i \dm \beta_j \int\limits_0^1 \dm r_j \dm \xi
             \frac{R\left(\frac{\beta_i}{\xi r_j}, \beta_i, \frac{\beta_j}{r_j},\beta_j,\gamma_h \right)}
             {(\beta_i\beta_j)^{n-1+2\ep}r_j^{3-n-2\ep} \xi^{2-n-\ep}}
             {}_2F_1\left( n,n+\ep;1-\ep; \xi \right) \\
           & \phantom{= {}} +  
             \int\limits_0^1 \dm \beta_i \dm \beta_j \int\limits_0^1 \dm r_i \dm \xi
             \frac{R\left(\frac{\beta_i}{r_i}, \beta_i, \frac{\beta_j}{\xi r_i},\beta_j,
             \gamma_h \right)}
             {(\beta_i\beta_j)^{n-1+2\ep}r_i^{3-n-2\ep} \xi^{2-n-\ep}}
             {}_2F_1\left( n,n + \ep;1-\ep; \xi \right).
  \end{split}
  \end{equation}

  % ttb
\item\label{item:ttb} 
finally, if $f_i = \theta(\alpha_i - \beta_i)$ and $ f_j =
  \theta(\beta_j - \alpha_j)$,  we replace $\alpha_i \to \beta_i/r_i, \beta_j\to \alpha_j/r_j$ and find 
  \begin{equation}
    \label{eq:reg1integrandTTB}
    J_{ij} =  
    \int\limits_0^1 \dm \beta_i \dm \alpha_j \int\limits_0^1 \dm r_i \dm r_j
    \frac{R\left(\frac{\beta_i}{r_i}, \beta_i, \alpha_j,\frac{\alpha_j}{r_j},\gamma_h\right)}
    {(\beta_i\alpha_j)^{n-1+2\ep}(r_i r_j)^{2-n-\ep} }\;
    {}_2F_1\left( n,n + \ep;1-\ep; r_i r_j \right).
  \end{equation}

\end{enumerate}

Further integrations are case specific. Often, integrations over variables $r$
and $\xi$ can be performed in terms of hypergeometric functions, but if mixed
propagators of the type $1/(\alpha_i + \alpha_j)$ appear, it becomes impossible
to do that. At any rate, the subsequent integrations are performed on the
case-by-case basis.

To illustrate the above discussion, we compute the $(m^2)^{-\ep}$ branch of the
following integral
% 
% j[Tpppb46tttarm,1,1,1,1,0,1,1,0,1,0,0,2]
% 
\begin{align}
  \label{eq:exampleIntNNNB-ttt-reg1}
  J = \int \frac{\dFtttB}{\left( k_{1} \cdot k_3 \right) \left( k_{2} \cdot n \right) \left( k_{123} \cdot \bar{n} \right)^2 \left(k_{123}^2+m^2\right)}.
\end{align}
This integral has three $\theta$-functions and contributes  to the $nn \bar n$ configuration.

A simple analysis shows that the leading $(m^2)^{-\ep}$ contribution originates
from the region where $\beta_3\sim m^2$; because of that, the integral
corresponds to the case \ref{item:tt} above with $i=1$ and $j=2$. The integrand
in eq.~\eqref{eq:exampleIntNNNB-ttt-reg1} has no scalar product $k_1 \cdot k_2$;
for this reason, no non-trivial angular integration has to be performed. We use
table~\ref{tab:densRescale1} to read off the simplified propagators, and apply
the variable transformations that are explained in item \ref{item:tt} above. We
then write the result as the sum of two terms
\begin{equation}
J = J_A + J_B,
\end{equation}
where 
\begin{align}
  \label{eq:reg1intSplit}
  J_A & = 2 \int \limits_0^\infty 
  \frac{\dm \beta_3}{\beta_3^{1+\ep}}\; 
  \int \limits_{0}^{1} \dm \beta_1\dm \beta_2 \;  \dm \alpha_3  \dm r_2 \dm \xi \frac{(\beta_1 \beta_2)^{-2 \ep}
        (\alpha_3 )^{-\ep} r_2^{1 + 2 \ep} \xi^{2 + \ep}
  \delta\left(1 - \beta_1 - \beta_2 - \alpha_3 \right)}{
        (\beta_1 + (\beta_2  + \alpha_3 r_2) \xi)^2 ((\beta_1  + \beta_2  \xi)\beta_3 + m^2 r_2 \xi)},
        \\
  J_B & = 2 \int \limits_0^\infty \frac{\dm \beta_3 }{\beta_3^{1+\ep}}
   \int \limits_{0}^{1}
  \dm \beta_1\dm \beta_2 \dm \alpha_3  \dm r_1 \dm \xi
   \frac{(\beta_1 \beta_2)^{-2 \ep} (\alpha_3)^{-\ep} r_1^{1 + 2 \ep} \xi^{1 + \ep}
        \delta(1 - \beta_1 - \beta_2 - \alpha_3 )}{
        (\beta_2 + (\beta_1 + \alpha_3 r_1) \xi)^2 ((\beta_2  + \beta_1 \xi)\beta_3 + m^2 r_1 \xi)}.
\end{align}
In both integrals, integration over $\beta_3$ can be immediately performed using
eq.~\eqref{eq:intgammahreg1ep}. Then, the zero-jettiness $\delta$ function is
removed by integrating over $\beta_2$ in $J_A$ and over $\beta_1$ in $J_B$.
After that, the variable transformation $\beta_{1,2} = t(1-\alpha_3)$ is
performed in integrals $J_{A,B}$, respectively.

To proceed further, it is convenient to treat both integrals on the same
footing. To do that, we introduce an auxiliary integral $J_n$
\begin{equation}
  \label{eq:exReg1auxInt}
  J_n = \frac{2 \Gamma(-\ep) \Gamma(1 + \ep)}{(m^2)^{1 + \ep} } \int\limits_0^1 \dm r \dm \xi \dm \alpha_3 \dm t
  \frac{\xi^n (1 - \alpha_3)^{1 - 3 \ep} \alpha_3^{-\ep} r^{\ep} (1 - t)^{-2 \ep}
  t^{-2 \ep}  (t - \xi - t \xi)^{\ep}}
  {(t - \alpha_3 t + \xi - \alpha_3 \xi + \alpha_3 r \xi - t \xi + \alpha_3 t \xi)^2},
\end{equation}
and notice that $J_1 = J_A$ and $J_0 = J_B$, provided that $r$ is identified
with $r_{1,2}$ in the two integrals, as appropriate.

To compute $J_n$, we change  the  integration  variables
\begin{equation}
  \label{eq:intReg1exVchange}
  \xi  = \frac{t u}{1 - u + t u}, \quad \alpha_3 = \frac{v}{u + v - u v},
\end{equation}
and obtain 
\begin{equation}
  \label{eq:exReg1auxIntSubs}
  J_n = \frac{2 \Gamma(-\ep) \Gamma(1 + \ep)}{(m^2)^{1 + \ep} } \int\limits_0^1 \dm r \dm t \dm u \dm v
  \frac{r^{\ep} (1 - t)^{-2 \ep} t^{-1 + n - \ep} u^{
  n - 3 \ep}  (1 - v)^{1 - 3 \ep}
  v^{-\ep} }{(1 - u + t u)^{n + \ep} (u + v - u v)^{1 - 4 \ep}(1 - v + r v)^2}.
\end{equation}
Integration over $v$ produces the Appell function $F_1$, see
eq.~\eqref{eq:hfIntReprF1}, which simplifies to $_2F_1$, c.f.\
eq.~\eqref{eq:F1toF21}. The integration over $t$ leads to a hypergeometric
function $_2F_1$. We obtain
\begin{equation}
\begin{split}
J_n & = 
  - \frac{2 
  \Gamma(2 - 3 \ep) \Gamma(1 - 2 \ep) \Gamma^2(1 - \ep)
      \Gamma(n - \ep)  \Gamma(1 + \ep)}{  \ep (m^2)^{1 + \ep} \Gamma(3 - 4 \ep) \Gamma(1 + n - 3 \ep)}
      \int\limits_0^1 \dm r \dm u
      \frac{r^{\ep}} {(1 - u)^{n + \ep} u^n}
      \\
      & \phantom{= {}} \times {}_2F_1\left( 2, 1 - \ep; 3 - 4 \ep; 1 - r u \right) {}_2F_1\left(n - \ep, n + \ep; 1 + n - 3 \ep;\frac{u}{u-1}  \right).
      \label{eq7.19}
      \end{split}
\end{equation}
It is possible to integrate over $r$, expressing the result in terms of the
hypergeometric function ${}_3F_2$. To do this, we employ
eqs.~(\ref{eq:f21tr2zb}) and (\ref{eq:hfIntReprFPQ}). We obtain the following
one-dimensional integral representation
\begin{align}
  J_n & =
        \frac{4  (1 - 4 \ep) 
        \Gamma(2 - 3 \ep) \Gamma(2 - 2 \ep) \Gamma(n - \ep) \Gamma^2(1 - \ep) \Gamma(1 + \ep)}
        {3 \ep^2 (m^2)^{1 + \ep}  ( 1 - 3 \ep) (1 + \ep)\Gamma(3 - 4 \ep) \Gamma(1 + n - 3 \ep) }\nonumber\\
      & \phantom{= {}} \times \int\limits_0^1 \dm u \;
        u^n {}_2F_1\left(1 - 2 \ep, n + \ep; 1 + n - 3 \ep; u  \right)
        {}_3F_2\left(2, 1 - \ep, 1 + \ep; 2 + \ep, 1 + 3 \ep; u  \right)\nonumber\\
  & \phantom{= {}} + \frac{2 \Gamma(2 - 3 \ep) \Gamma(1 - 2 \ep) \Gamma(n - \ep) \Gamma(-\ep) \Gamma(
   3 \ep) \Gamma(1 + \ep) }{
    (m^2)^{1 + \ep}  (1 - 2 \ep) \Gamma(1 + n - 3 \ep)}
   \int\limits_0^1 \dm u \; 
  u^{n - 3 \ep}
    \label{eq7.20} 
  \\
  & \phantom{= {}} \times {}_2F_1\left(1 - 2 \ep, n + \ep, 1 + n - 3 \ep, u  \right)
  {}_3F_2\left(1 - 4 \ep, 2 - 3 \ep, 1 - 2 \ep; 1 - 3 \ep, 2 - 2 \ep; u  \right),
  \nonumber 
\end{align}
where we also used eq.~\eqref{eq:f21trInv1} to simplify the $u$-dependent
hypergeometric function in eq.~\eqref{eq7.19}.

The integral representation for $J_n$, $n=0,1$ in eq.~\eqref{eq7.20} is
convenient because integration over $u$ is not singular. Hence, the integrand in
that equation can be expanded in powers of $\ep$, and integrated term by term
using \texttt{HyperInt}~\cite{Panzer:2014caa}. Upon doing that, we obtain the
final result for the integral in eq.~\eqref{eq:exampleIntNNNB-ttt-reg1}
\begin{equation}
  \label{eq:exampleIntNNNB-ttt-reg1-result}
  \begin{aligned}
  J = (m^2)^{-1-\ep}&\bigg[
  \frac{\pi^2}{3\ep^2}
  + \frac{10 \zeta_3}{\ep}
  + \frac{29\pi^4}{90}
  + \ep \left(208 \zeta_5 -\frac{38}{3} \pi^2 \zeta_3\right)
  \\
  & \phantom{= {}}
  + \ep^2 \left(\frac{2239}{5670} \pi^6 - 172 \zeta_3^2\right)
  + \mathcal{O}(\ep^{3})
  \bigg].
  \end{aligned}
\end{equation}
All integrals that possess the $\left( m^2 \right)^{-\ep}$-branch at large $m^2$
can be analyzed using similar steps once the large momentum component for one of
the partons is identified.

\subsection{$(m^{2})^{- \ep} \log{(m^{2})}$ boundary integral }

All branches of master integrals contain prefactors $(m^2)^{-n \ep}$, with $n =
0,1,2$. In addition, we have found that in case $n = 1$, one needs to account
for additional sub-branch $(m^2)^{-\ep}\log{(m^{2})}$, where the logarithm
appears \emph{before} expansion in $\ep$. This contribution may appear if, in a
given integral, two different integration regions over Sudakov parameters
contribute to the $(m^2)^{-\ep}$ branch. We will illustrate this phenomenon by
considering two integrals. The first integral is quite simple, it can be
computed in a closed form and used to understand the origin of such terms. The
second is more complex; we discuss it because it was actually used to determine
the required boundary condition.

We begin with the following integral
% 
% with numerator
% INT["Tpppb1dttarm", 6, 159, 6, 1, {1, 1, 1, 1, 1, 0, 0, 1, 0, -1, 0, 0}]
\begin{equation}
  \label{eq:R1logMMdef}
  R_1
  =
  \int \frac{\dFdttB \, \beta_2}  {(k_{1 2 3}^{2} + m^{2})}
  .
\end{equation}
It receives two non-vanishing contributions to the branch $(m^2)^{-\ep}$ from
the regions with $\alpha_2 \sim m^2$ and $\beta_3 \sim m^2$. Hence, we can write
\begin{equation}
R^{- \ep}_1 =
R_{1,\alpha_2} + R_{1,\beta_3},
\end{equation}
and a simple analysis shows that, individually, these contributions are not
regulated dimensionally.\footnote{To avoid confusion, we stress that this
integral is regularized dimensionally, in agreement with the discussion in
section~\ref{sec:unregDiv}. However, when we \emph{approximate} the integrand to
simplify the calculation of the $(m^2)^{-\ep}$ branch at large $m^2$, the
resulting expressions become ill-defined. } 

To proceed with the calculation of individual contributions to the
$(m^2)^{-\ep}$ branch, we introduce the factor $\beta_3^{-\nu}$ to the numerator
of the original integral eq.~\eqref{eq:R1logMMdef}. Considering the contribution
of the $\alpha_{2} \sim m^2$ region and making use of the integral
representation in eq.~\eqref{eq:reg1integrandDTB} with $n=0$, we find
\begin{equation}
  \label{eq:nuIntal2logMM}
  R_{1,\alpha_2} = \int\limits_0^\infty \dm \alpha_2 \int \limits_{0}^{1} 
  \dm \beta_1 \dm \beta_2 \dm \alpha_3  \dm r_3
  \left(\frac{r_3}{\alpha_3}\right)^\nu
  \frac{ \alpha_3^{1 - 2 \ep} \beta_1^{-2 \ep} \beta_2 (\alpha_2 \beta_2)^{-\ep}  
    \delta(1 - \alpha_3 - \beta_1 - \beta_2)}{r_3^{1-\ep} ( r_3 (\alpha_2 \beta_1 + m^2) + \alpha_2 \alpha_3)}.
\end{equation}
We continue with the integrations over $\alpha_2$ and $r_3$. We remove the
$\delta$-function by integrating over $\beta_1$, and change variables $ \alpha_3
= t (1 - \beta_2)$ after that. The integrations over $\beta_2$ and $t$ factorize
and we obtain
\begin{equation}
\begin{split}
\label{eq7.26}
  R_{1,\alpha_2}
  & = \frac{(1 - \ep) \Gamma(1 - 2 \ep) \Gamma^3(1 - \ep) \Gamma(1 + \ep)
    \Gamma(2 - 3 \ep - \nu)}
    {(m^2)^\ep \ep \nu \Gamma(2 - 3 \ep) \Gamma(4 - 4 \ep - \nu)}
  \\
  & \times {}_3F_2\left( 1 - 2 \ep, \nu, \ep + \nu; 2 - 3 \ep,1 + \nu; 1 \right).
\end{split}
\end{equation}

The $1/\nu$ pole in the above expression is compensated by the contribution from
the region $\beta_3 \sim m^2$, that we now discuss.

We use the 
representation in eq.~\eqref{eq:reg1integrandDT} to obtain
\begin{equation}
 R_{1,\beta_3} =  \int\limits_0^\infty \dm \beta_3
 \int \limits_{0}^{1} \dm \beta_1 \dm \beta_2 \dm \alpha_3  \dm r_3
  \frac{\beta_1^{-2 \ep} \beta_2^{2 - 2 \ep} \beta_3^{-\nu} (\alpha_3 \beta_3)^{-\ep} r_3^{-1 + \ep}
  \delta(1 - \alpha_3 - \beta_1 - \beta_2)}{ r_3 (\beta_1 \beta_3 + m^2) + \beta_2 \beta_3}.
\end{equation}
Performing the integrations, we find
\begin{equation}
\begin{split}
  R_{1,\beta_3} & =
    - \frac{1}{\nu} \frac{(1-\ep)\Gamma (1-2 \ep) \Gamma^2 (1-\ep)  
    \Gamma (-\ep-\nu +1) \Gamma (3 -3 \ep+\nu) \Gamma
    (\ep+\nu )}{(m^2)^{\ep+\nu} \Gamma(3-3 \ep) \Gamma(4 -4 \ep+\nu)} \\  
  & \phantom{= {}} \times {}_3F_2\left(1-2 \ep,\ep,-\nu ;3-3 \ep,1-\nu ;1\right).
    \label{eq7.28}
\end{split}
\end{equation}
Finally, combining $R_{1,\alpha_2}$ 
and $R_{1,\beta_3}$ and expanding 
in $\nu$ at fixed $\ep$, we find 
\begin{equation}
  \label{eq:nuIntal2logMMlogPart}
R_1^{-\ep} = 
R_{1,\alpha_2} + R_{1,\beta_3}  = 
 \frac{(1-\ep) \Gamma (1-2 \ep) \Gamma^3 (1-\ep) \Gamma (\ep)}{\Gamma (4-4 \ep)}\; (m^2)^{-\ep}\log{(m^{2})} + \dots, 
\end{equation}
where ellipses stand for contributions that do not contain the $(m^2)^{-\ep}
\log{(m^{2})}$ branch. We note that out of the two terms discussed above, only
$R_{1,\beta_3}$ produces a $\log{(m^{2})}$ contribution whereas another one is
only needed for removing the $1/\nu$ pole. This is a generic feature of such
integrals, that we exploit in the second example below.

% Real example
% INT["Tpppb11tttarm", 6, 2127, 6, 0, {1, 1, 1, 1, 0, 0, 1, 0, 0, 0, 0, 1}]

A more complicated integral that is actually employed for the calculation of the
relevant boundary constant reads
\begin{equation}
  \label{eq:R2logMMdef}
    R_2
    =
    \int \frac{\dFtttB}{(k_{1 2 3}^{2} + m^{2})(k_{23} \cdot \bar{n})}
    .
\end{equation}
To compute the $(m^2)^{-\ep}$ branch of this integral, we need to consider two
regions, $\alpha_1 \sim m^2$ and $\beta_3 \sim m^2$. We regulate both of these
contributions by multiplying the integrand in eq.~\eqref{eq:R2logMMdef} with
$\beta_3^{-\nu}$. Similar to the previous example, we find that the region
$\alpha_1 \sim m^2$ does not contribute to the $\log{(m^{2})}$ branch, whereas
the $\beta_3 \sim m^2$ region does. Both regions contribute to the coefficient
of the $1/\nu$ pole, where they cancel each other.

Focusing on the term $(m^2)^{-\ep} \log{(m^2)} $, we consider the region where $
\beta_3 \sim m^2$ and split the relevant integrations into two parts, similar to
the representation in eq.~\eqref{eq:reg1integrandTT}. We write $ \alpha_1 =
\beta_1/r_1 $ and $\alpha_2 = \beta_2/r_2$, consider two regions $r_1 < r_2$ and
$r_2 < r_1$ and find that only the first one contributes to the $(m^2)^{-\ep}
\log{(m^{2})}$ branch. Hence, changing variables $r_1 = \xi \; r_2$, we arrive
at the following representation of the relevant contribution\footnote{We
introduce notation $\tilde R_2$ to emphasize that the expression below only
gives the relevant $\log(m^2)$ term and not the whole integral $R_2$. } to the
integral
\begin{equation}
  \tilde{R}_2 = \int\limits_0^\infty \dm \beta_3 
  \int \limits_{0}^{1} 
  \dm \beta_1 \dm \beta_2  \dm \alpha_3 \dm r_2 \dm \xi 
  \frac{(\beta_1 \beta_2)^{1 - 2 \ep} \delta(1 - \alpha_3 - \beta_1 - \beta_2)}{
    \beta_3^{\nu} (\alpha_3 \beta_3)^{\ep} r_2^{3 - 2 \ep} \xi^{2 - \ep}
    \left(\alpha_3 + \frac{\beta_2}{r_2}\right) \left(m^2 + \frac{\beta_2 \beta_3}{r_2} + \frac{\beta_1 \beta_3}{r_2 \xi}\right)}.
\end{equation}
To compute this integral, we integrate over $\beta_3$, remove the
$\delta$-function by integrating over $\alpha_3$, and change variables as
follows, $\beta_2= t (1-\beta_1)$ and $\beta_1 =t (1 - v)/(t + v - t v)$.
Integrating over $v$ and $r_2$, we obtain the following two-dimensional integral
\begin{equation}
\begin{split}
\tilde R_2 & = \frac{ \Gamma (2-3 \ep) \Gamma (2-2 \ep) \Gamma (-\ep-\nu +1) \Gamma (\ep-\nu ) \Gamma (\ep+\nu)}{
    (m^2)^{\ep+\nu }  \Gamma (4-5 \ep) \Gamma (\ep-\nu +1)}
    \int\limits_0^1 \dm t \dm \xi \frac{(1-t)^{-\ep}} {t^{2-\ep} \xi ^{3 - 3 \ep+\nu}}\\
 & \phantom{= {}} \times 
  {}_2F_1\left(1,\ep-\nu ;1+\ep-\nu; 1- \frac{1}{t} \right)
  {}_2F_1\left(2-3 \ep,3-4 \ep+\nu;4-5 \ep;1- \frac{1}{t \xi }\right).
\end{split}
\end{equation}
Applying the transformations shown in
eqs.~(\ref{eq:f21trInv1},\ref{eq:f21trInv1mz}) to first and second
hypergeometric functions, respectively, we find that the integral splits into
two parts, where in each part integration over $\xi$ can be performed. Keeping
the part that provides the $(m^2)^{-\ep} \log{(m^{2})}$ contribution, we write
\begin{equation}
\tilde R_2 = 
-\frac{ \Gamma (2-3 \ep) (m^2)^{-\ep-\nu } \Gamma^2 (1-\ep) \Gamma (\ep )}{ \nu  \ep  \Gamma( 3 -4 \ep)}
    \int\limits_0^1 \dm x\, \frac{x^{1-2 \ep}} {(1-x)^{\ep} } {}_2F_1(1,1;1+ \ep;1-x)   + \dots ,
\end{equation}
Integrating over $x$ and extracting the $\log{(m^{2})} $ term, we obtain the
final result
\begin{equation}
  R_2 = 
  \frac{\Gamma (1-2 \ep) \Gamma^3 (1-\ep) \Gamma (1+\ep) \, }{2 \ep^2 (2-3 \ep)  \Gamma (2-4 \ep)} {}_3F_2(1,1,1-\ep;3-3 \ep,1+\ep;1)
 m^{-2\ep}\log{(m^{2})} + \cdots,
\end{equation}
where ellipses describe terms that do not contain the $\log{(m^{2})}$ term at
fixed $\ep$.

\subsection{Region $\left(m^2\right)^{-2\ep}$}
\label{sec:regmm2}

\begin{table}[t]
  \centering
  \renewcommand{\arraystretch}{1.2}
  \begin{tabular}{|l||c|c|c|c|c|}
    \hline
    $D_i$ & $\alpha_1 \sim \alpha_2 \sim  m^2 $ & $\alpha_2 \sim \alpha_3 \sim m^2 $ & $\alpha_1 \sim \alpha_3 \sim m^2$ \\
          & $q = k_1+k_2$ & $q = k_2+k_3$ &  $q = k_1+k_3$ \\
    \hline\hline
    $k_{12}\cdot n$  & $\beta_{12} = \beta_q$ & $\beta_{12} = \beta_2 + \beta_1$ & $\beta_{12} = \beta_1 + \beta_2$ \\
    $k_{23}\cdot n$  & $\beta_{23} = \beta_{2} + \beta_{3}$ & $\beta_{23} = \beta_{q}$ & $\beta_{23} = \beta_{3} + \beta_{2}$ \\
    $k_{13}\cdot n$  & $\beta_{13} = \beta_{1} + \beta_{3}$ & $\beta_{13} = \beta_{3} + \beta_{1}$ & $\beta_{13} = \beta_q$ \\
    $k_{123}\cdot n$ & $\beta_{123} = \beta_q + \beta_3$ & $\beta_{123} = \beta_q + \beta_1$  & $\beta_{123} = \beta_q + \beta_2$ \\
    \hline
    $k_{12}\cdot \bar{n}$ & $\alpha_{12} = \alpha_q$ & $\boldsymbol{\alpha_{2}}$ & $\boldsymbol{\alpha_{1}}$ \\
    $k_{23}\cdot \bar{n}$ & $\boldsymbol{\alpha_{2}}$ & $\alpha_{23} = \alpha_q$ & $\boldsymbol{\alpha_{3}}$ \\
    $k_{13}\cdot \bar{n}$ & $\boldsymbol{\alpha_{1}}$ & $\boldsymbol{\alpha_{3}}$ & $\alpha_{13} = \alpha_q$ \\
    $k_{123}\cdot \bar{n}$& $\boldsymbol{\alpha_{12}}= \alpha_q$ & $\boldsymbol{\alpha_{23}} = \alpha_q$ & $\boldsymbol{\alpha_{13}} = \alpha_q$ \\
    \hline
    $k_{12}^2$ &  $k_{12}^2 = q^2$ & $\boldsymbol{\alpha_{2}\beta_{1}}$ & $\boldsymbol{\alpha_{1}\beta_{2}}$\\
    $k_{23}^2$ & $\boldsymbol{\alpha_{2}\beta_{3}}$ & $k_{23}^2 = q^2$ & $\boldsymbol{\alpha_{3}\beta_{2}}$ \\
    $k_{13}^2$ & $\boldsymbol{\alpha_{1}\beta_{3}}$ & $\boldsymbol{\alpha_{3}\beta_{1}}$ & $k_{13}^2 = q^2$\\
    \hline
    $k_{123}^2$ & $k_{12}^2 + \boldsymbol{\alpha_{12} \beta_3} + m^2$ & $k_{23}^2 + \boldsymbol{\alpha_{23} \beta_1} + m^2$ & $k_{13}^2 + \boldsymbol{\alpha_{13} \beta_2} + m^2$\\
    \hline
  \end{tabular}
  \caption{Possible large integration parameters combinations giving
    contribution in the region $(m^2)^{-2\ep}$.  We do not include
    contribution of the region with $\alpha_1
    \sim \beta_3 \sim m^2 $  and  $\alpha_2 \sim \beta_3 \sim m^2$ resulting
    in scaleless integrals.
  }
  \label{tab:densRescale2}
\end{table}

Integrals where \emph{two} light-cone components of partons' momenta can become
large simultaneously may possess an $(m^2)^{-2\ep}$ branch.

Similarly to the $(m^2)^{-\ep}$ branch considered earlier, we compute it by
removing the $\theta$-functions which involve large light-cone components and
extend the relevant integrations to the interval $[0,\infty)$. Since in the
region where $\alpha_i \sim \alpha_j \sim m^2$, momenta $k_i$ and $k_j$ are not
anymore constrained by the $\theta$-functions, it is useful to introduce a new
vector $q = k_i + k_j$ and treat an integral over momenta $k_i,k_j$ as a normal
phase-space integral at fixed $q$.

All possible combinations of large Sudakov components, together with simplified
propagators, are summarized in table~\ref{tab:densRescale2}. The integration
measures for all the relevant cases can be simplified as follows
\begin{align}
  \alpha_2\sim \alpha_3\sim m^2: & \nonumber\\
  \dFdttA & \to  \frac1{\normNep} \dm^d q [\dm k_1] \delta\left(1 - \beta_q  - \beta_1 \right)\delta\left(\alpha_1 - \beta_1 \right) \dm \Phi(q,k_2,k_3),\\
    \alpha_1\sim \alpha_3\sim m^2: & \nonumber\\
  \dFtdtA & \to  \frac1{\normNep} \dm^d q [\dm k_2] \delta\left(1 - \beta_q  - \beta_2 \right)\delta\left( \alpha_2 - \beta_2 \right)
 \dm \Phi(q,k_1,k_3),
 \\
  \alpha_1\sim \alpha_2\sim m^2: & \nonumber\\
  \dFttdA & \to \frac1{\normNep} \dm^d q [\dm k_3] \delta\left(1 - \beta_q  - \beta_3 \right)\delta\left( \alpha_3 - \beta_3 \right)
    \dm \Phi(q,k_1,k_2), \label{eq:dPhiRegmm2ttdA}\\
  \alpha_1\sim \alpha_2\sim m^2: & \nonumber\\
  \dFttdB & \to \frac1{\normNep}\dm^d q [\dm k_3] \delta\left(1 - \beta_q  - \alpha_3 \right)\delta\left( \beta_3 - \alpha_3 \right)
  \dm \Phi(q,k_1,k_2),
   \label{eq:dPhiRegmm2ttdB}\\
  \alpha_1\sim \alpha_2\sim m^2: & \nonumber\\
  \dFtttB & \to \frac1{\normNep}\dm^d q [\dm k_3] \delta\left(1 - \beta_q  - \alpha_3 \right)\theta\left( \beta_3 - \alpha_3 \right)
  \dm \Phi(q,k_1,k_2),
\label{eq:dPhiRegmm2tttB}
\end{align}
where 
\begin{equation}
  \dm \Phi(q,k_i,k_j) =
  \frac{1}{\normNep^2} [\dm k_i] [\dm k_j ] \; \delta^{(d)}\left( q - k_{ij}\right).
\end{equation}

Since integrations over two partons' momenta that add up to $q$ correspond to
standard phase-space integrals, it is useful to discuss how to simplify them. To
study the most general form of such integrals, we consider the case with $q =
k_1 + k_2$, and note that all other momenta assignments can be analyzed
similarly. Hence, we consider the following family of integrals
\begin{equation}
  \label{eq:Intk1k2mm2ep}
  J_{a_1\dots a_6} = \frac{1}{\normNep^2} \int  \frac{[\dm k_1] [\dm k_2] \delta^{(d)}\left(q-k_1-k_2\right)}{\left(k_1\cdot n  \right)^{a_1} \left(k_2\cdot n  \right)^{a_2} \left(k_1\cdot \bar{n}  \right)^{a_3} \left(k_2\cdot \bar{n}  \right)^{a_4}
  \left(k_1\cdot n  + \beta_3\right)^{a_5} \left(k_2\cdot n  + \beta_3\right)^{a_6}}.
\end{equation}
One can use the integration-by-parts technology to reduce them to the minimal
set of master integrals. This approach was employed in the previous
(same-hemisphere) calculation as described in
refs.~\cite{Baranowski:2021gxe,Baranowski:2022khd}. We have also used it in the
current computation alongside with an alternative approach that we describe
below.

Since the non-trivial part of the integral in eq.~\eqref{eq:Intk1k2mm2ep} is the
integration over azimuthal angles, we start with discussing it. It is convenient
to consider the reference frame where $q = (q_0, \vec 0)$, parametrize momenta
of partons $1$ and $2$ using energies and angles, and integrate over energies
removing the $\delta$-functions. As we explain below, remaining integrals over
angles can be related to cases familiar from earlier
studies~\cite{Somogyi:2011ir,Lyubovitskij:2021ges}.

Using energies and angles to parametrize light-like vectors, we  write  the required inverse propagators 
as follows
\begin{equation}
  \label{eq:betaM0}
  \begin{split}
k_1 \cdot n & = \frac{\beta_q}{2}\left( \rho_1 \cdot \rho_n \right),
  \quad
 k_2 \cdot n  = \frac{\beta_q}{2}\left( \rho_1 \cdot \bar{\rho}_n \right),\\
  % \quad
 k_1 \cdot \bar{n} & = \frac{\alpha_q}{2}\left( \rho_1 \cdot \rho_{\bar{n}} \right),
  \quad
  k_2 \cdot \bar{n}  = \frac{\alpha_q}{2}\left( \rho_1 \cdot \bar{\rho}_{\bar{n}} \right).
  \end{split}
\end{equation}
In the above expressions, we introduced the light-like vectors $\rho_1 = ( 1,
\vec{k}_1/|\vec{k}_1|)$, $\rho_n = \left( 1, \vec{n}/|\vec{n}|\right)$, and
$\rho_{\bar{n}} = \left( 1, \vec{\bar{n}}/|\vec{\bar{n}}|\right)$, as well as
conjugate vectors $ \bar \rho_n$ and $\bar \rho_{\bar n}$ which are defined as
parity-transformed versions of $\rho_n$ and $\rho_{\bar n}$.

Inverse propagators that contain further Sudakov parameters in addition to
scalar products shown in eq.~\eqref{eq:betaM0} can be written as scalar products
of a light-like vector and a time-like vector. To see this, we introduce two
four-vectors $\rho_m = \left( 1, \lambda \vec{n}/|\vec{n}| \right)$ and
$\bar{\rho}_m = \left( 1, - \lambda \vec{n}/|\vec{n}| \right)$, with
\begin{equation}
  \label{eq7.35}
  \lambda = \frac{\beta_q}{\beta_q + 2 \beta_3},
  \end{equation}
  and  write 
  \begin{equation}
    \label{eq:betaMM}
    k_1 \cdot n + \beta_3  = \left(\frac{\beta_q}{2} + \beta_3  \right)\left( \rho_1 \cdot \rho_m \right),
    \quad
    k_2 \cdot n + \beta_3  = \left(\frac{\beta_q}{2} + \beta_3  \right)\left( \rho_1 \cdot \bar{\rho}_m \right).
  \end{equation}
The integral in eq.~\eqref{eq:Intk1k2mm2ep} can be written as follows
\begin{equation}
  \label{eq:Jk1k2Ang}
  J_{a_1\dots a_6} =
  \frac{1}{\normNep^2}
  \frac{\Omega^{(d-1)}}{(8\pi^2)^{d-1}}
  \frac{\left( q^2 \right)^{d/2-2} \, 2^{a}}{
    \beta_q^{a_1+a_2} \alpha_q^{a_3+a_4}
  \left(\beta_q + 2 \beta_3  \right)^{a_5+a_6}} \Omega^{(d-1)}_{a_1\dots a_6}\left ( \rho_n, \bar{\rho}_n, \rho_{\bar{n}}, \bar{\rho}_{\bar{n}},\rho_{m},\bar{\rho}_{m} \right ),
\end{equation}
where $a=\sum\limits_{i=1}^6 a_i$ and the angular integral over directions of a
light-like vector $k$ with $k_0=1$ is defined by the following formula
\begin{equation}
  \label{eq:angIntDef}
  \Omega^{(d-1)}_{a_1\dots a_n}\left[ v_1,\dots, v_n \right] = \frac{1}{\Omega^{(d - 1)}} \int \frac{\dm \Omega^{(d-1)}_{k}}{
    \left( k\cdot v_1 \right)^{a_1}
    \left( k\cdot v_2 \right)^{a_2}\dots
    \left( k\cdot v_n \right)^{a_n}
  }.
\end{equation}
General results for such integrals, alongside with many specific cases, are
discussed in refs.~\cite{Somogyi:2011ir,Lyubovitskij:2021ges,Ahmed:2024pxr}.

For our purposes angular integrals can be further simplified using e.g.\ the
relation between $ k \cdot \rho$ and $k \cdot \bar \rho$ scalar products. The
required partial fractioning rules can be found in appendix~\ref{sec:angIntApp}.
Once these relations are applied, the final set of $\Omega$-integrals consists
of integrals with at most two denominators. Such integrals have been computed in
ref.~\cite{Somogyi:2011ir} and we provide explicit expressions for them in
appendix~\ref{sec:angIntApp}. Furthermore, since angular integrals with
arbitrary powers of the denominators are known in terms of hypergeometric
functions, linear relations between integrals with different powers can easily
be constructed ~\cite{Lyubovitskij:2021ges}, and used to simplify the resulting
expressions without the need to express hypergeometric functions through
elementary functions.

After the partial fractioning and the simplification of angular integrals, the
original integral whose $(m^2)^{-2\ep}$ branch defines the required boundary
value is given by a linear combination of several simple integrals. Further
integrations over components of the vector $q$ can be often performed following
the discussion of the $nnn$ case in ref.~\cite{Baranowski:2021gxe}. However, new
challenges arise for the $nn \bar n$ case, as we explain below.

The first point is that in the same-hemisphere configuration the zero-jettiness
constraint $\delta(1 - \beta_{123})$ implies that $\beta_3 = 1- \beta_{12} =
1-\beta_q$ and, therefore, entries in eq.~\eqref{eq:betaMM} simplify. The second
point is that in the same hemisphere configuration, integrals with three
$\theta$-functions do not appear as master integrals. For the choice of large
Sudakov components that we discuss, this implies that not only $\beta_3$ but
also $\alpha_3$ is equal to $1-\beta_q$. All in all, this implies that the
integration over $k_3$ can be easily performed, and only the integration over
$q$ remains. For $nn \bar n$ integrals with three $\theta$-functions -- the
examples of which are shown in eq.~\eqref{eq:dPhiRegmm2tttB} -- the required
computation is more complex. Below we consider an example which illustrates how
such integrals can be calculated.

We are interested in the $(m^2)^{-2\ep}$ branch of the following integral
%$\theta\theta\theta$ integral
% INT["Tpppb6tttarm", 7, 2639, 8, 0, {1, 1, 1, 1, 0, 0, 2, 0, 0, 1, 0, 1}]
\begin{equation}
  \label{eq:exampleReg2int}
  J = \int\frac{\dFtttB}{
    \left(k_2\cdot \bar{n}  \right)
    \left( k_{13}\cdot n \right)
    \left(k_{123}^2 + m^2  \right)^2}.
\end{equation}
The relevant contribution comes from the region $\alpha_1\sim\alpha_2\sim m^2$.
The integration measure in this case is given by eq.~\eqref{eq:dPhiRegmm2tttB}.
Simplifying the propagators in the limit $\alpha_{1,2} \sim m^2 \to \infty$, we
obtain
\begin{equation}
  \label{eq:exampleReg2int-J12}
  J \sim  \frac{1}{\normNepTrip}\int \dm^d q [ \dm k_3 ] \; \frac{\delta\left(1 - \beta_q  - \alpha_3 \right)\theta\left( \beta_3 - \alpha_3 \right)}{\left(q^2 + \alpha_{12}\beta_3 + m^2  \right)^2}
  \int [ \dm k_1 ] [ \dm k_2 ] \frac{\delta^{(d)}\left( q - k_{12}\right)}{\left( k_2\cdot \bar{n} \right)\left( k_1\cdot n  \ + \beta_3\right)}.
  \end{equation}
Using eqs.~(\ref{eq:Intk1k2mm2ep},\ref{eq:Jk1k2Ang},\ref{eq:miAngInt000xx0}), we write the  integral over momenta $k_{1,2}$  as follows 
  \begin{equation}
    J_{000110}
    =
    \frac{1}{\normNep^2}
    \frac{\Omega^{(d-1)}}{(8\pi^2)^{d-1}}
    \frac{4 \left( q^2 \right)^{-\ep}}{\alpha_q \left( \beta_q+2 \beta_3 \right)} I_{d-1;1,1}^{(1)}\left(\bar{\rho}_{\bar{n}}\cdot \rho_{m},\rho_m^2 \right).    
  \end{equation}
Scalar products that appear as arguments of the function $I_{d-1;1,1}^{(1)}$ are given by 
\begin{equation}
\rho_{\bar{n}}\cdot \rho_m  = 1-\lambda(1-2 u), \;\;\; \rho_m^2  = 1-\lambda^2, 
\end{equation}
where $\lambda$ can be found in  eq.~\eqref{eq7.35}
and $u$ reads 
\begin{equation}
u  = \frac{q^2}{(q \cdot n) (q \cdot \bar n)} = \frac{q^2}{\alpha_q \beta_q}.
\end{equation}
Finally, using the expression for the angular integral in
eq.~\eqref{eq:angInt-2-0m} and applying the transformation of the Appell
function shown in eq.~\eqref{eq:appelF1tr26}, we obtain the following result for
the angular integral
\begin{equation}
\begin{split}
 I_{d-1;1,1}^{(1)}\left(\bar{\rho}_{\bar{n}}\cdot \rho_{m},\rho_m^2 \right) & = 
 \frac{( 2 \ep-1) (\beta_q + 2 \beta_3) }{4 \ep (\beta_q + \beta_3)}
 \\
& \phantom{= {}} \times  F_1\left(1; 1, -\ep; 1 - 2 \ep; \frac{q^2}{\left( \beta_3 + \beta_q  \right) \alpha_q},
   \frac{\beta_q}{\beta_3 + \beta_q}  \right),
   \end{split}
\end{equation}
so that the complete integral over momenta $k_1,k_2$ becomes
\begin{equation}
  J_{000110} =
      \frac{1}{\normNep^2}
    \frac{\Omega^{(d-1)}}{(8\pi^2)^{d-1}}
 \frac{( 2 \ep-1) \left( q^2 \right)^{-\ep} }{ \ep (\beta_q + \beta_3)\alpha_q}
 F_1\left(1; 1, -\ep; 1 - 2 \ep; \frac{q^2}{\left( \beta_3 + \beta_q  \right) \alpha_q},
   \frac{\beta_q}{\beta_3 + \beta_q}  \right).
\end{equation}

For the remaining integrations over $q$ and $k_3$, we employ the Sudakov
decomposition for both of these momenta. The integration measure reads
\begin{equation}
  \label{eq:dqdk3Sudakov}
  \frac{1}{\normNep} \dm^d q \; [\dm k_3] = 
  \frac{\Omega^{(d-2)}}{4}
  \frac{\dm \alpha_3 \dm \beta_3 \dm \alpha_q \dm \beta_q \dm q_{\perp}^2}{\left( \alpha_3 \beta_3 q_{\perp}^2 \right)^\ep}, 
\end{equation}
where we used the fact that the integrand is independent of the directions of
$q^\mu_\perp$ and $k^\mu_{3,\perp}$. We then change the variable $\beta_3 =
\alpha_3/r_3$ eliminating the Heaviside function $\theta(\beta_3 - \alpha_3)$,
and write $q^2 = \alpha_q \beta_q t, q_\perp^2 = \alpha_q\beta_q (1-t)$. We then
find that the integral in eq.~\eqref{eq:exampleReg2int-J12} becomes
\begin{equation}
    \begin{split} 
  \label{eq:exampleReg2int-final}
  J & \sim 
      \frac{2   \Gamma^2(1 - \ep) \Gamma(1 + 2 \ep)}{(m^2)^{1 + 2 \ep}}
  \int\limits_0^1 \dm \alpha_3 \dm \beta_q  \delta(1 - \alpha_3 - \beta_q) (\alpha_3 \beta_q)^{1 - 2 \ep} \\
  \times & \int\limits_0^1 \dm t \dm r_3 \dm w \frac{  r_3^{-\ep} (\alpha_3 + \beta_q r_3)^{-\ep} 
(\alpha_3 + \beta_q r_3 t)^{-1 + 2 \ep} (1 - w)^{-1 - 2 \ep} (\alpha_3 + \beta_q r_3 - \beta_q r_3 w)^{\ep} }
{(\alpha_3 + \beta_q r_3 - \beta_q r_3 t w) (1 - t)^{\ep} t^{\ep}},
\end{split}
\end{equation}
where we integrated over $\alpha_q$ and introduced the integral representation
for the Appell function, c.f.\ eq.~\eqref{eq:hfIntReprF1}, using an auxiliary
variable $w$. Next, we remove the $\delta$-function integrating over $\alpha_3$,
and change the variable $\beta_q = y/(r_3 + y - r_3 y)$. Integration over $r_3$
can be expressed through the hypergeometric function ${}_2F_1$, c.f.\
eq.~\eqref{eq:hfIntReprF21}. Applying the transformation in
eq.~\eqref{eq:f21trInv1}, we find
\begin{equation}
\begin{split}
J & \sim \frac{2  \Gamma^2(1 - \ep)
  \Gamma(1 + 2 \ep) }{(m^2)^{1 + 2 \ep} (1 - \ep)}
   \int\limits_0^1 \dm w\dm y  \frac{(1 - y)^{1 - 2 \ep}         
            {}_2F_1\left(1 - \ep, \ep; 2 - \ep; 1 - y  \right)
  }{y^{\ep} (1 - w)^{1 + 2 \ep} 
    (1 - w y)^{-\ep}
  }
  \\  
  & \phantom{\sim {}}  \times 
  \int\limits_0^1 \dm t \;
   \frac{
           (1 - t)^{-\ep} t^{-\ep}
  }{(1 - w t y)  (1 - y(1-t))^{1 - 2 \ep}}.
\end{split}
\end{equation}
The integral over $t$ can be expressed through the first Appell
function~\eqref{eq:hfIntReprF1}, which reduces to the ${}_2F_1$ function thanks
to eq.~\eqref{eq:F1toF21}. Using the transformation eq.~\eqref{eq:f21trInv1}, we
derive the two-dimensional integral representation for the $(m^2)^{-2\ep}$
branch of $J$
\begin{equation}
\begin{split}
J & \sim  \frac{2  \Gamma^4(1 - \ep) \Gamma(1 + 2 \ep) }{(m^2)^{1 + 2 \ep} (1 - \ep) \Gamma(2 - 2 \ep) }
  \int\limits_0^1 \dm w \dm y  \frac{(y(1-y))^{1 - 2 \ep}}{(1 - w)^{1+2\ep}}  \\
  & \phantom{\sim {}} \times
  {}_2F_1\left(1, 2 - 2 \ep; 2 - \ep; 1 - y\right)
  {}_2F_1 \left(1 - 2 \ep, 1 - \ep; 2 - 2 \ep; y (1 + w(1-y))\right).
\end{split}
\end{equation}
The only divergence of the integral occurs at the point $w\to 1$, which can be
easily subtracted. The subtracted integral is finite, can be expanded in $\ep$
under the integral sign and integrated using
\texttt{HyperInt}~\cite{Panzer:2014caa}. The final result reads
\begin{equation}
  \begin{split}
    J &\sim
      - \frac{\pi^2}{12 \ep}
      - \left(\frac{\pi^2}{6} + \frac{\log{(2)} \pi^2}{4} + \frac{7}{8} \zeta_3\right)\\
    & - \ep \left(
      \frac{\pi^2}{3}
      + \frac{\pi^2}{2} \log{(2)}
      + \frac{7}{4} \zeta_3
      - \frac{3}{8}\log{(2)}^4
      + \frac{\pi^2}{2}\log{(2)}^2
      + \frac{17}{160} \pi^4
      - 9 \textrm{Li}_{4} \left( 1/2 \right)
      \right)\\
    & -\ep^2 \bigg(
      \frac{2}{3} \pi^2
      + \log{(2)} \pi^2
      + \frac{7}{2} \zeta_3
      + \frac{17}{80} \pi^4
      + \pi^2  \log^2{(2)}
      - \frac{3}{4} \log^4{(2)}
      - 18 \textrm{Li}_{4}\left( 1/2 \right)
      \\
    &
      + \frac{211}{480}  \pi^4 \log{(2)}
      + \frac{\pi^2}{6}\log^3{(2)}       
      - \frac{3}{40} \log^5{(2)}
      - \frac{43}{96} \pi^2 \zeta_3
      - \frac{1023}{64} \zeta_5
      + 9 \textrm{Li}_{5} \left( 1/2 \right)    
      \bigg)
    \\
    % ep^3 part
    & - \ep^3 \bigg(
      \frac{4}{3}\pi^2
      +2 \pi^2 \log{(2)}
      +7 \zeta_3
      +\frac{17}{40}\pi^4
      -\frac{3}{2}\log^4{(2)}
      +2 \pi^2 \log^2{(2)}
      - 36 \textrm{Li}_{4}\left( 1/2 \right)
      \\
    & 
      -\frac{3}{20}\log^5{(2)}
      +\frac{1}{3}\pi^2 \log^3{(2)}
      +\frac{211}{240}\pi^4\log{(2)}
      -\frac{1023}{32}\zeta_5
      -\frac{43}{48}\pi^2\zeta_3
      +18 \textrm{Li}_{5}\left( 1/2 \right)
    \\
    & 
      +\frac{221}{960}\pi^4 \log^2{(2)}
      -\pi^2 \zeta_3 \log{(2)} 
      -\frac{375}{16}\zeta_3^2    
      +\frac{121}{2016}\pi^6
      -\frac{1}{80}\log^6{(2)}
      +\frac{1}{32}\pi^2\log^4{(2)}
      \\
    &
      -\frac{1}{4}\pi^2\textrm{Li}_{4}\left( 1/2 \right)
      +\frac{159}{4} \zeta_{-5,-1}
      -9 \textrm{Li}_{6}\left( 1/2 \right)
      \bigg)
      + \mathcal{O}(\ep^4).        
  \end{split}
\end{equation}

The $(m^2)^{-2\ep}$ branch in all integrals can be calculated once a pair of
large parameters is identified, the expansion of denominators is performed
according to table~\ref{tab:densRescale2}, angular integrals as in
eq.~\eqref{eq:Jk1k2Ang} introduced and computed, and subsequent integrations
over $q$ and the remaining Sudakov parameters are performed.

\section{Numerical checks}
\label{sec:numchecks}
Calculation of the soft function requires computing a large number of
complicated unconventional integrals. Because of this, checking them using
alternative methods is critical. Techniques that are usually employed for this
purpose include the sector-decomposition \cite{Borowka:2017idc} and
Mellin-Barnes methods~\cite{Czakon:2005rk,Belitsky:2022gba}. In our case,
however, the complex structure of divergencies of the relevant integrals, as
well as high perturbative order, makes direct application of public codes based
on these methods unfeasible. Hence, care is required to organize the numerical
checks to make them manageable. Below we discuss how this can be achieved.

All integrals that we computed numerically can be split into three groups.
First, there are integrals without propagators that involve scalar products of
soft-parton momenta; such integrals are calculated using the standard sector
decomposition approach. Second, there are master integrals with complicated
propagators that can be efficiently dealt with using the Mellin-Barnes method.
The third group includes integrals that depend on $1/k_{123}^2$ and they pose
the biggest challenge. The problem is to find a suitable parametrization that
minimizes the number of sectors that appear, if overlapping singularities are
treated using the sector decomposition method, or a suitable compact
representation if the Mellin-Barnes technology is employed.

All integrals that we may want to check numerically can be written as follows
\begin{equation}
  \label{eq:miIIrep}
  \mathcal{I}\left[ \{f_1,f_2,f_3\}, \{a_1\dots a_7\}, \{b_1\dots b_7\},\{c_1\dots c_4\} \right]
  = \int \frac{\textrm{d}\Phi_{f_1 f_2 f_3}}{A_1^{a_1}\dots A_7^{a_7} B_1^{b_1}\dots B_7^{b_7}C_1^{c_1}\dots C_4^{c_4}},
\end{equation}
where the integration measure is determined by the vector $\{f_1,f_2,f_3\}$
\begin{equation}
  \label{eq:dPhiDefNumInt}
  \textrm{d}\Phi_{f_1 f_2 f_3} = \frac{1}{\normNepTrip}[\dm  k_1 ] [\dm k_2 ]
  [ \dm  k_3 ]f_1(\bar{\gamma}_1 - \gamma_1)f_2(\bar{\gamma}_2 - \gamma_2)f_3(\bar{\gamma}_3 - \gamma_3)\delta(1-\gamma_1-\gamma_2-\gamma_3),
\end{equation}
with $\gamma_i = \beta_i, \bar{\gamma}_i = \alpha_i$ for $f_i=\delta,\theta$ and
$\gamma_i = \alpha_i, \bar{\gamma}_i = \beta_i$ for $f_i=\bar{\theta}$. We have
split all appearing propagators into three groups. Propagators in the first
group involve vector $n$
\begin{align}
  \label{eq:denomintarosIIn}
  A_1 & = k_1\cdot n, \quad A_2 = k_2\cdot n,   \quad A_3 = k_3\cdot n,\nonumber\\
  A_4 & = k_{12}\cdot n,   \quad A_5 = k_{23}\cdot n,   \quad A_6 = k_{31}\cdot n, \quad 
  A_7 = k_{123} \cdot n.
\end{align}
The second group contains propagators that depend on vector $\bar{n}$
\begin{align}
  \label{eq:denomintarosIInb}
  B_1 & = k_1\cdot \bar{n}, \quad   B_2 = k_2\cdot \bar{n}, \quad  B_3 = k_3\cdot \bar{n}, \nonumber\\
  B_4 & = k_{12}\cdot \bar{n}, \quad  B_5 = k_{23}\cdot \bar{n}, \quad  B_6 = k_{31}\cdot \bar{n}, \quad  B_7 = k_{123} \cdot \bar{n}.
\end{align}
The last group contains propagators that depend on the scalar products of soft momenta
\begin{align}
  \label{eq:denomintarosIIsp}
  C_1 & = k_1 \cdot k_2, \quad C_2 = k_2 \cdot k_3, \quad C_3 = k_1 \cdot k_3, \quad  C_4 = k_{123}^2.
\end{align}

\subsection{Integrals without the $1/k_{123}^2$ propagator}
\label{sec:no123intsNum}

We will discuss first the numerical computation of integrals in
eq.~\eqref{eq:miIIrep} with $c_3=c_4=0$. This assignment covers all integrals in
the soft function that do not contain the propagator $1/k_{123}^2$.
To derive a suitable representation for such integrals, we change integration
variables $\bar{\gamma}_i=\gamma_i/s_i^2$, $ 0 < s_i < 1$, to resolve all
$\theta$-functions' constraints.\footnote{The same variable transformations can
be applied to integrals with $\delta$-function constraints. We explain below how
the corresponding integral representations can be constructed. } Two inverse
denominators that involve scalar products between momenta of soft partons read
\begin{align}
  f_i&=\theta, f_j = \theta: & k_{ij}^2 &= \frac{\gamma_i\gamma_j}{s_i^2 s_j^2}\left( (s_i - s_j)^2  + 4 s_i s_j \frac{t_{ij}^2}{ t_{ij}^2 + \bar{t}_{ij}^2 } \right),  \label{eq:spAng01sVarsTT}\\
  f_i&=\theta, f_j = \bar{\theta}: &k_{ij}^2 &= \frac{\gamma_i\gamma_j}{s_i^2 s_j^2}\left( (1-s_i s_j)^2  + 4 s_i s_j \frac{t_{ij}^2}{ t_{ij}^2 + \bar{t}_{ij}^2 } \right),  \label{eq:spAng01sVarsTTB}
\end{align}
where we introduce angle variables $t^2_{ij}/(t^2_{ij} + {\bar t}_{ij}^2) =
\sin^2\left( \phi_{ij}/2\right) \in [0,1]$ and the notation $\bar{x}=1-x$ for
all variables $ 0 < x < 1$ appearing in this section. To satisfy the
$\delta$-function constraint $\delta(1-\gamma_1-\gamma_2-\gamma_3)$ we apply the
variable transformation
\begin{equation}
  \label{eq:gammaPerm}
 \{\gamma_{\sigma(1)},\gamma_{\sigma(2)},\gamma_{\sigma(3)}\} \to \{x,y
\bar{x},\bar{y} \bar{x}\},   
\end{equation}
where $\sigma$ is a permutation of the set $\{1,2,3\}$. Different permutations
of the set $\{\gamma_1,\gamma_2,\gamma_3\}$ lead to different integrand
expressions, which, after the integration, should give identical results,
providing useful internal consistency checks of the numerical calculation.

Combining the Jacobians of all transformations described above, we obtain the
following expression for the integration measure, where integrations on the
right-hand side are confined to a unit hypercube
\begin{equation}
\begin{split}
  \label{eq:dPhi01no123}
  & \dm \Phi_{f_1f_2f_3}
   =  \dm x \dm y  \prod_{i=1}^3   \frac{2 \gamma_i
    \dm s_i }{s_i^3} \; \times \,\bar{x} 
    \left( \frac{s_1 s_2 s_3}{ x y \bar{y} \bar{x}^2} \right)^{2\ep}
    \\
  &
    \quad \times  \frac{\Omega^{(d-3)}}{\Omega^{(d-2)}} \frac{\dm t_{12}}{\left( t_{12}\bar{t}_{12} \right)^{2\ep}} \left( \frac{ 2}{t_{12}^2 + \bar{t}_{12}^2} \right)^{1-2\ep}
    \frac{\Omega^{(d-3)}}{\Omega^{(d-2)}}  \frac{\dm t_{23}}{\left( t_{23}\bar{t}_{23} \right)^{2\ep}} \left( \frac{ 2}{t_{23}^2 + \bar{t}_{23}^2} \right)^{1-2\ep}.
\end{split}
\end{equation}
For each constraint $f_i=\theta$ or $f_i=\bar{\theta}$ we replace $\gamma_i$
according to eq.~\eqref{eq:gammaPerm} choosing a particular permutation
$\sigma$, and for each $f_i=\delta$ we set $s_i =1$ and drop the integration
over $s_i$ by setting $\gamma_i \dm s_i \to 1$ in eq.~\eqref{eq:dPhi01no123}.
The propagators $1/A_i$ and $1/B_i $, $i = 1,\dots,7$, are expressed in terms of
$x,y,s_i$ following the variable change specified by eq.~\eqref{eq:gammaPerm}.
The two propagators $1/C_{1,2}$ are written as in
eqs.~(\ref{eq:spAng01sVarsTT},\ref{eq:spAng01sVarsTTB}) depending on the
relevant constraints. If $f_i=\delta$, the two expressions become equal once
$s_i=1$ is employed.

Using the measure in eq.~\eqref{eq:dPhi01no123} we can, in principle, compute
integrals defined in eq.~\eqref{eq:miIIrep} without the propagator $1/C_4$,
using public codes \texttt{FIESTA}~\cite{Smirnov:2021rhf} or
\texttt{pySecDec}~\cite{Borowka:2017idc}, which implement the sector
decomposition approach. However, since the inverse propagator
eq.~\eqref{eq:spAng01sVarsTT} has a (line) singularity inside the integration
region at $s_i = s_j$, and since such singularities are not automatically found
by any of the public codes, we need to consider the cases $s_i < s_j$ and $s_j <
s_i$ separately. If we do this, all singularities appear at the boundary of the
integration domain, and such cases \emph{are} processed automatically by
\texttt{FIESTA}~\cite{Smirnov:2021rhf} and
\texttt{pySecDec}~\cite{Borowka:2017idc}. Nevertheless, since integrals that we
consider are multidimensional, it turns out to be beneficial to analyze the
existence of potential singularities at the upper integration boundaries
\emph{manually}, avoiding the automatic option for doing that in public codes.
We have mostly used \texttt{FIESTA}~\cite{Smirnov:2021rhf} to compute integrals
with the analytic regulator and we emphasize that the introduction of the
analytic regulator $\dm\Phi \to \dm\Phi
\left(\gamma_1\gamma_2\gamma_3\right)^\nu$ does not impact our discussion. Using
these methods, we have checked numerically both regular integrals calculated in
section~\ref{sec:no123intsNoNu} and also the $1/\nu$ divergent parts of
integrals $I^{1 / \nu}_2$ and $I^{1 / \nu}_4$ considered in
section~\ref{sec:divergent-integral}.

\subsection{Calculation of integrals with $1/k_{123}^{2}$ using the Mellin-Barnes representation}
\label{sec:ints123num}

It turns out to be very difficult to perform numerical checks for integrals with
the propagator $1/k_{123}^2$ using the sector decomposition approach because
suitable parametrization of angular variables is hard to find. We avoid this by
constructing Mellin-Barnes representations for such integrals and using them for
the numerical evaluation.

Since the dependence on the momenta directions in the transverse space in such
integrals comes only from propagators which contain scalar products of soft
momenta, we focus on this part first. The most general form of the considered
integral reads
\begin{equation}
  \label{eq:intTrSplitMB}
  \int [\dm k_1] [\dm  k_2] [\dm  k_3]
  \frac{R\left( \left\{k_i\cdot n  \right\} ,  \left\{k_i\cdot \bar{n}  \right\} \right)}{(k_1\cdot k_2)^{a_{12}} (k_2\cdot k_3)^{a_{23}} (k_3\cdot k_1)^{a_{31}} (k_1\cdot k_2 + k_2\cdot k_3 + k_3\cdot k_1)^{a_{123}}},
\end{equation}
where the rational function $R$ contains propagators $1/A_{1,..,7}$ and
$1/B_{1,..,7}$, c.f.\ eqs.~(\ref{eq:denomintarosIIn},\ref{eq:denomintarosIInb}).
We focus on the case $a_{123} > 0$ and, as the first step, we split the
$1/k_{123}^2$ propagator using the Mellin-Barnes representation
\begin{equation}
  \label{eq:k123combMB}
  \frac{1}{(k_1\cdot k_2 + k_2\cdot k_3 + k_3\cdot k_1)^{\lambda}} =
  \frac{1}{\Gamma(\lambda)}\int\limits_{c-i\infty}^{c+i\infty} \frac{\dm z_1 \dm z_2}{\left(  2\pi i\right)^2}
  \frac{\Gamma(\lambda+z_{12})\Gamma(-z_1)\Gamma(-z_2)}{
    (k_1\cdot k_2)^{z_{12}+\lambda} (k_2\cdot k_3)^{-z_1} (k_3\cdot k_1)^{-z_2}
  },
\end{equation}
where $z_{12} = z_1 + z_2$. After this transformation, integration over
directions of soft momenta in the transverse space reduces to the calculation of
the following integral
\begin{equation}
  \label{eq:I123base}
  I(n_1,n_2,n_3) = \int 
  \frac{
  \dm \Omega_{1}^{(d-2)} \dm \Omega_{2}^{(d-2)} \dm \Omega_{3}^{(d-2)}
 }{(k_1\cdot k_2)^{n_1} (k_2\cdot k_3)^{n_2} (k_3\cdot k_1)^{n_3}},
\end{equation}
where powers $n_{1,2,3}$ are arbitrary.

To integrate over angles in the transverse space, it is useful to consider
auxiliary $(d-1)$-dimensional (Minkowski) vectors, which contain
$(d-2)$-dimensional transverse momenta. We write
\begin{equation}
  \label{eq:rhoDef}
  \rho_i = \left(1, \frac{\vec{k}_{i,\perp}}{|\vec{k}_{i,\perp}|} \right),
  \quad 
  \rho_i \cdot \rho_j = 1 - \frac{\vec{k}_{i,\perp} \cdot \vec{k}_{j,\perp}} {|\vec{k}_{i,\perp}||\vec{k}_{j,\perp}|},
  \quad
  \rho_i^2  = 0.
\end{equation}
The scalar products in eq.~\eqref{eq:I123base} take the form
\begin{equation}
  \label{eq:sp2rho}
    (k_{i}\cdot k_{j}) = \frac{\left(\sqrt{\alpha_i\beta_j}  - \sqrt{\alpha_j\beta_i}\right)^2}{2} + \sqrt{\alpha_i\beta_i\alpha_j\beta_j} (\rho_i\cdot \rho_j).
  \end{equation}
If $f_i = f_j = \delta$, the first term in the above equation vanishes. However,
in general this does not happen, and we need to split the scalar product
further. We write
  \begin{equation}
    \label{eq:kikjAngSplitMB}
    \frac{1}{(k_{i}\cdot k_{j})^\lambda } = \frac{1}{\Gamma(\lambda)}\int\limits_{c-i\infty}^{c+i\infty} \frac{\dm z}{2\pi i}
    \Gamma(-z)\Gamma(\lambda+z) \frac{2^{-z}\left(\sqrt{\alpha_i\beta_j}  - \sqrt{\alpha_j\beta_i}\right)^{2z}}{\left( \alpha_i\beta_i\alpha_j\beta_j \right)^{z/2 + \lambda/2} }\frac{1}{(\rho_{i}\cdot \rho_{j})^{z+\lambda}}.
  \end{equation}

Hence, to proceed further, we require the following integral
 \begin{equation}
  \label{eq:angintw12w23w31def}
  J(w_1,w_2,w_3)  = \left( \frac{1}{\Omega^{(d-2)}} \right)^3 \int \frac{\dm \Omega_{1}^{(d-2)} \dm \Omega_{2}^{(d-2)} \dm \Omega_{3}^{(d-2)}}{(\rho_1\cdot \rho_2)^{w_1} (\rho_2\cdot \rho_3)^{w_2} (\rho_3\cdot \rho_1)^{w_3}},
\end{equation}
for arbitrary values of $w_{1,2,3}$. Since, to the best of our knowledge, such
integral is not available in the literature, we briefly explain how to compute
it. First, we apply eq.~\eqref{eq:angInt-2-00} to integrate over $\rho_3$. The
hypergeometric function that appears in that equation depends on $1-\rho_1 \cdot
\rho_2 /2$ and obeys the standard series representation
$\sum\limits_{n=0}^{\infty} c_n \left(1 - (\rho_1\cdot \rho_2)/2 \right)^n$.
Writing
\begin{equation}
\sum\limits_{n=0}^{\infty} c_n \left(1 - (\rho_1\cdot \rho_2)/2 \right)^n = 
\sum\limits_{n=0}^{\infty} c_n \sum\limits_{k=0}^{\infty}
\frac{\Gamma(k-n)}{\Gamma(-n) k!}\left( \frac{\rho_1\cdot\rho_2}{2} \right)^k,
\end{equation}
we can integrate over directions of $\rho_2$ using eq.~\eqref{eq:angInt-1-0}.
Finally, we sum over $k$ and $n$ using the definition of a hypergeometric
function, and simplify the result to $\Gamma$-functions since after each
summation, the obtained hypergeometric function is evaluated at one. We find
\begin{equation}
  \label{eq:angintw12w23w31}
J(w_1,w_2,w_3)  = \frac{\Gamma^3(1-\ep)\Gamma\left( 1 - 2\ep -w_{123}\right) }{\pi^{3/2} 2^{6\ep+w_{123}}\Gamma(1-2\ep)}
  \frac{\prod\limits_{i=1}^3\Gamma\left( \frac{1}{2}-\ep -w_i\right)}{\prod\limits_{i=1}^{3}\prod\limits_{j=1}^{i-1}\Gamma\left(1-2\ep -w_{ij}\right)},
\end{equation}
where $w_{ij} = w_i + w_j$ and $w_{123} = w_1+w_2+w_3$. The integral
$J(w_1,w_2,w_3)$ is a symmetric function of its arguments; it is normalized to
give $J(0,0,0) = 1$ and it reduces to known expressions when any of its
arguments vanishes.

The Mellin-Barnes representation that we derived so far is five-dimensional
since we required two integrations for $1/k_{123}^2$ and then one for each of
the three terms $1/k_i k_j$.\footnote{If some of the constraints are
$\delta$-functions, the number of Mellin-Barnes integrations is smaller.} For
each constraint $f_i=\theta$ or $f_i=\bar{\theta}$, we change the large variable
$\bar{\gamma_i} = \gamma_i/s_i^2$ and obtain
\begin{equation}
  \label{eq:sPar}
  \int\limits_0^{\infty} \dm \bar{\gamma}_i \, \delta(1-\gamma_i- \cdots ) \theta(\bar{\gamma}_i-\gamma_i) f(\bar{\gamma}_i)= \int\limits_0^1 d s_i \left( \frac{2\gamma_i}{s_i^3} \right)\delta(1-\gamma_i- \cdots ) f\left( \frac{\gamma_i}{s_i^2} \right).
\end{equation}
To resolve the $\delta$-function constraint, we apply a variable transformation
from eq.~\eqref{eq:gammaPerm} choosing a particular permutation. We find
\begin{equation}
  \label{eq:delta01int}
  \int\limits_0^1 \textrm{d} \gamma_1\textrm{d} \gamma_2\textrm{d} \gamma_3 \delta(1-\gamma_{1}-\gamma_{2}-\gamma_{3}) f(\gamma_{1},\gamma_{2},\gamma_{3}) =
  \int\limits_0^1 \textrm{d} x \textrm{d} y \, \bar{x} f(x,y\bar{x},\bar{x}\bar{y}).
\end{equation}

It remains to convert the remaining part of the integral into a Mellin-Barnes
representation suitable for numerical integration. To do that, we enable
integrations over $x,y$ and $s_i$ by introducing additional Mellin-Barnes
integrations to convert sums of these variables, that may appear in the inverse
propagators from groups $A$ and $B$, into variables' products. From
eq.~\eqref{eq:sp2rho} one notices that for the $nnn$ case, i.e. when
$f_i=f_j=\theta$, the representation of $k_i \cdot k_j$ leads to the factor
$\left( s_i - s_j \right)^p$. We would like to avoid such quantities since,
after the application of the Mellin-Barnes splitting formula, they produce a
factor $(-1)^z$ which conflicts with the $z$-integration contour that runs to
complex infinity. The solution is to split the integral into two parts where
$s_i<s_j$ and $s_i > s_j$.

The remaining integral consists of polynomial factors in variables $x,y,s_1\dots
s_n$ raised to some powers. We change the integration variables $v_i\to
u_i/(1+u_i)$ for all $v_i \in \{x,y,s_1\dots s_n\}$, mapping all $[0,1]$
integration intervals to $[0,\infty)$ intervals. Integrals that we are left with
have the following generic representation
\begin{equation}
  \label{eq:zInfIntForMBcreate}
  \int\limits_0^\infty \prod\limits_{i=1}^{n+2}\dm u_i \prod\limits_{j=0}^m P_j(\vec{u})^{w_j},
\end{equation}
where $P_j(\vec u)$ are polynomials of variables $u_1,u_2,...,u_n$. For such
integrals, the Mellin-Barnes representations with the minimal number of
integrations can be constructed using the package
\texttt{MBcreate}~\cite{Belitsky:2022gba}. This package employs the recursive
splitting of polynomials $P_j$ into two parts by introducing a single
Mellin-Barnes integration at each step, until all $u_i$-integrations can be
performed using the following formula
\begin{equation}
  \label{eq:gamMB}
  \int\limits_0^\infty \dm u\, u^p(a u +b )^q = \frac{\Gamma(p+1)\Gamma(-p-q-1)}{\Gamma(-q)} a^{-p-1} b^{p+q+1}.
\end{equation}
We have found that for successful application of the package \texttt{MBcreate}
to integrals of interest, we need to extend it by an additional (simple)
integration rule
\begin{equation}
  \label{eq:gamMBadd}
  \int\limits_0^\infty \dm u \frac{u^p (1+u)^q} {((1+u)^2 + u^2)^r}  =
  \int\limits_{c-i\infty}^{c+i\infty} \frac{\textrm{d} z}{2\pi i}
  \frac{\Gamma(z+r)\Gamma(-z)\Gamma(1+p+2z)\Gamma(2 r-p-q-1)}{\Gamma(r)\Gamma(2z+2r-q)}.
\end{equation}
In practice, we use \texttt{MBcreate}~\cite{Belitsky:2022gba} to construct the
representation for all possible permutations $\sigma$ corresponding to different
ways of resolving the $\delta$-function constraints in
eq.~\eqref{eq:delta01int}, and select one with the minimal number of MB
integrations.

The above steps are sufficient to derive the minimal Mellin-Barnes
representations for all required integrals. We note that we work with
analytically regularized integrals which allows us to check that our master
integrals are regulated dimensionally. To find suitable integration contours and
to perform analytic continuation of Mellin-Barnes representations, we use the
package \texttt{MBresolve}~\cite{Smirnov:2009up}, which is applicable also when
the analytic regulator is present. Furthermore, we find it useful to introduce
additional parameter to the angular integral in eq.~\eqref{eq:angintw12w23w31}
which ensures that all poles of the Mellin-Barnes representation are separated.
We do this by shifting all arguments in $J(w_1,w_2,w_3)$ by an infinitesimal
positive-definite quantity, which we take to be equal to the analytic regulator.
Integrals, analytically continued and expanded in all small parameters, are
integrated numerically using the package
\texttt{MB}~\cite{Czakon:2005rk}.\footnote{We have modified the \texttt{MB}
package in such a way that a \texttt{C} code suitable for numerical integration
is generated. The modified package can be downloaded from
\url{https://github.com/apik/mbc}.}

We will illustrate this discussion by describing    numerical computation of the integral
 \begin{align}
J=\int \frac{\dFdttB}{\left(k_2\cdot k_3\right) \ \left(k_{123}^2\right) \left(k_3\cdot \bar{n}\right) \left(k_{123}\cdot \bar{n}\right)}.
 \end{align}
 As the first step, we introduce the factor
$\beta_1^{\nu}\beta_2^{\nu}\alpha_3^{\nu}(\rho_1 \cdot \rho_2)^{-\nu}(\rho_1
\cdot \rho_3)^{-\nu} (\rho_2 \cdot \rho_3)^{-\nu}$, $\nu \to 0^{+}$, into the
integrand to achieve the pole separation in the Mellin-Barnes integral. Then,
following the steps outlined in this section we obtain the Mellin-Barnes
representation
 \begin{align}
     J= \int\limits_{c-i\infty}^{c+i\infty} \; \left ( \prod_{n=1}^{10} 
 \frac{\textrm{d} z_n}{2\pi i} \right) \; 
 \left(J_a( \vec z) +J_b(\vec z )\right),
 \end{align}
 where 
 \begin{align}
 \begin{split}
J_a&=2^{3 \nu-4 \ep-2 z_3-2 z_4-2 z_5}  \frac{\Gamma (1-\ep)^2 \Gamma (-z_{10}) \Gamma (-z_3) \Gamma (-z_4) \Gamma (-z_5) \Gamma (-z_6) \Gamma (z_6+1) }{\pi  \Gamma \left(\frac{1}{2}-\ep\right) \Gamma (-2 z_4) \Gamma (3 \nu-6 \ep)\Gamma (z_1+z_2+2) \Gamma (-2 z_5-z_7) } \\
& \phantom{= {}} \times \frac{\Gamma (-z_7)\Gamma (-z_8) \Gamma (z_8+1) \Gamma (z_1+z_2+1) \Gamma (z_3-z_1) \Gamma (z_5-z_2) \Gamma (z_7-2 z_4) }{\Gamma (2 \nu-2 \ep-z_1-z_4-z_5-1) \Gamma (2 \nu-2 \ep-z_2-z_3-z_4-1) }\\
& \phantom{= {}} \times\frac{\Gamma (3 \nu -6 \ep+z_9)\Gamma (z_1+z_2+z_4+2) \Gamma (2 z_3+2 z_4-z_7+1)}{ \Gamma (2 \ep-2 z_{10}+z_2+z_3+z_4+1) \Gamma (2 \nu-2 \ep+z_1+z_2-z_3-z_5+1)}\\
& \phantom{= {}} \times \frac{ \Gamma \left(\nu-\ep+z_1-z_3+\frac{1}{2}\right) \Gamma (\nu-2 \ep+z_{10}-z_2)\Gamma \left(\nu-\ep+z_2-z_5+\frac{1}{2}\right) }{\Gamma (2 \ep+z_1-z_4+z_5-2 z_6+z_7+1) }\\
& \phantom{= {}} \times \frac{\Gamma \left(\nu-\ep-z_1-z_2-z_4-\frac{3}{2}\right)\Gamma (-2 \nu+4 \ep-z_1+z_6-z_9) }{\Gamma (4 \nu-4 \ep+z_1-z_4+z_5-2 z_6+z_7+z_8+2 z_9+2)} \\
& \phantom{= {}} \times \Gamma (3 \nu-2 \ep-z_3-z_4-z_5-1) \Gamma (2 \nu-4 \ep+z_1+z_{10}-z_6+z_9+1)\\
& \phantom{= {}} \times \Gamma (2 \ep-2 z_{10}+z_2-z_3-z_4+z_7) \Gamma (4 \nu-4 \ep+z_1-z_4-z_5-2 z_6+2 z_9)\\
& \phantom{= {}} \times \Gamma (2 \ep+z_1-z_4+z_5-2 z_6+z_7+z_8+1) \Gamma (-2 z_5-z_7-z_8-1) \\
&\phantom{= {}} \times\Gamma (-\nu+2 \ep-z_{10}+z_2-z_9) \Gamma (2 z_5+z_7+z_8+2),
\end{split}\\
\begin{split}
J_b&=2^{3 \nu-4 \ep-2 z_3-2 z_4-2 z_5} \frac{\Gamma (1-\ep)^2 \Gamma (-z_{10}) \Gamma (-z_3) \Gamma (-z_4) \Gamma (-z_5) \Gamma (-z_6) \Gamma (z_6+1)}{\pi  \Gamma \left(\frac{1}{2}-\ep\right) \Gamma (-2 z_5) \Gamma (3 \nu-6 \ep) \Gamma (z_1+z_2+2) }  \\
& \phantom{= {}} \times \frac{\Gamma (-z_7) \Gamma (-z_8) \Gamma (z_8+1) \Gamma (z_1+z_2+1) \Gamma (z_3-z_1) \Gamma (z_5-z_2) \Gamma (z_7-2 z_5)}{ \Gamma (-2 z_4-z_7) \Gamma (2 \nu-2 \ep-z_1-z_4-z_5-1) \Gamma (2 \nu-2 \ep-z_2-z_3-z_4-1)  } \\
& \phantom{= {}}  \times \frac{\Gamma (z_1+z_2+z_4+2) \Gamma (2 z_3+2 z_5-z_7+1) \Gamma (-2 z_4-z_7-z_8-1) }{\Gamma (2 \nu-2 \ep+z_1+z_2-z_3-z_5+1) \Gamma (2 \ep+z_1+z_4-z_5+2 z_6+z_7+3)} \\
& \phantom{= {}} \times \frac{\Gamma \left(\nu-\ep+z_1-z_3+\frac{1}{2}\right) \Gamma \left(\nu-\ep+z_2-z_5+\frac{1}{2}\right)}{\Gamma (4 \ep-z_1+2 z_{10}+z_4-z_5+z_7+z_8+2)} \\
& \phantom{= {}} \times \frac{\Gamma \left(\nu-\ep-z_1-z_2-z_4-\frac{3}{2}\right) \Gamma (3 \nu-2 \ep-z_3-z_4-z_5-1) }{\Gamma (-4 \nu+10 \ep+z_1+z_2+z_3+z_5+2 z_6-2 z_9+3)}  \\
& \phantom{= {}} \times  \Gamma (-\nu+2 \ep+z_1-z_{10}+z_6-z_9+1) \Gamma (-2 \nu+4 \ep+z_1+z_2+z_6-z_9+2) \\
& \phantom{= {}} \times \Gamma (2 \nu-4 \ep-z_1+z_{10}-z_2-z_6+z_9-1) \Gamma (2 z_4+z_7+z_8+2)\\
& \phantom{= {}} \times \Gamma (-4 \nu+10 \ep+z_1+z_2-z_3-z_5+2 z_6+z_7-2 z_9+2)\Gamma (3 \nu-6 \ep+z_9) \\
& \phantom{= {}} \times \Gamma (2 \ep+z_1+z_4-z_5+2 z_6+z_7+z_8+3)\Gamma (4 \ep-z_1+2 z_{10}-z_4-z_5) \\
& \phantom{= {}} \times \Gamma (\nu-2 \ep-z_1+z_{10}-z_6-1).
 \end{split}
\end{align}
 We use the MB package to expand the above integrand in the parameters $\nu$ and
$\ep$, and take the limit $\nu \to 0^{+}$, $\ep \to 0$ afterwards. Integrating
over Mellin-Barnes parameters, the following numerical result is obtained
\begin{align}
J_{\textrm{MB}}&=\frac{0.104167}{\ep^4}+\frac{0.25}{\ep^3}+\frac{1.7820119(14) }{\ep^2}-\frac{2.077362(8) }{\ep} \nonumber \\
& \phantom{= {}} -100.8647(18) 
-1049.07(4) \ep 
-8238(2) \ep^2.
\end{align}
Upon computing the same integral by solving the differential equations in the
auxiliary parameter, as described in section~\ref{sec:DEs}, we find
\begin{align}
J_{\textrm{DE}}&=\frac{0.104167}{\ep^4}+\frac{0.250000}{\ep^3}+\frac{1.78201}{\ep^2}-\frac{2.07736}{\ep}
\nonumber \\
& \phantom{= {}} -100.863
-1048.99 \ep-8238.03 \ep^2.
\end{align}
The two results are in very good agreement with each other and, in fact, this
level of agreement is typical for the many comparisons that we performed.

Out of 139 master integrals with the $1/k_{123}^2$ propagator 107 are checked
through all required orders in $\ep$ with at most a $5 \%$ deviation for the
central value. For the remaining 32 integrals the precision rapidly deteriorates
once we move to higher terms in the $\ep$ expansion. The total number of
evaluations used to obtain the value of a particular integral by means of
Monte-Carlo integration varies greatly from integral to integral leading to
vastly different runtimes needed for obtaining accurate numerical results.

\begin{figure}[h]
  \centering
  \includegraphics[width=0.8\textwidth]{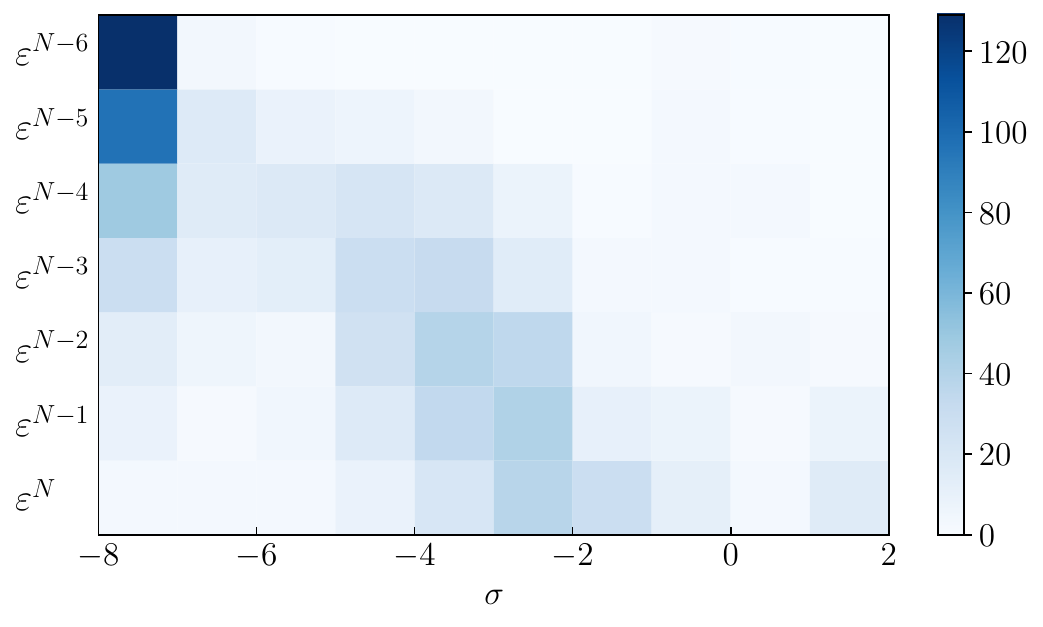}
  \caption{
  Number of master integrals with propagator $1/k_{123}^2$ that have  specific relative errors as defined in eq.~\eqref{eq:MBrelativerror}, calculated for each order in $\ep$. Here $\ep^{N}$ denotes the highest order in $\ep$ required for each individual integral in the final expression of the soft function.}
  \label{fig:mbIntsPlot}
\end{figure}

The numerical results are summarized in the density plot
fig.~\ref{fig:mbIntsPlot} where the relative error
\begin{align}
    \sigma= \log_{10}\left|\frac{J_{\textrm{DE}}-J_{\textrm{MB}}}{J_{\textrm{DE}}} \right|,
    \label{eq:MBrelativerror}
\end{align}
is calculated for each order in the $\ep$-expansion. We note that results for
all integrals expanded through all but the very last order in the
$\ep$-expansion are indirectly checked by our ability to reproduce divergencies
of the soft function. On the other hand, we do not see anything particular when
moving from the $\ep^{N-1}$-row to the $\ep^{N}$-row in
fig.~\ref{fig:mbIntsPlot} strongly suggesting that standard deterioration of the
numerical precision occurs for higher orders in the expansion when the
Mellin-Barnes method is employed.

\subsection{Numerical checks at finite $m^2$}
\label{sec:ints123numFinM}

Determination of complicated integrals from the differential equations at finite
$m^2$, as discussed in section~\ref{sec:DEs}, is a critical element of our
approach to compute the zero-jettiness soft function. Hence, high-quality
numerical checks of integrals at finite $m^2$ are very much needed. Among other
things, they give us confidence in the correctness of the filtered system of
differential equations and of the boundary conditions at $m^2 = \infty$. The
goal of this section is to discuss how such checks are performed.

We note that integrals where a mass-like parameter is introduced into a
propagator $1/k_{123}^2$ are less divergent than the original ones, since $m^2$
acts as an infra-red regulator. It is possible to construct the Mellin-Barnes
representations for such integrals, following the discussion in the previous
section. However, quite often, this does not allow high-precision evaluation of
 sufficiently many terms in the $\ep$-expansion, because such
representations contain one additional Mellin-Barnes integration in comparison
to the $m^2 = 0$ case. Because of this, we look for a way to perform such checks
for numerous $m^2$-dependent integrals, using the sector decomposition methods.
The idea is to numerically compute many \emph{nearly-finite} integrals, for
which the proliferation of sectors can be avoided, and then use the
integration-by-parts reduction to relate them to the master integrals that we
actually employ and would like to check.

To explain how this works, we introduce three angles between vectors $k_{1,2,3}$
in the $(d-2)$-dimensional transverse space. The scalar products read
\begin{align}
  \label{eq:angles}
  % 12
  (k_{1}\cdot k_{2})  & = \frac{\left(\sqrt{\alpha_1\beta_2}  - \sqrt{\alpha_2\beta_1}\right)^2}{2} + \sqrt{\alpha_1\beta_1\alpha_2\beta_2}\left( 1-\cos{\theta_{12}} \right),\nonumber\\
  % 23
  (k_{2}\cdot k_{3})  & = \frac{\left(\sqrt{\alpha_2\beta_3}  - \sqrt{\alpha_3\beta_2}\right)^2}{2} + \sqrt{\alpha_2\beta_2\alpha_3\beta_3}\left( 1-\cos{\theta_{23}} \right),\\                       
  % 13
  (k_{1}\cdot k_{3})  & = \frac{\left(\sqrt{\alpha_1\beta_3}  - \sqrt{\alpha_3\beta_1}\right)^2}{2} + \sqrt{\alpha_1\beta_1\alpha_3\beta_3}\left(1 - \cos{\theta_{12}} \cos{\theta_{23}} - \sin{\theta_{12}} \sin{\theta_{23}} \cos{\theta_{13}}  \right),\nonumber                             
\end{align}
where $\theta_{ij}$ are the relevant angles. We change variables to map the
three angles onto a unit hypercube; the variables are introduced in such a way
that the appearance of square roots in the integrand is avoided. Explicit
formulas read
\begin{alignat}{3}
  \label{eq:angles01}
  \sin \theta_{12} & = \frac{2 t_1 \bar{t}_1}{\bar{t}_1^2 + t_1^2}, \quad
  & \cos \theta_{12} &= \frac{\bar{t}_1 - t_1}{\bar{t}_1^2 + t_1^2},\\
  \sin \theta_{23} & = \frac{2 t_2 \bar{t}_2}{\bar{t}_2^2 + t_2^2}, \quad
  & \cos \theta_{23} &= \frac{\bar{t}_2 - t_2}{\bar{t}_2^2 + t_2^2},\\
  \sin \theta_{13} & = 2\sqrt{\lambda_3\bar{\lambda}_3}, \quad
  & \cos \theta_{13} & = 1 - 2\lambda_3,
\end{alignat}
where, as before, $\bar x = 1- x$, for all variables. The integration measure
becomes
\begin{align}
  \label{eq:dk123transAngs}
  \left( \Omega^{(d-2)} \right)^{-3}
  &  \dm \Omega_{1}^{(d-2)} \dm \Omega_{2}^{(d-2)} \dm \Omega_{3}^{(d-2)}= \\
  \frac{\Omega^{(d-3)}\Omega^{(d-4)}}{\left( \Omega^{(d-2)} \right)^2}
  & 
    \frac{\dm t_1  \, 2^{1-2\ep} }{\left( t_1 \bar{t}_1 \right)^{2\ep}\left( t_1^2 + \bar{t}_1^2 \right)^{1-2\ep}}
    \frac{\dm t_2  \, 2^{1-2\ep} }{\left( t_2 \bar{t}_2 \right)^{2\ep}\left( t_2^2 + \bar{t}_2^2 \right)^{1-2\ep}}
      \frac{\dm \lambda_3 \, 2^{-1-2\ep} }{\left( \lambda_3 \bar{\lambda}_3 \right)^{1+\ep}},\nonumber
  \end{align}
  where all integrations on the right -hand side are restricted to the $[0,1]$
intervals. The remaining integrals over the light-cone components are simplified
using eq.~\eqref{eq:sPar} to resolve $\theta$-function constraints, and
eq.~\eqref{eq:delta01int} to eliminate the jettiness $\delta$-function. For each
pair of same-hemisphere emissions with $f_i=f_j=\theta$, we split the
integration into two regions $s_i < s_j$ and $s_j < s_i$ to avoid the line
singularity. Since in the most complex integrals with three $\theta$-functions,
only two partons can be emitted into the same hemisphere, there is only one
scalar product where such a splitting is required. Performing these steps, we
obtain an eight-dimensional integral over a unit hyper-cube in variables
$x,y,s_1,\dots ,s_n,t_1,t_2,\lambda_3$. In general, these integrals have a
complicated structure of infra-red divergences and even if some divergences are
``screened'' because of the $m^2$ parameter in the $1/k_{123}^2$ propagator,
they are still very difficult to calculate. Part of the reason for this is an
interplay of ultra-violet and infra-red divergences in many master integrals,
which becomes hidden if all integration variables are mapped onto a unit
hypercube. This observation suggests that one possibility to simplify the
computation of mass-dependent integrals, is to choose them to be
ultra-violet-finite.

Hence, instead of calculating $m^2$-dependent master integrals, we find it more
beneficial to find a (large) set of less divergent integrals, compute them
numerically with high precision, and 
express them through master integrals to obtain constraints on the latter. In
fact, by calculating a large enough number of such quasi-finite integrals and
comparing them with the results of the reduction and with the solutions of the
differential equations, we obtain a very powerful tool for checking master
integrals at finite $m^2$.

Because we need many integrals for such checks, it is useful to develop an
automated procedure to construct them. To this end, we note that since we have
access to numerical solutions for all master integrals with $m^2$, we can select
a subset of integrals $\vec{F}$ which are $\ep$-finite. Because we work at
finite $m^2$, any derivative of any such integral with respect to $m^2$, is also
finite. Hence, it can be ``easily'' computed numerically and also expressed
through the master integrals using the integration-by-parts reduction.

Although the procedure described above is relatively straightforward, it is
important to notice that the construction of the $\ep$-expansion requires care
since the integration measure in eq.~\eqref{eq:dk123transAngs} exhibits
artificial divergences at $\lambda_3 \to 0$ and $\lambda_3 \to 1$. We have found
it to be convenient to subtract these divergences from the integrands and add
integrated subtracted terms back. Properly subtracted integrands can be expanded
in $\ep$ and integrated numerically with \texttt{CUBA}~\cite{Hahn:2004fe}.
Results for finite integrals obtained in this way provide the most accurate
numerical checks of the finite-$m^2$ master integrals, especially for higher
orders of the $\ep$-expansion.

Two comments are in order here. First, one can extend the list of integrals
$\vec F$ by including there also integrals with simple divergences, e.g.\ the
ultra-violet ones. Second, it is convenient to express the derivatives of
integrals $\vec F$ through master integrals recursively, using the differential
equation, instead of performing the real reduction, see eq.~\eqref{eq6.13}

Finally, we note that for some integrals and their derivatives, that enter the
differential equations, it was not possible to find finite or easily-calculable
divergent integrals that would allow the above checks. In such cases we had to
work with integrals whose divergence structure is complex, and involves
overlapping divergences. Unfortunately, the numerical integration rapidly
becomes very challenging due to a large number of terms that appear after the
sector decomposition of all possible overlapping divergences.

To reduce the complexity of the integration, it is crucial to find integrals
with as simple a divergence structure as possible. We do such an analysis before
the actual integral calculation using the following technique. Our starting
point is the parametrization of the $(d-2)$-dimensional transverse momenta
integration measure shown in eq.~\eqref{eq:dk123transAngs}, complemented with
integrations over other variables over the unit hypercube. After splitting the
integration domain in case of the same-hemisphere emissions to avoid the ``line
singularities'', all divergencies of the integral are located at the boundaries
of the integration domain, i.e. either at zero or at one.

Our approach to detecting potential divergences of the integrals is inspired by
techniques developed for the analytical regularization of parametric
integrals~\cite{Panzer:2014gra} and for finding finite
integrals~\cite{vonManteuffel:2014qoa}. Suppose we integrate over a set of
variables $\{v_i\}$ over the unit hypercube. It is useful to switch to new
variables $v_i = z_i/(1+z_i)$ remapping all divergences from $v_i = 1$ to $z_i =
\infty$. Integrals of interest become
\begin{equation}
  \label{eq:infIntPolys}
  \int\limits_0^\infty \prod\limits_{i=1}^n \frac{\dm z_i}{z_i^{1-\nu_i}} \; \prod\limits_{j=1}^m P_j^{a_j}\left(z_1,\dots,z_n \right).
\end{equation}
Potential divergences of such integrals can be identified by considering subsets
of variables $\{z_1,...,z_n\}$ and studying limits when the variables from the
subset approach zero or infinity. Specifically, we consider two non-intersecting
subsets of $Z_n = \{z_1,\dots,z_n\}$, that we will refer to as $Z_0$ and
$Z_\infty$, such that $Z_0 \cup Z_\infty \in Z_n$. We then rescale all variables
$z_i \in Z_0$ by $z_i \to \lambda z_i$ and all $z_i \in Z_\infty$ by $z_i \to
z_i / \lambda$, and consider the limit $\lambda \to 0$. Upon doing this, the
integral receives an overall factor $\lambda^w$. The integral is divergent in
the above limit if
\begin{equation}
  \label{eq:divCritZinf}
  w \le 0.
\end{equation}
By considering all possible subsets $\{Z_0,Z_\infty\}$, it is possible to detect
all potential divergences of the integral. If divergences with a non-empty set
$Z_\infty$ are detected, we split the integration domain for each variable
$z_i\in Z_\infty$ and  remap divergences
that arise at $z_i \to \infty$ to zero by an appropriate change of variables.
Continuing this procedure, we obtain the set of integrals where \emph{all}
potential divergencies are located at the origin.

To select candidates that are best suited for numerical evaluation, we choose
integrals for which \emph{the minimal number of such splittings is needed}.
Since the remaining integrals after the splitting can only have divergences at
the origin, we prefer integrals with the smallest number of shortest $Z_0$
lists. The resulting integrals are computed numerically with
\texttt{FIESTA}~\cite{Smirnov:2021rhf}.\footnote{ We note that some finite-$m^2$
master integrals have also been checked with a patched version of
\texttt{pySecDec}~\cite{Borowka:2017idc}.}

\section{Soft function renormalization and checks from the renormalization group equation}
\label{sec:results}

The result for the N3LO soft function has been already presented in
ref.~\cite{Baranowski:2024vxg}. In this section and in a few appendices, we
collect all the different contributions that are needed to obtain this result.

We start with the bare soft function. Restoring the dependencies on $\tau$,
$s_{ab} = 2 p_a \cdot p_b$, and the normalization factor $P$, we write
\begin{equation}
  \label{eq:SFbareN3LO}
  S_{\tau,B} = \delta(\tau)
  + \frac{A_s}{\tau} \left( \frac{s_{ab}}{P^2 \tau^2} \right)^{\ep} S_1
  + \frac{A_s^2}{\tau} \left( \frac{s_{ab}}{P^2 \tau^2} \right)^{2\ep} S_2
  + \frac{A_s^3}{\tau} \left( \frac{s_{ab}}{P^2 \tau^2} \right)^{3\ep} S_3
  + \mathcal{O}\left( A_s^4 \right),
\end{equation}
where the expansion parameter reads 
\begin{equation}
\label{eq9.2}
A_s = \frac{g_s^2}{(4\pi)^{2-\ep} \Gamma(1-\ep)},
\end{equation} 
and $g_s$ is the bare strong coupling constant. The NLO and NNLO contributions
$S_{1,2}$, needed through higher orders in $\ep$, have been computed in
refs.~\cite{Monni:2011gb,Kelley:2011ng,Baranowski:2020xlp}, and can be found in
eq.~\eqref{eq:SBnnlo}. The N3LO part $S_3$ is comprised out of three terms
\begin{equation}
  \label{eq:S3rvSplit}
  S_3 = S_{\tau,B}^{\textrm{RVV}} + S_{\tau,B}^{\textrm{RRV}} + S_{\tau,B}^{\textrm{RRR}},
\end{equation}
which refer to the two-loop corrections to single-real emission, the one-loop
corrections to double-real emission, and the triple-real emission, respectively.

It is straightforward to compute the two-loop correction to the single-gluon
emission because the corresponding soft current is simple and because the
integration over the single-gluon phase space is easy. This contribution was
calculated in ref.~\cite{Chen:2020dpk}; we present its independent computation
in appendix~\ref{sec:RVVcorrections}.

One-loop corrections to real-emission amplitudes with two soft partons ($q \bar
q$ and $gg$) were calculated in refs.~\cite{Chen:2020dpk,Baranowski:2024ene}.
The RRV contribution in eq.~\eqref{eq:S3rvSplit} can be written as
\begin{equation}
S_{\tau,B}^{\textrm{RRV}} = \cos{(\pi \ep)} \left( S_{\textrm{RRV},gg}^{(3)} +
S_{\textrm{RRV},q\bar{q}}^{(3)} \right),
\end{equation}
  where the two quantities $S_{\textrm{RRV},gg}^{(3)},
S_{\textrm{RRV},q\bar{q}}^{(3)}$ can be found in ancillary files to
ref.~\cite{Baranowski:2024ene}.

Triple-real emission contribution to the N3LO soft function,
$S_{\tau,B}^{\textrm{RRR}}$, whose calculation is described in this paper, is
quite demanding. To describe it, we note that each of the perturbative
contributions in eq.~\eqref{eq:SFbareN3LO} admits a Laurent expansion in $\ep$
\begin{equation}
  \label{eq:SbareEpExp}
  S_n = \sum\limits_{k=1-2n}^\infty \Sbc{n,k} \; \ep^k.
\end{equation}
We also note that it is possible to predict all contributions to the N3LO soft
function $S_3$ through $S_{3,0}$ from the renormalization group equation that
the soft function satisfies, c.f.\ eq.~\eqref{eq:SlapRen}. Hence, using these
predictions, the results for $S_3^{\rm RVV}$ and $S_3^{\rm RRV}$, and
eq.~\eqref{eq:S3rvSplit}, it is possible to predict all terms in the expansion
of $S_{\tau,B}^{\textrm{RRR}}$ in $\ep$ through ${\cal O}(\ep^0)$. For this
reason, we do not show these terms and only say that our calculation of the
triple-real contribution does reproduce them. The coefficient
$S_{3,1}^{\textrm{RRR}}$, which cannot be predicted from the renormalization
group equation and the knowledge of other contributions to the soft function,
reads
\begin{equation}
\begin{split}
  \label{eq:SrrrBare31}
  S_{3,1}^{\textrm{RRR}}
  & = \CR^3
    \biggl(
    131072 \zeta_3^2
    -\frac{156928}{945}\pi^6
    \biggr)
    - \CR^2 \nf \TF
    \bigg(
    \frac{277376}{243}
    -\frac{21440}{81}\pi^2    
  \\  
  & \phantom{= {}}
    +\frac{70912}{9}\zeta_3
    +\frac{12544}{135}\pi^4
    -\frac{12800}{3}\pi^2 \zeta_3
    +61184 \zeta_5
    \bigg)
    - \CR^2\CA \bigg(
    \frac{540224}{243}
  \\
  & \phantom{= {}}
    - \frac{10496}{81}\pi^2    
    - \frac{218048}{9}\zeta_3
    - \frac{9112}{27}\pi^4
    + \frac{35200}{3}\pi^2 \zeta_3
    - 168256 \zeta_5
    - \frac{13928}{105}\pi^6
  \\
  & \phantom{= {}} 
    + 73104 \zeta_3^2    
    \bigg)
    - 18285.58462095074 \CR\CA^2    
    +  3809.242380482391 \CR\CA\nf\TF \\
  & \phantom{= {}} - 556.5414878890519 \CR\CF\nf\TF.
\end{split}
\end{equation}
We note that the $\CR^3$ and $\CR^2$ terms in eq.~\eqref{eq:SrrrBare31}
originate from iterated emissions; their calculation is discussed in
appendix~\ref{sec:iter-real-emiss}. Combining $S_{3,1}^{\rm RRR}$ with
contributions from the RVV result in \eqref{eq:Srvv3}, and the RRV result
from~\cite{Baranowski:2024ene}, we obtain
\begin{equation}
  \label{eq:Sbare31}
  \begin{split}
    \Sbc{3,1}
    & = \CR^3 \bigg(
      131072 \zeta_3^2
      -\frac{156928}{945}\pi^6
      \bigg)
      -\CR^2 \nf \TF
      \bigg(
      \frac{277376}{243}
      -\frac{21440}{81}\pi^2
      +\frac{70912}{9}\zeta_3
    \\
    & \phantom{= {}}
      +\frac{12544}{135}\pi^4
      -\frac{12800}{3}\pi^2\zeta_3
      +61184 \zeta_5      
      \bigg)
      -\CR^2 \CA \bigg(
      \frac{540224}{243}
      -\frac{10496}{81}\pi^2
    \\
    & \phantom{= {}}
      -\frac{218048}{9} \zeta_3
      -\frac{9112}{27}\pi^4
      +\frac{35200}{3} \pi^2\zeta_3
      -168256 \zeta_5
      +\frac{1024}{35}\pi^6
      +13104 \zeta_3^2
      \\
    & \phantom{= {}}
      +\CR \nf^2 \TF^2 \bigg(
      \frac{224512}{81}\zeta_3
      -\frac{839552}{2187}
      -\frac{1664}{27}\pi^2
      -\frac{9664}{405}\pi^4
      \bigg)
    \\    
    & \phantom{= {}}
      -8308.6818721153875355162222571 \CR \CA^2       
    \\
    & \phantom{= {}}
      +4352.6418957721480642594288751 \CR \CA \nf \TF
    \\
    & \phantom{= {}}
      -158.21578438939415938180801214 \CR \CF \nf \TF.
  \end{split}
\end{equation}

We need to renormalize the bare soft function and the bare strong coupling
constant that appear in eq.~\eqref{eq:SFbareN3LO}. The renormalization of the
strong coupling constant amounts to expressing the bare coupling $A_s$ through
the $\overline {\rm MS}$ coupling $\alpha_s(\mu)$ with the help of the following
formula
\begin{equation}
  \label{eq:AsRen}
  A_s = a_s(\mu)  \frac{\mu^{2\ep} e^{\ep \gamma_E}}{\Gamma(1-\ep)} Z_{a_s},
  \quad
  Z_{a_s} = 1 - \as \frac{\beta_0}{\ep} + \as^2 \left( \frac{\beta_0^2}{\ep^2} - \frac{\beta_1}{2\ep} \right) + \mathcal{O}\left( \as^3 \right),
\end{equation}
where $\beta_{0,1}$ are the expansion coefficients of the QCD $\beta$-function
that can be found in appendix~\ref{sec:RGandS}, and $a_{s}(\mu) =
\alpha_{s}(\mu) / (4 \pi)$.

The remaining divergences of the soft function are removed by a dedicated
$\overline {\rm MS}$ renormalization. It is convenient to perform this
renormalization working with the Laplace-transformed soft function, defined as
\begin{equation}
  \label{eq:SbareLap}
  \tilde{S}_B\left(a_s(\mu),L_S\right) = \int\limits_0^\infty \dm \tau e^{-\tau u} S_{\tau,B}(a_s(\mu)),
\end{equation}
In the above equation 
$L_S = \log{\left( \mu \bar{u} \sqrt{s_{ab}}/P \right)}$ and  $\bar{u} = u e^{\gamma_E}$.

The renormalized soft function reads
\begin{equation}
  \label{eq:SlapRen}
  \tilde{S}\left( a_s,L_S \right) = Z_s\left( a_s,L_S \right) \tilde{S}_{B}\left( a_s,L_S \right),
\end{equation}
where both $a_s$ and $L_S$ depend on the scale $\mu$. The constant $Z_s$
fulfills the renormalization group
equation~\cite{Korchemsky:1993uz,Korchemsky:1992xv,Monni:2011gb,Billis:2019vxg}
\begin{equation}
  \label{eq:GamSdef}
  \left( \frac{\partial}{\partial L_S} + \beta_{a_s} \frac{\partial}{\partial a_s} \right) \log{Z_s\left( a_s, L_S \right)} 
  = -4 \Gamma_{\textrm{cusp}}(a_s) L_S -2\gamma^s (a_s),
\end{equation}
where the $\beta$-function and the anomalous dimensions can be found in
appendix~\ref{sec:RGandS}. Solving the renormalization group equation, we find
the renormalization constant $Z_s$; the result is given in
appendix~\ref{sec:RGandS}. Using it in eq.~\eqref{eq:SlapRen}, we obtain the
renormalized soft function.

It is convenient to write the renormalized soft function in an exponential form.
To this end, we define the logarithm of the soft function $\tilde{s}(\Ls) =
\log{\tilde{S}(\Ls)}$ and write its perturbative expansion as
\begin{equation}
  \label{eq:SlogsExp}
  \tilde{s}(\Ls) = \log \left [ {\tilde S}(L_S) \right ]=
  \sum\limits_{i=1}^\infty \as ^i \sum\limits_{j=0}^{i+1} C_{i,j} L_S^j \,.
\end{equation}
Coefficients of terms with non-vanishing powers of logarithm $L_s$ in
eq.~\eqref{eq:SlogsExp} follow from the renormamlization group analysis and can
be calculated from the following recurrence relation
\begin{equation}
  \label{eq:CijLogsRec}
  C_{i,j} = \left( \frac{2}{j} \sum\limits_{k=1}^{i-1} C_{k,j-1} \beta_{i-1-k}  \right) - 2\left( \delta_{j,2} \Gcusp_{i-1} + \delta_{j,1} \Gsoft_{i-1} \right).
\end{equation}
In eq.~\eqref{eq:CijLogsRec}, $\beta_i$, $\Gcusp_i$ and $\Gsoft_i$ are the
$A_s$-expansion coefficients of the $\beta$-function, cusp anomalous dimension
and soft anomalous dimension, respectively. They can be found in
appendix~\ref{sec:RGandS}, c.f.\
eqs.~(\ref{eq:betaQCDcoeffs},\ref{eq:gamCuspCoeff},\ref{eq:gamSCoeff}). Explicit
expressions for coefficients $C_{ij}$ in eq.~\eqref{eq:CijLogsRec} are presented
in appendix~\ref{sec:RGandS} as well.

The coefficients $C_{i,0}$, $i=1,2,3$, have already been provided in
ref.~\cite{Baranowski:2024vxg}. We repeat them here for completeness
\begin{equation}
\begin{split}
  \label{eq:SrenC0}
  C_{1,0} & = -\CR \pi^2, \;\;\;\;
               C_{2,0}  = \CR \Bigg [ \nf \TF \left(\frac{80}{81}
              +\frac{154 \pi^2}{27}
              - \frac{104 \zeta_3}{9} \right)
             \\
              & - \CA \left( \frac{2140}{80}
              +  \frac{871\pi^2}{54}
              -  \frac{286\zeta_3}{9}
              -  \frac{14\pi^4}{15}\right)\Bigg ],
             \\
  C_{3,0} & = \CR \Bigg [ \nf^2\TF^2 \left(
              \frac{265408}{6561}
              - \frac{400\pi^2}{243}
              - \frac{51904\zeta_3}{243}
              + \frac{328\pi^4}{1215}
              \right)
             \\
              % numerical parts
              & \phantom{= {}} + \nf \TF \left(\CF X_{FF}
              +  \CA X_{FA}\right)
              + \CA^2 X_{AA}
              \Bigg ],
\end{split}
\end{equation}
The numerical constants truncated to sixteen significant digits are given
by~\cite{Baranowski:2024vxg}

\begin{alignat}{3}
  &X_{FF} && = &    68&.9425849800376,
                   \nonumber \\
  &X_{FA} && =  &  839&.7238523813981, 
                   \label{eq:Xnum} \\
  &X_{AA} && =  &-753&.7757872704537.
                   \nonumber
\end{alignat}

\section{Conclusion}
\label{sec:concl}

Recently~\cite{Baranowski:2024vxg}, we have presented the calculation of N3LO
QCD corrections to the zero-jettiness soft function. The technical details of
the computation were not discussed in that reference. The goal of this paper is
to fill this gap and to provide a detailed discussion of the theoretical methods
employed and developed by us in the course of that computation. A particularly
challenging contribution at N3LO QCD is the triple real-emission correction
since in this case the phase space includes one Heaviside functions per soft
parton making it especially complex.

Methods discussed in this paper encompass the extension of reverse
unitarity~\cite{Anastasiou:2002yz} to real-emission integrals with
$\theta$-function constraints, the need to introduce an analytic regulator and
the idea of ``filtering'', which allows us to remove this regulator from the
properly-constructed integration-by-parts identities, computation of phase-space
integrals using differential equations obtained by introducing an auxiliary
parameter, calculation of boundary conditions as well as numerical computation
of zero-jettiness phase-space integrals which turns out to be quite demanding.
We hope that theoretical methods developed by us in the context of the
zero-jettiness soft function computation and presented in this and earlier
papers~\cite{Baranowski:2022khd,Baranowski:2022vcn,Baranowski:2024ene}, will be
useful for extending the $N$-jettiness slicing scheme to arbitrary number of
hard partons at N3LO in perturbative QCD.

\section*{Acknowledgments}
We acknowledge useful conversation with with Arnd Behring and Fabian Lange.
Parts of the computation reported in this paper were performed at the HPC system
at the Max Planck Computing and Data Facility (MPCDF) in Garching.

The research of KM and AP was partially supported by the Deutsche
Forschungsgemeinschaft (DFG, German Research Foundation) under grant
396021762-TRR~257. The research of MD was supported by the European Research
Council (ERC) under the European Union’s research and innovation programme grant
agreement 949279 (ERC Starting Grant HighPHun). The work of DB is supported in
part by the Swiss National Science Foundation (SNSF) under contract
200020$\mathrm{\_}$219367.

\appendix

\section{Perturbative expansion coefficients}
\label{sec:RGandS}

In this appendix, we collect the various quantities that are needed for the
calculation of the N3LO zero-jettiness soft function, see e.g.\
ref.~\cite{Billis:2019vxg}. The QCD $\beta$-function is defined as follows
\begin{equation}
\beta_{a_s} = -2\ep a_s -2
a_s^2\sum\limits_{k=0}^\infty \beta_k a_s^k,
\end{equation}
where $a_s = \alpha_s/(4 \pi)$.
The two coefficients, relevant for the computation of the soft function at N3LO, read 
\begin{equation}
  \label{eq:betaQCDcoeffs}
  \beta_0 =  \frac{11}{3}\CA - \frac{4}{3}\nf\TF, \quad 
  \beta_1 =  \frac{34}{3} \CA^2 - \frac{20}{3} \CA \nf \TF - 4 \CF \nf \TF.
\end{equation}

We write the $\alpha_s$ expansion of the  cusp  anomalous dimension as 
\begin{equation}
\Gamma_{\textrm{cusp}} =  \as \left[
\Gcusp_0 + \as \Gcusp_1 + \as^2 \Gcusp_2 + \mathcal{O}\left( \as^3 \right)\right],
\end{equation}
with the expansion coefficients 
\begin{equation}
  \label{eq:gamCuspCoeff}
  \begin{split}   
    \Gcusp_{0} & =  4 \CR,\\
    \Gcusp_{1} & =  \CR \CA \left(\frac{268}{9}-\frac{4}{3}\pi ^2\right) 
                 -\frac{80}{9} \CR \nf \TF,\\
    \Gcusp_{2} & =  \CR \CA^2  \left(
                 \frac{88}{3}\zeta_3
                 +\frac{490}{3}
                 -\frac{536}{27}\pi^2
                 +\frac{44}{45} \pi^4
                 \right)\\
               & \phantom{= {}} +\CR  \CA \nf \TF \left(
                 - \frac{224}{3}\zeta_3
                 - \frac{1672}{27}
                 + \frac{160}{27}\pi^2
                 \right) \\
               & \phantom{= {}} +\CR \CF \nf \TF \left(
                 64 \zeta_3
                 -\frac{220}{3}
                 \right)
                 -\frac{64}{27} \CR \nf^2 \TF^2.
  \end{split}
\end{equation}
The soft anomalous dimension $\gamma^s$ reads 
\begin{equation}
\gamma^s =  \as \left[
\Gsoft_0 + \as \Gsoft_1 + \as^2 \Gsoft_2 + \mathcal{O}\left( \as^3 \right)\right],
\end{equation}
where 
\begin{equation}
  \label{eq:gamSCoeff}
  \begin{split}
  \Gsoft_{0} & = 0,\\
  \Gsoft_{1} & = \CR \CA \left(
                -28 \zeta_3
                +\frac{808}{27}
                -\frac{11}{9} \pi^2
                \right)
                +\left(
                \frac{4}{9} \pi^2
                -\frac{224}{27}
                \right) \CR \nf \TF,
                \\
  \Gsoft_{2} & =  \CR \CA^2 
                 \left(
                 -\frac{1316}{3}\zeta_3
                 +\frac{88}{9}\pi ^2\zeta_3
                 +192 \zeta_5
                 +\frac{136781}{729}
                 -\frac{6325}{243}\pi^2
                 +\frac{88}{45}\pi^4
                 \right)
                 \\
                 & \phantom{= {}} +\CR \CA \nf \TF \left(
                   \frac{1456}{27}\zeta_3
                   -\frac{23684}{729}
                   +\frac{2828}{243}\pi^2
                   -\frac{16}{15}\pi^4
                   \right)\\
              & \phantom{= {}} + \CR \CF \nf \TF \left(
                \frac{608}{9}\zeta_3
                -\frac{3422}{27}
                +\frac{4}{3}\pi^2
                +\frac{16}{45}\pi^4
                \right)\\
             & \phantom{= {}} +\CR \nf^2 \TF^2
                \left(
                \frac{448}{27}\zeta_3
                -\frac{8320}{729}
                -\frac{80}{81}\pi^2
                \right).
  \end{split}
\end{equation}
Using these quantities and the renormalization group equation, we compute the
soft-function renormalization constant. We find
\begin{equation}
  \label{eq:ZsNNNLO}
  \begin{split}
    Z_s & = 1  +
              \as \biggl[  
              \frac{\Gcusp_{0}}{\ep^2}
              +\frac{2 \Gcusp_{0}}{\ep}  \Ls
              \biggr]
              +\as^2 \biggl[ 
              \frac{\Gcusp_{0}^2}{2 \ep^4}
              -\frac{3 \beta _0 \Gcusp_{0}}{4\ep^3}
              +\frac{\Gcusp_{1}}{4 \ep^2}
              +\frac{\Gsoft_{1}}{2 \ep}
              +\biggl(
              \frac{2 \Gcusp_{0}^2}{\ep^3}
              -\frac{\beta_0\Gcusp_{0}}{\ep^2}
    \\
        & \phantom{= {}}
                        +\frac{\Gcusp_{1}}{\ep}
              \biggr)\Ls          
              +\frac{2 \Gcusp_{0}^2}{\ep^2}\Ls^2
              \biggr]
              +  \as^3 \biggr[
              \frac{\Gcusp_{0}^3}{6 \ep^6}
              -\frac{3 \beta _0 \Gcusp_{0}^2}{4 \ep^5}
          +\biggl( \frac{11}{18} \beta_0^2 \Gcusp_{0}+\frac{1}{4} \Gcusp_{0} \Gcusp_{1} \biggr)\frac{1}{\ep^4}
                              \\
        & \phantom{= {}}
          - \biggl(\frac{4}{9} \beta _1 \Gcusp_{0}+\frac{5}{18} \beta _0 \Gcusp_{1}
                 -\frac{1}{2} \Gcusp_{0} \Gsoft_{1}  \biggr)\frac{1}{\ep^3}
                 + \biggl(\frac{\Gcusp_{2}}{9}-\frac{1}{3} \beta _0 \Gsoft_{1}  \biggr)\frac{1}{\ep^2}
              +\frac{\Gsoft_{2}}{3 \ep}
              +\bigg(
          \frac{\Gcusp_{0}^3}{\ep^5}
          \\
        & \phantom{= {}} 
              -\frac{5 \beta _0 \Gcusp_{0}^2}{2 \ep^4}
              +\biggl(\frac{2}{3} \beta _0^2 \Gcusp_{0}+\frac{3}{2} \Gcusp_{0} \Gcusp_{1}  \biggr)\frac{1}{\ep^3}
              -\biggl(
              \frac{2}{3} \beta _1 \Gcusp_{0}
              +\frac{2}{3} \beta _0 \Gcusp_{1}
              -\Gcusp_{0} \Gsoft_{1} \biggr)\frac{1}{\ep^2}
              + \frac{2 \Gcusp_{2}}{3\ep}
          \bigg)\Ls
          \\
            & \phantom{= {}} 
              +\bigg(
              \frac{2\Gcusp_{0}^3}{\ep^4}
              -\frac{2 \beta _0 \Gcusp_{0}^2}{\ep^3}
              +\frac{2 \Gcusp_{1} \Gcusp_{0}}{\ep^2}
              \bigg)\Ls^2
              +\frac{4 \Gcusp_{0}^3}{3 \ep^3}\Ls^3
              \biggr] + \mathcal{O}(\as^4).
  \end{split}
\end{equation}

The last two ingredients that one needs are the soft functions at NLO and NNLO.
The coefficients of the $\ep$-expansion through the required orders
are~\cite{Baranowski:2020xlp}
\begin{equation}  
  \label{eq:SBnnlo}
  \begin{split}
    \Sbc{1,-1} & = 8 \CR,
                  \\
    \Sbc{2,-3} & = -32 \CR^2, \\
    \Sbc{2,-2} & = -\frac{16}{3}\CR \nf \TF + \frac{44}{3}\CR \CA, \\
    \Sbc{2,-1} & = \frac{64}{3}\pi^2\CR^2
                 -\frac{80}{9}\CR \nf \TF
                 + \CR \CA \left(
                 \frac{268}{9}
                 -\frac{4}{3} \pi^2
                 \right),
                 \\
    \Sbc{2,0} & = 512 \CR^2 \zeta_3
                +\CR \nf \TF \left(
                \frac{16}{9} \pi^2
                -\frac{448}{27}
                \right)
                + \CR \CA \left(
                -56\zeta_3
                +\frac{1616}{27}
                -\frac{44}{9} \pi^2 
                \right),
                \\
    \Sbc{2,1} & = \frac{64 }{5}\pi^4\CR^2
                + \CR \nf \TF  \left(
                \frac{320}{3} \zeta_3
                -\frac{320}{81}
                -\frac{112}{9} \pi^2
                \right)
                \\
               & \phantom{= {}}
                 + \CR \CA \left(
                 -\frac{880}{3}  \zeta_3
                 +\frac{8560}{81} 
                 -\frac{98}{45} \pi^4 
                 +\frac{268}{9} \pi^2 
                 \right),
                \\
  \Sbc{2,2} & =  \CR^2\left(
                6144  \zeta_5
                -\frac{1024}{3} \pi^2 \zeta_3
                \right)
                 \\
               & \phantom{= {}}
                 + \CR \nf \TF \left(
                 -\frac{1408}{3}  \zeta_3
                 +\frac{8576}{243} 
                +\frac{208}{45} \pi^4 
                 +\frac{640}{81} \pi^2 
                 \right)
                \\
               & \phantom{= {}}
                 + \CR \CA \left(
                -680  \zeta_5
                + 1072  \zeta_3
                + \frac{32}{3} \pi^2 \zeta_3
                +\frac{49664}{243} 
                -\frac{572}{45} \pi^4 
                -\frac{1472}{81} \pi^2
                \right)
                 ,\\
  \Sbc{2,3} & =
                \CR^2\left(
                \frac{10112}{945}\pi^6
                -4096  \zeta_3^2
              \right)
              + \CR \nf \TF \bigg(
                 \frac{9088}{3} \zeta_5
                 +\frac{13952}{27} \zeta_3
              -\frac{1280}{9} \pi^2 \zeta_3              
                 \\
               & \phantom{= {}}
                 +\frac{138688}{729}
                 -\frac{832}{45} \pi^4 
                 -\frac{4000}{81} \pi^2 
                 \bigg)
                 + \CR \CA \bigg(
                 -\frac{24992}{3} \zeta_5
                 -504 \zeta_3^2
                 -\frac{31456}{27}\zeta_3
                 \\
               & \phantom{= {}}
                 +\frac{3520}{9} \pi^2 \zeta_3
                 +\frac{270112}{729} 
                 -\frac{40}{63} \pi^6 
                 +\frac{1876}{45} \pi^4
                 +\frac{9664}{81} \pi^2
                 \bigg).
  \end{split}
\end{equation}
Divergent contributions to 
the soft function are related to anomalous dimensions and 
$\beta$-function coefficients. We summarize  these relations below 
\begin{equation}
  \label{eq:SpoleEquations}
  % NLO
  \begin{split}
    \Sbc{1,-1} & = 2 \Gcusp_{0},
                 \quad 
                 \Sbc{1,0}= 0,\\
                 % NNLO
    \Sbc{2,-3} & =  -2 \Gcusp_{0}^2,
                 \quad
                 \Sbc{2,-2}=  \beta_0 \Gcusp_{0},
    \\
    \Sbc{2,-1} & = -2 \boxS{\Sbc{1,1}} \Gcusp_{0}
                 +\frac{4}{3} \pi^2 \Gcusp_{0}^2+\Gcusp_{1}                
                 ,\\
    \Sbc{2,0} & = - 2 \boxS{\Sbc{1,2}} \Gcusp_{0}
                + 2 \beta_0 \boxS{\Sbc{1,1}}
                -\frac{1}{6} \pi^2 \beta_0 \Gcusp_{0} 
                + 32 \zeta_3 \Gcusp_{0}^2+2 \Gsoft_{1},\\
                % NNNLO
    \Sbc{3,-5} & = \Gcusp_{0}^3,
                 \quad
                 \Sbc{3,-4}= -\frac{3}{2} \beta_0 \Gcusp_{0}^2,
    \\
    \Sbc{3,-3} & =  \frac{3}{2} \boxS{\Sbc{1,1}} \Gcusp_{0}^2
                 +\frac{2}{3} \beta_0^2 \Gcusp_{0}
                 -2 \pi^2 \Gcusp_{0}^3-\frac{3}{2}
                 \Gcusp_{1} \Gcusp_{0}
                 ,\\
    \Sbc{3,-2} & = \frac{3}{2} \boxS{\Sbc{1,2}} \Gcusp_{0}^2
                 -\frac{15}{4} \beta_0 \boxS{\Sbc{1,1}} \Gcusp_{0}
                 + \frac{9}{4} \pi^2 \beta_0 \Gcusp_{0}^2+\frac{1}{3} \beta_1 \Gcusp_{0}
                 +\frac{4}{3} \beta_0 \Gcusp_{1}-3 \Gcusp_{0} \Gsoft_{1}
                 -64 \zeta_3 \Gcusp_{0}^3,\\
    \Sbc{3,-1} & = -\frac{3}{2} \boxS{\Sbc{2,1}} \Gcusp_{0}
                 -\frac{3}{2} \boxS{\Sbc{1,3}} \Gcusp_{0}^2
                 -\frac{3}{4} \beta_0 \boxS{\Sbc{1,2}} \Gcusp_{0}
                 + \boxS{\Sbc{1,1}} \biggl(
                 -\frac{3}{4} \Gcusp_{1}
                 +3 \beta_0^2 
                 -\pi^2  \Gcusp_{0}^2
                 \biggr)
    \\
               & \phantom{= {}}
                 -\frac{1}{3} \pi^2 \beta_0^2 \Gcusp_{0}
                 -\frac{4}{15} \pi^4 \Gcusp_{0}^3
                 +2 \pi^2 \Gcusp_{1} \Gcusp_{0}
                 +\frac{2}{3}\Gcusp_{2}
                 +72 \beta_0 \zeta_3 \Gcusp_{0}^2
                 +4 \beta_0 \Gsoft_{1}
                 ,\\
    \Sbc{3,0} & = -\frac{3}{2} \boxS{\Sbc{2,2}} \Gcusp_{0}
                +3 \beta_0 \boxS{\Sbc{2,1}}
                -\frac{3}{2} \boxS{\Sbc{1,4}} \Gcusp_{0}^2
                +\frac{21}{4} \beta_0 \boxS{\Sbc{1,3}} \Gcusp_{0}
                + \boxS{\Sbc{1,2}} \biggl(
                -\pi^2 \Gcusp_{0}^2
    \\
               & \phantom{= {}}
                 -\frac{3}{4}  \Gcusp_{1}
                 -3 \beta_0^2
                 \biggr)
                 % \\
                 % &
                 + \boxS{\Sbc{1,1}} \biggl(
                 +\frac{9}{8} \pi^2 \beta_0 \Gcusp_{0}
                 -48 \zeta_3  \Gcusp_{0}^2
                 -\frac{3}{2}  \Gsoft_{1}+
                 \frac{3}{2} \beta_1 
                 \biggr)
\\
               & \phantom{= {}}
                 +\frac{1}{5} \pi^4 \beta_0 \Gcusp_{0}^2
                  -\frac{1}{6} \pi^2 \beta_1 \Gcusp_{0}
                -\frac{1}{6} \pi^2 \beta_0 \Gcusp_{1}
                +4 \pi^2 \Gcusp_{0} \Gsoft_{1}
                +\frac{2}{3} \beta_0^2 \zeta_3 \Gcusp_{0}
                -960 \zeta_5 \Gcusp_{0}^3
                \\
               & \phantom{= {}} + 96 \pi^2 \zeta_3 \Gcusp_{0}^3
                 +72 \zeta_3 \Gcusp_{1} \Gcusp_{0}
                + 2 \Gsoft_{2}.
\end{split}
\end{equation}
Contributions  that cannot be obtained from the renormalization group equation alone are marked  
with boxes. 

 Terms in the soft function that are multiplied by   
powers of the logarithm $L_S$ 
are 
\begin{align}
    C_{1,1} & =  0,
\nonumber     \\
    C_{1,2} & =  -8 \CR,
\nonumber     \\
    C_{2,1} & =  \CA \CR \left(56 \zeta_3-\frac{1616}{27}-\frac{44}{9}\pi^2\right)
              +\CR \nf \TF\left(\frac{448}{27}+\frac{16}{9}\pi^2\right),
\nonumber \\
    C_{2,2} & =  \CR \CA\left(\frac{8}{3}\pi^2-\frac{536}{9}\right)
              + \frac{160}{9}\CR \nf \TF ,
\nonumber \\
    C_{2,3} & = -\frac{176}{9} \CR \CA + \frac{64}{9} \CR \nf \TF,
\nonumber \\
    C_{3,1} & =  \CA^2 \CR \biggl(
              \frac{36272}{27}\zeta_3
              -\frac{176}{9}\pi^2 \zeta_3
              -384 \zeta_5
              -\frac{556042}{729}
              -\frac{50344}{243} \pi^2
              +\frac{88}{9} \pi^4
              \biggr)
\nonumber \\
            & \phantom{= {}} +\CR \CA \nf \TF \biggl(
              -\frac{12064}{27}\zeta_3
              +\frac{160648}{729}
              +\frac{38816}{243} \pi^2
              -\frac{128}{45} \pi^4
              \biggr)
   \label{eq:logSlogs}   \\
            & \phantom{= {}} + \CR \CF \nf \TF \biggl(
              -\frac{1216}{9} \zeta_3
              +\frac{6844}{27}
              +\frac{16}{3} \pi^2
              -\frac{32}{45} \pi^4
              \biggr)
\nonumber \\
            & \phantom{= {}} +\CR \nf^2 \TF^2 \biggl(
              \frac{256}{9}\zeta_3
              +\frac{12800}{729}
              -\frac{256}{9}\pi^2
              \biggr),
\nonumber \\
    C_{3,2} & =  \CA^2 \CR \biggl(
              352 \zeta_3
              -\frac{62012}{81}
              +\frac{104}{27}\pi^2
              -\frac{88}{45}\pi^4
              \biggr)
              + \CR \CA \nf \TF \biggl(\frac{32816}{81}
\nonumber \\
            & \phantom{= {}}
              +\frac{128}{9}\pi^2\biggr)
              + \CR \CF \nf \TF \biggl(\frac{440}{3}-128 \zeta_3\biggr)
              + \CR \nf^2\TF^2\biggl(-\frac{3200}{81}-\frac{128}{27}\pi^2\biggr) ,
\nonumber \\
    C_{3,3} & = \CR \CA^2 \biggl(\frac{352}{27}\pi^2-\frac{28480}{81}\biggr) 
              + \CR \CA \nf \TF \biggl(\frac{18496}{81}-\frac{128}{27}\pi^2\biggr)
\nonumber \\
            & \phantom{= {}} + \frac{64}{3} \CR \CF \nf \TF -\frac{2560}{81} \CR \nf^2\TF^2,
\nonumber \\
    C_{3,4} & =  - \frac{1936}{27} \CR \CA^2
              + \frac{1408}{27} \CR \CA \nf\TF
              - \frac{256}{27} \CR \nf^2 \TF^2.
\nonumber 
  \end{align}

\section{Single soft-gluon emission corrections}
\label{sec:RVVcorrections}

\begin{figure}[t]
  \begin{equation*}
    \sum\limits_{f=\theta,\bar{\theta}}\int \dm \Phi_f \textrm{Re}\left[\boldsymbol{J}_{\mu,a}^{(l)\dagger}\boldsymbol{J}_{\mu,a}^{(m)}\right] \sim 
    \vcenter{\hbox{\begin{tikzpicture}[use Hobby shortcut, scale=1]
          \coordinate (L) at (-1.5,0);
          \coordinate (R) at (1.5,0);
          \coordinate (U) at (0,1);
          \coordinate (D) at (0,-1);
          \coordinate (k) at (0,0);
          \draw[sglu] (L) -- (R);
          \node[anchor=north east] at (k) {\small$k$};
          \draw[WLBC] (-1.5,0.2)--(U);
          \draw[-{Stealth[scale=0.7]}] (-0.8,0.8)  --  node[pos=0.3, anchor = south] {\small$n$} (-0.3,1.0666)  {};
          \draw[WLBC] (U)-- (1.5,0.2);
          \draw[WLBC] (1.5,-0.2)-- (D);
          \draw[-{Stealth[scale=0.7]}] (-0.8,-0.8)  --  node[pos=0.3, anchor = north] {\small$\bar{n}$} (-0.3,-1.0666)  {};
          \draw[WLBC] (D)-- (-1.5,-0.2);
          \draw[higgs] (-2.3,0) -- (L);
          \draw[higgs] (2.3,0) -- (R);
          \draw [black, line width = 0.5pt, dotted] (0,-1.2) to (0,1.2);
          % Blobs
          \filldraw[fill=white, draw=black] (L) circle [radius=4mm];
          \node at (L) {\tiny \texttt{l-loop}};
          \filldraw[fill=white, draw=black] (R) circle [radius=4mm];
          \node at (R) {\tiny \texttt{m-loop}};
          \path[use as bounding box] (-2.5,-1.2) rectangle (2.5,1.2);
      \end{tikzpicture}}}
  \end{equation*}
  \caption{Soft amplitude squared $(l\times m)$-loop contribution with single
    soft gluon emission.}
  \label{fig:SggSccSqq}
\end{figure}

In this appendix we compute the single-gluon N3LO QCD contribution to the
zero-jettiness soft function. It follows from eq.~\eqref{eq:generic-eikonal-def}
that in order to do that, we need to integrate the square of the single-soft
eikonal current, truncated at ${\cal O}(\alpha_s^3)$, over single-gluon phase
space subject to the zero-jettiness constraint.

To facilitate this computation, we write  perturbative corrections to the soft current 
\begin{equation}
  \label{eq:JpertExp}
  \boldsymbol{J}_{\mu,a}(k) = g_s \sum\limits_{l=0}^\infty g_s^{2l} \boldsymbol{J}_{\mu,a}^{(l)}(k),
\end{equation}
and define the product of $l$- and $m$-loop currents
\begin{equation}
  \label{eq:wJJdef}
  w_{l,m}(k) = -g^{\mu\nu} \textrm{Re}\left[\boldsymbol{J}_{\mu,a}^{(l)\dagger}(k)\boldsymbol{J}_{\nu,a}^{(m)}(k)\right],
\end{equation}
where summation over all color degrees of freedom, as well as over gluon
polarizations, is assumed. Thanks to the simple dependence of the soft currents
on the external momenta, this quantity reads
\begin{equation}
  \label{eq:wlmMomPart}
  w_{l,m}(k) = \mathcal{C}_{l,m} \textrm{Re}\left[ e^{-i \pi l\ep} e^{i \pi m\ep}\right] \left(\frac{n\cdot \bar{n}}{2 (k\cdot n) (k\cdot \bar{n}) } \right)^{1+(l+m)\ep},
\end{equation}
where the constant $\mathcal{C}_{l,m}$ contains 
all the information about colors and 
other quantities that depend on $l$ and $m$  but not on momenta $k$, $n$ and $\bar n$.

To calculate soft-function contribution $s_{l,m}$ from diagrams with $l \times m$ loops we
sum over two hemisphere  configurations and  integrate over the gluon momentum $k$
\begin{equation}
\begin{split}
  \label{eq:SgLMcontribution}
  s_{l,m} & = \normNep \sum\limits_{f=\theta,\bar{\theta}}\int \dm \Phi_f w_{l,m}(k)
            \\
          & = \normNep \mathcal{C}_{l,m} \cos{[(l-m)\pi\ep]} \sum\limits_{f=\theta,\bar{\theta}}\int \dm \Phi_f \left( \frac{n\cdot \bar{n}}{2 (k\cdot n) (k\cdot \bar{n}) } \right)^{1+(l+m)\ep}
          \\
          & = 2 \normNep \cos{[(l-m)\pi\ep]} \mathcal{C}_{l,m} \;  F(l+m) \;\left( \frac{n\cdot \bar{n}}{2 } \right)^{1+(l+m)\ep}.
\end{split}
\end{equation}
Using the following expressions for the integration measures, 
\begin{equation}
  \label{eq:dPhiGdef}
  \dm \Phi_\theta = \frac{1}{\normNep}[\dm k] \delta(1- k\cdot n) \theta(k\cdot \bar{n} - k\cdot n),
  \quad
  \dm \Phi_{\bar{\theta}} = \frac{1}{\normNep}[\dm k] \delta(1- k\cdot \bar{n}) \theta( k\cdot n - k\cdot \bar{n}),
\end{equation}
and the fact that the integrand is symmetric under $n \leftrightarrow \bar n$
replacement, we find
\begin{align}
  \label{eq:JintCalc}
  F(m) = \int \frac{\textrm{d} \alpha \textrm{d} \beta} {\alpha^\ep \beta^\ep} \delta(1-\beta) \theta(\alpha -\beta)  \frac{1} {\alpha^{1+m \ep} \beta^{1+m \ep}}
  = \int\limits_1^\infty \frac{\textrm{d} \alpha}{\alpha^{1+\ep+m\ep}} = \frac{1}{(m+1)\ep}.
\end{align}
The final expression for $n\cdot \bar{n} = 2$ reads
\begin{equation}
  \label{eq:sPartFinal}
  s_{l,m} = 2 \normNep \cos\left[(l-m)\pi \ep  \right] \frac{\mathcal{C}_{l,m}}{(1+l+m)\ep}.
\end{equation}
To find corrections at N3LO, we require $s_{0,2},\; s_{2,0}$ and $s_{1,1}$. To
fix the normalization, we start with the lowest order contribution. Writing the
leading order current as
\begin{equation}
  \label{eq:J0Def}
  \boldsymbol{J}_{\mu,a}^{(0)}(k)= \boldsymbol{T}_a \left(\frac{n_\mu}{k\cdot n} -
    \frac{\bar{n}_\mu}{k\cdot \bar{n}}\right),
\end{equation}
and using it in eqs.~(\ref{eq:JpertExp},\ref{eq:wlmMomPart}), it is easy to fix
the constant $\mathcal{C}_{0,0} = 4 \CR$. The one-loop current was computed in
ref.~\cite{Catani:2000pi} and we borrow it from that reference. It
reads\footnote{Since we consider $n$, $\bar n$ and $k$ to be final-state
momenta, the $(-1)^{\ep}$ factor should be 
understood as $e^{+i \pi \ep}$, see
ref.~\cite{Catani:2000pi}.}
\begin{equation}
  \label{eq:J1Def}
  \boldsymbol{J}_{\mu,a}^{(1)}(k) 
   =
  \normNep \mathcal{X} \left[\frac{-(n\cdot \bar{n}) }{2 (k\cdot n) (k\cdot \bar{n}) } \right]^{\ep} \boldsymbol{J}_{\mu,a}^{(0)}(k),
\end{equation}
with
\begin{equation}
  \label{eq:J1coeff}
  \mathcal{X} = - \CA \frac{\Gamma^4(1 - \ep) \Gamma^2(1 + \ep)}{\ep^2 \Gamma(1 - 2 \ep)}.
\end{equation}
Then,  using eq.~\eqref{eq:J0Def} and eq.~\eqref{eq:J1Def}, we  calculate the NLO contribution
\begin{equation}
w_{1,0} = w_{0,1} = w_{0,0} \ \normNep \mathcal{X} \left[\frac{ - n\cdot \bar{n}}{2 (k\cdot n) (k\cdot \bar{n}) } \right]^{\ep},
\end{equation}
and use eq.~\eqref{eq:wlmMomPart} to fix
the following constants. 
\begin{equation}
  \label{eq:CnloFix}
  \mathcal{C}_{0,1} = \mathcal{C}_{1,0} = \normNep \mathcal{X} \mathcal{C}_{0,0}.
\end{equation}
Similarly, using eq.~\eqref{eq:J1Def} we can calculate  $w_{1,1}$ and extract the 
corresponding constant
\begin{equation}
  \label{eq:C11nnnlo}
  \mathcal{C}_{1,1} = \left( \normNep \mathcal{X} \right)^2 \mathcal{C}_{0,0}.
\end{equation}
The result for the two-loop contribution
is~\cite{Badger:2004uk,Duhr:2013msa,Li:2013lsa}. We use ref.~\cite{Duhr:2013msa}
and eq.~\eqref{eq:wlmMomPart} to fix two remaining constants
\begin{equation}
  \label{eq:C20C02}
  \mathcal{C}_{0,2} = \mathcal{C}_{2,0} = 4 \CR \left(
   \frac{  \normNep \Gamma^3(1-\ep) \Gamma(1+\ep)}{\Gamma(1-2\ep)}
  \right)^2 r_{\textrm{soft}}^{(2)}.
\end{equation}
The constant  $r_{\textrm{soft}}^{(2)}$ is defined in~\cite{Duhr:2013msa}; it reads 
\begin{equation}
  \label{eq:r2soft-def}
  r_{\textrm{soft}}^{(2)} = 2 \CA\nf\TF {\cal R}_1 + \CA^2 {\cal R}_2,
\end{equation}
where
\begin{align}
  \begin{split}
    {\cal R}_1
    & =
      \frac{1}{6 \ep^3}
      +\frac{5}{18 \ep^2}
      +
      \biggl(\frac{19}{54}
      +\frac{\pi ^2}{18}  \biggr)
      \frac{1}{\ep}
    \\
    & \phantom{= {}}
      +\biggl(
      \frac{65}{162}
      +\frac{5}{54}\pi^2
      -\frac{8}{3}\zeta_3
      \biggr)
      +\ep \biggl(
      \frac{211}{486}
      -\frac{4}{81}\pi^2
      -\frac{40}{9}\zeta_3
      -\frac{\pi^4}{15}
      \biggr)
    \\
    & \phantom{= {}}        
      -\ep^2 \biggl(
      \frac{287 \zeta_3}{27}
      +\frac{8\pi ^2 \zeta_3}{9}
      +32 \zeta_5
      -\frac{665}{1458}
      +\frac{151 \pi ^2}{486}
      +\frac{\pi ^4}{9}
      \biggr)
      + \mathcal{O}\left(\ep^3\right),
  \end{split}\\
  \begin{split}
    {\cal R}_2
    & =
      \frac{1}{2 \ep^4}
      -\frac{11}{12 \ep^3}
      -\biggl(
      \frac{67}{36}
      -\frac{1}{4}\pi^2
      \biggr)\frac{1}{\ep^2}
      - \biggl(
      \frac{193}{54}
      + \frac{11}{36}\pi^2
      - \frac{1}{2}\zeta_3
      \biggr)\frac{1}{\ep}
    \\
    &\phantom{= {}}
      -\biggl(
      \frac{571}{81}
      +\frac{67}{108}\pi^2
      - \frac{44}{3}\zeta_3
      -\frac{29}{360}\pi^4
      \biggr)
      -\ep \biggl(
      \frac{3410}{243}
      +\frac{83 }{81}\pi^2
      - \frac{268}{9}\zeta_3
      -\frac{11}{30}\pi^4
    \\
    &\phantom{= {}}
      -\frac{2 }{3}\pi^2\zeta_3
      +\frac{37}{2} \zeta_5
      \biggr)
      -\ep^2 \biggl(
      \frac{20428}{729}
      +\frac{1007}{486} \pi^2
      - \frac{1679}{27}\zeta_3
      -\frac{67}{90}\pi ^4
      -\frac{44}{9}\pi^2 \zeta_3
    \\
    & \phantom{= {}}
      -176 \zeta_5
      +\frac{451}{11340}\pi ^6
      +\frac{29}{2}\zeta_3^2
      \biggr)
      + \mathcal{O}\left(\ep^3\right).
  \end{split}
\end{align}

Finally, we use the above results 
to compute  the single-gluon contribution 
to the soft function through N3LO. Writing
\begin{equation}
  \label{eq:SgPertExp}
  S_g = A_s S_g^{(1)} + A_s^2 S_g^{(2)} + A_s^3 S_g^{(3)} + \mathcal{O}\left( A_s^4 \right),
\end{equation}
where the strong coupling constant $A_s$ is defined in eq.~\eqref{eq9.2}, we
obtain
\begin{align}
  S_g^{(1)} & = \frac{s_{0,0}}{\normNep} = \frac{8}{\ep}\CR ,   \label{eq:S1res} \\
  S_g^{(2)} & = \frac{1}{\normNep^2}\left( s_{1,0} + s_{0,1} \right) = \frac{8}{\ep} \CR \cos{(\pi \ep)} \mathcal{X}, \label{eq:S2res}\\
  S_g^{(3)} & = \frac{1}{\normNep^3}\left(s_{2,0} + s_{1,1} + s_{0,2}  \right)\nonumber\\
            & = \frac{8}{3\ep}\CR \left(\mathcal{X}^2 + 2 \cos{(2\pi \ep)}
              \left(\frac{ \Gamma^3(1-\ep) \Gamma(1+\ep)}{\Gamma(1-2\ep)}\right)^2 r_{\textrm{soft}}^{(2)} \right).
  \label{eq:S3res}
\end{align}
To conclude, we explicitly write the result for $S_3^{RVV} = S_g^{(3)}$ expanded
through the required order in $\ep$
\begin{equation}
  \label{eq:Srvv3}
  \begin{split}
    S_3^{\textrm{RVV}}
    & =  S_g^{(3)} 
      =
      \frac{16 }{3 \ep^5}\CR\CA^2
      + \biggl( 
      \frac{16}{9} \CA \CR \nf \TF
      -\frac{44 }{9} \CA^2\CR
      \biggr)\frac{1}{\ep^4}
    \\
    & \phantom{= {}}      
      - \biggl(
      \biggl(\frac{268}{27}+\frac{28 \pi ^2}{9}\biggr) \CA^2 \CR
      -\frac{80}{27} \CA \CR \nf\TF
      \biggr)
      \frac{1}{\ep^3}
    \\
    & \phantom{= {}}      
      - \biggl(
      \CA^2 \CR \biggl(
      \frac{1544}{81}
      -\frac{220}{27}\pi^2
      +\frac{56}{3}\zeta_3
      \biggr)
      - \CR \CA \nf \TF \biggl(
      \frac{304}{81}
      -\frac{80}{27}\pi^2
      \biggr) 
      \biggr)
      \frac{1}{\ep^2}
      \\
    & \phantom{= {}} -
      \biggl(
      +\CA^2 \CR \biggl(
      \frac{9136}{243}
      -\frac{1340}{81}\pi^2
      -\frac{880}{9}\zeta_3
      +\frac{86}{135}\pi^4
      \biggr)
      + \CA \CR \nf \TF \biggl(
      \frac{320}{9} \zeta_3
      \\  
    & \phantom{= {}}
      - \frac{1040}{243}
      +\frac{400}{81}\pi^2\biggr)
      \bigg)
      \frac{1}{\ep}
      - \biggl(
      \CR  \CA^2 \biggl(
      \frac{54560}{729}
      -\frac{7936}{243}\pi^2
      -\frac{5360}{27}\zeta_3
      -\frac{308}{135}\pi^4
      \\
    & \phantom{= {}}    
      -\frac{32}{3}\pi^2 \zeta_3
      +\frac{488}{3}\zeta_5
      \biggr)
      -\CR \CA \nf \TF \biggl(
      \frac{3376}{729}
      -\frac{1952}{243}\pi^2
      -\frac{1600}{27}\zeta_3
      -\frac{112}{135}\pi^4\biggr)
      \biggr)
      \\
    & \phantom{= {}} -\ep \biggl(
      \CR \CA^2 \biggl(
      \frac{326848}{2187}
      -\frac{46760}{729}\pi^2
      -\frac{33040}{81}\zeta_3
      -\frac{1876}{405}\pi^4
      +\frac{4400}{27}\pi^2\zeta_3
      \\
    & \phantom{= {}}
      -\frac{2992}{3}\zeta_5
      +\frac{2524}{8505}\pi^6
      +\frac{136}{3}\zeta_3^2
      \biggr)
      - \CA \CR \nf \TF
      \biggl(
      \frac{10640}{2187}
      -\frac{8656}{729}\pi^2
      -\frac{10400}{81}\zeta_3
      \\
    & \phantom{= {}}
      -\frac{112}{81}\pi^4
      +\frac{1600}{27}\pi^2\zeta_3
      -\frac{1088}{3}\zeta_5    
      \biggr)
      \biggr)+\mathcal{O}\left(\ep^2\right).
  \end{split}
\end{equation}

\section{Iterated real-emission contributions}
\label{sec:iter-real-emiss}

As we explain in this appendix, some of the real-emission contributions to  the N3LO soft function can be computed with a relative ease. 
To this end, 
we consider the real-emission contribution to the $n$-th 
order term in the expansion of the  
soft function in $\alpha_s$, c.f.\ eq.~\eqref{eq:soft-function-Sn-def}, and 
write it as 
\begin{equation}
  \label{eq:SnXiDef}
  S_\tau^{(n)} = \int \prod_{i=1}^{n} \; [\mathrm{d}k_i] \;   \delta(\tau-\mathcal{T}_0(n)) \; \xi_n(k_1,\dots,k_n).
\end{equation}
The functions $\xi_n$ contain contributions from all possible final states and
color structures and, in general, they cannot be predicted from similar
functions at lower orders. However, for certain color structures, their form
simplifies. Following refs.~\cite{Catani:2019nqv,Catani:2022hkb}, we write
\begin{align}
  \label{eq:SnXi123}
  \xi_1(k_1)
  & = \CR w_g(k_1),\\
  \xi_2(k_1,k_2)
  & = \frac{1}{2!}  \left( \CR^2w_g(k_1) w_g(k_2) + \CR \CA  w_{gg}(k_1,k_2) \right)  + \CR \nf \TF w_{q\bar{q}}(k_1,k_2),\\
  \xi_3(k_1,k_2,k_3)
  & =  \frac{1}{3!} \bigg(
    \CR^3 w_g(k_1) w_g(k_2) w_g(k_3) 
    \nonumber\\
  & \phantom{= {}} +
    \CR^2 \CA
    \left[
    w_g(k_1) w_{gg}(k_2,k_3) + (1\leftrightarrow 2) + (1\leftrightarrow 3)
    \right]
    \bigg)
      \label{eqc.4}
    \\
  & \phantom{= {}} + \CR^2 \nf \TF w_g(k_1) w_{q\bar{q}}(k_2,k_3) + \dots, \nonumber 
\end{align}
where ellipses stand for contributions that are unrelated to lower order ones. The functions 
$w_g(k_1)$ and $w_{gg}(k_1,k_2)$ are called $w_{BC}^{(1)}(k_1)$ and
$w_{BC}^{(2)}(k_1,k_2)$ in 
ref.~\cite{Catani:2019nqv}; the function $w_{q\bar{q}}(k_1,k_2)$ is called
$w_{BC}^{(2)}(k_1,k_2)$ in ref.~\cite{Catani:2022hkb};
they can be extracted from those references.
The common feature of  the contributions to $\bar \zeta_3$ shown explicitly in eq.~\eqref{eqc.4} is 
that the 
single-gluon emission contribution $w_g$ factorizes as least once.

Since the zero-jettiness 
${\cal T}_0$ is a \emph{linear} 
function of the soft partons' momenta, the factorization property can be exploited by computing the Laplace transform of the soft function. We introduce it by considering 
the $C_R^3$ contribution to 
$S_\tau^{(3)}$. We write 
\begin{equation}
 S_u^{(3),C_R^3} = 
 \frac{1}{3!}\int \dm \tau \; 
 e^{-u \tau} 
 \prod \limits_{i=1}^{3} 
 [\dm k_i] \delta(\tau-
 {\cal T}_0) \; 
 w_g(k_1) w_g(k_2) w_g(k_3).
\end{equation}
Integrating over $\tau$ and using the fact that ${\cal T}_0 = {\cal T}_0(k_1) + {\cal T}_0(k_2) + {\cal T}_0(k_3)$, we obtain 
\begin{equation}
  \label{eq:SiterCR3lap}
  S_u^{(3),C_R^3} = 
  \frac{1}{3!} \left [ S_u^{(1),\CR} \right ]^3.
\end{equation}

Subleading contributions $C_R^2 C_A$ and $C_R^2 \nf \TF$ are obtained in the same way. Computing 
the Laplace transform and accounting for the permutations, we find 
\begin{equation}
  S_u^{(3),C_R^2 C_A }  = S_u^{(1),\CR} S_u^{(2),C_R C_A},\;\;\;\;
  S_u^{(3),C_R^2 \nf \TF }  = S_u^{(1),\CR} S_u^{(2),C_R \nf\TF}.
  \label{eq:SiterCR2lapTF}
\end{equation}
Because the scaling of the soft function $S^{(n)}_\tau$ with respect to $\tau$
is fully determined by $n$, the following relation between the soft function and
its Laplace image holds
\begin{equation}
  \label{eq:SnLapAndInv}
  S_u^{(n) } = \Gamma(-2n\ep) \; u^{2 n \ep}
  S_{\tau}^{(n)} \Bigg |_{\tau = 1}.
\end{equation}
Hence, once the Laplace imagine is known, it is straightforward to reconstruct
the $\tau$-dependent soft function from the above equation. When writing the
soft function $S_{\tau}$ below, we will always assume that $\tau$ is taken to be
one and we will not indicate this further.

Required NLO contribution to eq.~\eqref{eq:SiterCR2lapTF} can be easily
calculated from the exact result for $S_\tau^{(1)}$
\begin{equation}
  \label{eq:S1lap}
  S_\tau^{(1),\CR } = \frac{8}{\ep} \;\; \rightarrow  \;\; S_u^{(1),\CR } = \frac{8\Gamma(-2\ep)}{\ep} u^{2 \ep}. 
\end{equation}
Using eq.~\eqref{eq:SiterCR3lap} and eq.~\eqref{eq:SnLapAndInv}, we find the  part of
the N3LO soft function 
proportional to $C_R^3$ to be
\begin{equation}
  \label{eq:SrrrCR3}
  S_{\tau}^{(3),\CR^3} = \frac{\Gamma^3 (-2 \ep)}{6 \Gamma (-6 \ep)} \left[ S_{\tau}^{(1),\CR} \right]^3
  =  \frac{256 \Gamma^3 (-2 \ep)}{3 \ep^3 \Gamma (-6 \ep)}.
\end{equation}

To calculate contributions proportional to $C_R^2$ 
using eq.~\eqref{eq:SiterCR2lapTF}
, we can extract the required parts
$S_u^{(2),C_R C_A}$ and $S_u^{(2),C_R \nf\TF}$ from the  NNLO
soft function 
in eq.~\eqref{eq:SBnnlo}. To get the real-emission contribution for  the final-state gluons, we need to 
 subtract virtual corrections 
 from the NNLO soft
function. On the contrary, in case of  
the $q \bar q$, no subtraction is needed, and 
we can use the NNLO contribution $S_{\tau}^{(2),\CR\nf\TF}$ directly.   We obtain 
\begin{equation}
\begin{split}
  \label{eq:SrrrCR2nfTF}
  S_{\tau}^{(3),\CR^2\nf\TF}
  & =
    \frac{\Gamma (-2 \ep) \Gamma (-4 \ep)}{\Gamma (-6 \ep)} S_{\tau}^{(1),\CR} S_{\tau}^{(2),\CR\nf\TF}
   \\
  & = \frac{32}{\ep^4}+\frac{160}{3 \ep^3}
    +\frac{1}{\ep^2}\left(\frac{896}{9}-\frac{160}{3} \pi^2 \right)
    +\frac{1}{\ep}\left( -2176 \zeta_3+\frac{640}{27}+\frac{32}{9}\pi^2 \right)
   \\
  &  +\left(256 \zeta_3-\frac{17152}{81}-\frac{4864}{27}\pi^2-\frac{3424}{45}\pi^4\right)
    +\bigg(-\frac{70912}{9}\zeta_3
    \\
    & 
        +\frac{12800}{3}\pi^2\zeta_3
    -61184 \zeta_5
    -\frac{277376}{243}
    +\frac{21440}{81}\pi^2
    -\frac{12544}{135}\pi^4
    \bigg)\ep 
   + \mathcal{O}\left(\ep^2\right).
\end{split}
\end{equation}

As we mentioned earlier, to 
obtain the $\CR^2\CA$ part of the soft function, 
we need to subtract the real-virtual corrections
 from the ${\cal O}(\CR \CA) $ term in the NNLO soft function.
 The real-virtual contribution reads
\begin{equation}
  \label{eq:SnnloRV}
  S_{\tau,\textrm{RV}}^{(2)}  = -\frac{8 \Gamma^5(1-\ep) \Gamma^3(1+\ep)}{\ep^3 \Gamma^2(1-2 \ep) \Gamma (1 + 2\ep)}, 
\end{equation}
where the color 
factor $\CR \CA$ is not shown.
By combining the relevant contributions, we obtain the final result  for the 
coefficient of the $C_R^2 C_A$
color structure 
\begin{equation}
\begin{split}
  \label{eq:SrrrCR2CA}
  S_{\tau}^{(3),\CR^2\CA}
  & = \frac{\Gamma (-2 \ep) \Gamma (-4 \ep)}{\Gamma (-6 \ep)} S_{\tau}^{(1),\CR} \left[  S_{\tau}^{(2),\CR\CA} - S_{\textrm{RV}}^{(2)}\right]
   \\
  & = -\frac{48}{\ep^5}
    -\frac{88}{\ep^4}
    +\frac{1}{\ep^3}\left(88 \pi^2-\frac{536}{3}  \right)
    +\frac{1}{\ep^2}\left( 2736 \zeta_3-\frac{3232}{9}+\frac{440}{3} \pi^2\right)
\\
  & +\left( 5984 \zeta_3-\frac{17120}{27}+\frac{536}{9} \pi^2+\frac{388}{5}\pi^4 \right)\frac{1}{\ep}
    +\bigg(2144 \zeta_3 -4896 \pi^2 \zeta_3 
  \\
  & + 68880 \zeta_5-\frac{99328}{81}+\frac{15872}{27} \pi^2+\frac{9416}{45}\pi^4\bigg)
    + \bigg(\frac{218048}{9}\zeta_3 - \frac{35200}{3} \pi^2 \zeta_3
    \\
  & 
      -73104 \zeta_3^2  +168256 \zeta_5
    -\frac{540224}{243}
    +\frac{10496}{81} \pi^2
    +\frac{9112}{27}\pi^4
\\
    & +\frac{13928}{105}\pi^6\bigg)\ep 
    + \mathcal{O}\left(\ep^2\right).
    \end{split}
\end{equation}

\section{Angular integrations}
\label{sec:angIntApp}

Partial fractioning identities

\begin{align}
  \label{eq:pfAngDenDen}
    % nb-nb
    \frac{1}{\left( \rho_1 \cdot \rho_{\bar{n}} \right)\left( \rho_1 \cdot \bar{\rho}_{\bar{n}} \right)}
    & = \frac{1}{2}\frac{1}{\left( \rho_1 \cdot \rho_{\bar{n}} \right)} + \frac{1}{2}\frac{1}{\left( \rho_1 \cdot \bar{\rho}_{\bar{n}} \right)},\\
    % n-n
    \frac{1}{\left( \rho_1 \cdot \rho_n \right)\left( \rho_1 \cdot \bar{\rho}_{n} \right)}
    & = \frac{1}{2}\frac{1}{\left( \rho_1 \cdot \rho_n \right)} + \frac{1}{2}\frac{1}{\left( \rho_1 \cdot \bar{\rho}_{n} \right)},
      \\
    \frac{1}{\left( \rho_1 \cdot \rho_n \right)\left( \rho_1 \cdot \rho_{m} \right)}
    & = \frac{1}{1-\lambda}\frac{1}{\left( \rho_1 \cdot \rho_n \right)} - \frac{\lambda}{1-\lambda} \frac{1}{\left( \rho_1 \cdot \rho_{m} \right)},
      \\
    \frac{1}{\left( \rho_1 \cdot \rho_n \right)\left( \rho_1 \cdot \bar{\rho}_{m} \right)}
    & = \frac{1}{1+\lambda} \frac{1}{\left( \rho_1 \cdot \rho_n \right)} + \frac{\lambda}{1+\lambda} \frac{1}{\left( \rho_1 \cdot \bar{\rho}_{m} \right)},
      \\
    \frac{1}{\left( \rho_1 \cdot \bar{\rho}_n \right)\left( \rho_1 \cdot \rho_{m} \right)}
    & = \frac{1}{1+\lambda} \frac{1}{\left( \rho_1 \cdot \bar{\rho}_n \right)} + \frac{\lambda}{1+\lambda} \frac{1}{\left( \rho_1 \cdot \rho_{m} \right)},
      \\
    \frac{1}{\left( \rho_1 \cdot \bar{\rho}_n \right)\left( \rho_1 \cdot \bar{\rho}_{m} \right)}
    & = \frac{1}{1-\lambda}\frac{1}{\left( \rho_1 \cdot \bar{\rho}_n \right)} - \frac{\lambda}{1-\lambda} \frac{1}{\left( \rho_1 \cdot \bar{\rho}_{m} \right)},
      \\
    \frac{1}{\left( \rho_1 \cdot \rho_m \right)\left( \rho_1 \cdot \bar{\rho}_{m} \right)}
    & = \frac{1}{2}\frac{1}{\left( \rho_1 \cdot \rho_m \right)} + \frac{1}{2}\frac{1}{\left( \rho_1 \cdot \bar{\rho}_{m} \right)}.
\end{align}

\begin{align}
  \label{eq:pfAngNumDen}
  %   nb-nb
  \frac{\left( \rho_1 \cdot \rho_{\bar{n}} \right)}{\left( \rho_1 \cdot \bar{\rho}_{\bar{n}} \right)}
  & = -1 + 2 \frac{1}{\left( \rho_1 \cdot \bar{\rho}_{\bar{n}} \right)},
  &
    \frac{\left( \rho_1 \cdot \bar{\rho}_{\bar{n}} \right)}{\left( \rho_1 \cdot \rho_{\bar{n}} \right)}
  & = -1 + 2 \frac{1}{\left( \rho_1 \cdot \rho_{\bar{n}} \right)},
  \\
  % n-n
  \frac{\left( \rho_1 \cdot \rho_n \right)}{\left( \rho_1 \cdot \bar{\rho}_{n} \right)}
  & = -1 + 2 \frac{1}{\left( \rho_1 \cdot \bar{\rho}_{n} \right)},
  &
    \frac{\left( \rho_1 \cdot \bar{\rho}_{n} \right)}{\left( \rho_1 \cdot \rho_n \right)}
  & = -1 + 2 \frac{1}{\left( \rho_1 \cdot \rho_n \right)},
  \\
  \frac{\left( \rho_1 \cdot \rho_n \right)}{\left( \rho_1 \cdot \rho_{m} \right)}
  & = \frac{1}{\lambda} - \frac{1-\lambda}{\lambda}\frac{1}{\left( \rho_1 \cdot \rho_{m} \right)},
  &
    \frac{\left( \rho_1 \cdot \rho_{m} \right)}{\left( \rho_1 \cdot \rho_n \right)}
  & = \lambda + (1-\lambda)\frac{1}{\left( \rho_1 \cdot \rho_n \right)},
  \\
  \frac{\left( \rho_1 \cdot \rho_n \right)}{\left( \rho_1 \cdot \bar{\rho}_{m} \right)}
  & = - \frac{1}{\lambda} + \frac{1+\lambda}{\lambda}   \frac{1}{\left( \rho_1 \cdot \bar{\rho}_{m} \right)},
  &
    \frac{\left( \rho_1 \cdot \bar{\rho}_{m} \right)}{\left( \rho_1 \cdot \rho_n \right)}
  & = -\lambda + (1+\lambda) \frac{1}{\left( \rho_1 \cdot \rho_n \right)},
  \\
  \frac{\left( \rho_1 \cdot \bar{\rho}_n \right)}{\left( \rho_1 \cdot \rho_{m} \right)}
  & = - \frac{1}{\lambda} + \frac{1+\lambda}{\lambda} \frac{1}{\left( \rho_1 \cdot \rho_{m} \right)},
  &
    \frac{\left( \rho_1 \cdot \rho_{m} \right)}{\left( \rho_1 \cdot \bar{\rho}_n \right)}
  & =  -\lambda + (1+\lambda)  \frac{1}{\left( \rho_1 \cdot \bar{\rho}_n \right)},
  \\
  \frac{\left( \rho_1 \cdot \bar{\rho}_n \right)}{\left( \rho_1 \cdot \bar{\rho}_{m} \right)}
  & = \frac{1}{\lambda} - \frac{1-\lambda}{\lambda} \frac{1}{\left( \rho_1 \cdot \bar{\rho}_{m} \right)},
  &
    \frac{\left( \rho_1 \cdot \bar{\rho}_{m} \right)}{\left( \rho_1 \cdot \bar{\rho}_n \right)}
  & = \lambda + (1-\lambda)  \frac{1}{\left( \rho_1 \cdot \bar{\rho}_n \right)},
  \\
  \frac{\left( \rho_1 \cdot \rho_m \right)}{\left( \rho_1 \cdot \bar{\rho}_{m} \right)}
  & = -1 + 2 \frac{1}{\left( \rho_1 \cdot \bar{\rho}_{m} \right)},
  &
    \frac{\left( \rho_1 \cdot \bar{\rho}_{m} \right)}{\left( \rho_1 \cdot \rho_m \right)}
  & = -1 + 2 \frac{1}{\left( \rho_1 \cdot \rho_m \right)}.
\end{align}
Integrals defined in~\eqref{eq:angIntDef} after partial fractioning
identities application
\begin{align}
  \Omega^{(d-1)}_{000000} & = 1,\label{eq:miAngInt000000}\\
  % One denominator, massless
  \Omega^{(d-1)}_{a_100000} & = I_{d-1;a_1}^{(0)},\\
  \Omega^{(d-1)}_{0a_20000} & = I_{d-1;a_2}^{(0)},\\
  \Omega^{(d-1)}_{00a_3000} & = I_{d-1;a_3}^{(0)},\\
  \Omega^{(d-1)}_{000a_400} & = I_{d-1;a_4}^{(0)},\\
    % One denominator, massive
  \Omega^{(d-1)}_{0000a_50} & = I_{d-1;a_5}^{(1)}\left(\rho_m^2 \right),\\
  \Omega^{(d-1)}_{00000a_6} & = I_{d-1;a_6}^{(1)}\left(\bar{\rho}_m^2 \right),\\
    % Two denominators, massless
  \Omega^{(d-1)}_{a_10a_3000} & = I_{d-1;a_1,a_3}^{(0)}\left(\rho_n\cdot \rho_{\bar{n}} \right),\\
  \Omega^{(d-1)}_{a_100a_400} & = I_{d-1;a_1,a_4}^{(0)}\left(\rho_n\cdot \bar{\rho}_{\bar{n}} \right),\\
  \Omega^{(d-1)}_{0a_2a_3000} & = I_{d-1;a_2,a_3}^{(0)}\left(\bar{\rho}_n\cdot \rho_{\bar{n}} \right),\\
  \Omega^{(d-1)}_{0a_20a_400} & = I_{d-1;a_2,a_4}^{(0)}\left(\bar{\rho}_n\cdot \bar{\rho}_{\bar{n}} \right),\\
  % Two denominators, one massive and one massless
  \Omega^{(d-1)}_{00a_30a_50} & = I_{d-1;a_3,a_5}^{(1)}\left(\rho_{\bar{n}}\cdot \rho_{m},\rho_m^2 \right),\\
  \Omega^{(d-1)}_{000a_4a_50} & = I_{d-1;a_4,a_5}^{(1)}\left(\bar{\rho}_{\bar{n}}\cdot \rho_{m},\rho_m^2 \right),\label{eq:miAngInt000xx0}\\
  \Omega^{(d-1)}_{00a_300a_6} & = I_{d-1;a_3,a_6}^{(1)}\left(\rho_{\bar{n}}\cdot \bar{\rho}_{m},\bar{\rho}_m^2 \right),\\
  \Omega^{(d-1)}_{000a_40a_6} & = I_{d-1;a_4,a_6}^{(1)}\left(\bar{\rho}_{\bar{n}}\cdot \bar{\rho}_{m},\bar{\rho}_m^2 \right).\label{eq:miAngInt000x0x}
\end{align}

All considered angle integrals are normalized by full solid angle, so
\begin{equation}
  \label{eq:angIntNormOne}
  I_0^{(0)} =
  I_0^{(1)}\left( \rho^2 \right)=
  I_{0,0}^{(0)}\left( \rho_1\cdot \rho_2 \right) =
  I_{0,0}^{(1)}\left( \rho_1\cdot \rho_2, \rho_2^2 \right) = 1.
\end{equation}

\begin{align}
  I_{d-1;n}^{(0)}
  & = \frac{1}{\Omega^{(d-1)}}\int \frac{\dm \Omega_v^{(d-1)}}{\left( v\cdot \rho \right)^n},
    \quad \rho^2 = 0,
    \label{eq:In0def}\\
  I_{d-1;n}^{(1)}\left(\rho^2 \right)
  & = \frac{1}{\Omega^{(d-1)}}\int \frac{\dm \Omega_v^{(d-1)}}{\left( v\cdot \rho \right)^n},
    \quad \rho^2 \neq 0,\label{eq:In1def}\\
  I_{d-1;a,b}^{(0)}\left(\rho_1\cdot \rho_2 \right)
  & = \frac{1}{\Omega^{(d-1)}}\int \frac{\dm \Omega_v^{(d-1)}}{\left( v\cdot \rho_1 \right)^a \left( v\cdot \rho_2 \right)^b}, \quad \rho_1^2=\rho_2^2=0,\label{eq:Iab0def}\\
  I_{d-1;a,b}^{(1)}\left(\rho_1\cdot \rho_2, \rho_2^2 \right)
  & = \frac{1}{\Omega^{(d-1)}}\int \frac{\dm \Omega_v^{(d-1)}}{\left( v\cdot \rho_1 \right)^a \left( v\cdot \rho_2 \right)^b}, \quad \rho_1^2=0, \rho_2^2 \neq 0,\label{eq:Iab1def}
\end{align}
where $v$ is a light-like vector. 
\begin{align}
  &I_{d-1;n}^{(0)}
  = \frac{\Gamma\left(2-2\ep  \right) \Gamma\left(1-\ep -n  \right)}{\Gamma\left(1-\ep \right) \Gamma\left( 2-2\ep-n \right)}  \label{eq:angInt-1-0},\\
  &I_{d-1;n}^{(1)}\left(\rho^2 \right)
   = \left( 1 + \beta \right)^{-n}  {}_2F_1\left(n,1-\ep,2-2\ep; \frac{2 \beta}{1+\beta}  \right), \quad \beta = \sqrt{1-\rho^2} \label{eq:angInt-1-m},\\
  &I_{d-1;a,b}^{(0)}\left(\rho_1\cdot \rho_2 \right)
   = \frac{\Gamma\left(2-2\ep\right)\Gamma\left(1-\ep -a \right)\Gamma\left(1-\ep -b \right)}{2^{a+b}\Gamma^2\left(1-\ep\right)\Gamma\left(2-2\ep - a -b \right)}
    {}_2F_1\left(a,b, 1-\ep; 1-\frac{\rho_1\cdot \rho_2}{2} \right), \label{eq:angInt-2-00}\\
  &I_{d-1;a,b}^{(1)}\left(\rho_1\cdot \rho_2, \rho_2^2 \right)
   = \frac{2^a}{\left( \rho_1\cdot \rho_2 \right)^b} \frac{\Gamma\left(2-2\ep\right)\Gamma\left(1-\ep-a \right)}{\Gamma\left(1-\ep \right)\Gamma\left(2-2\ep-a  \right)} \times \nonumber\\
  & \quad F_1\left(
    b, 1-\ep-a, 1-\ep-a,2-2\ep-a
    ;1-\frac{1+\beta}{\rho_1\cdot \rho_2}, 1-\frac{1-\beta}{\rho_1\cdot \rho_2} \right), \quad \beta = \sqrt{1-\rho_2^2}.\label{eq:angInt-2-0m}
\end{align}

\section{Special functions}
\label{sec:SpecFuncs}
 In this appendix we collect various identities for (generalized) hypergeometric functions,  see e.g.\ ref.~\cite{Abramowitz:1974}, that we have used in this paper.
\paragraph{Transformation rules}
\begin{align}
  & {}_2F_1\left(a,b;c;z  \right) =
    \left( 1-z \right)^{-a}
    \frac{\Gamma\left( b-a \right) \Gamma\left(c \right)}{\Gamma\left( c-a \right) \Gamma\left(b \right)}
    \,{}_2F_1\left(a,c-b;1+a-b;\frac{1}{1-z}  \right)\nonumber\\
  & \qquad + \left( 1-z \right)^{-b}
    \frac{\Gamma\left( a-b \right) \Gamma\left(c \right)}{\Gamma\left( c-b \right) \Gamma\left(a \right)}
    \,{}_2F_1\left(b,c-a;1-a+b;\frac{1}{1-z}  \right),  \label{eq:f21trInv1mz}\\
    % z2zb
    & {}_2F_1\left(a,b;c;z  \right) =
    \frac{\Gamma\left( c-a-b \right) \Gamma\left(c \right)}{\Gamma\left( c-a \right) \Gamma\left(c-b \right)}
    \,{}_2F_1\left(a,b;a+b+1-c;1-z\right)\nonumber\\
  & \qquad +
  \left( 1-z \right)^{c - a - b}
    \frac{\Gamma\left( a+b-c \right) \Gamma\left(c \right)}{\Gamma\left( a \right) \Gamma\left(b \right)}
    \,{}_2F_1\left(c-a,c-b;c+1-a-b;1-z\right),  \label{eq:f21tr2zb}\\
  &{}_2F_1\left(a,b;c; z  \right) =
   \left(1-z  \right)^{-b}
    {}_2F_1\left(c-a,b;c; \frac{z}{z-1}  \right)\label{eq:f21trInv1},\\
  &{}_2F_1\left(a,b;2b; z  \right) =
   \left(\frac{1+\sqrt{1-z}}{2}  \right)^{-2a}
    {}_2F_1\left(a,a-b+\frac{1}{2};b+\frac{1}{2}; \frac{1-\sqrt{1-z}}{1+\sqrt{1-z}}  \right),\label{eq:f21tr69quad}\\
  & F_1\left( a; b, b^\prime; c; w, z \right) =
   (1 - w)^{-a} F_1 \left(a; c - b - b^\prime, b^\prime; c; \frac{w}{w - 1},\frac{z - w}{1 - w} \right),  \label{eq:appelF1tr26}\\
  % 7.2.4-63
  & F_1\left( a; b, b^\prime; b + b^\prime; w, z \right) =
   \left( 1-z \right)^{-a}\, {}_2F_1\left( a,b;b+b^\prime;\frac{w-z}{1-z} \right).  \label{eq:F1toF21}
\end{align}
\paragraph{Integral representations}
\begin{align}
  {}_2F_1\left( a,b;c;z \right) & = \frac{\Gamma(c)}{\Gamma(b)\Gamma(c-b)} \int\limits_0^1 \dm t\,t^{b-1} (1-t)^{c-b-1} (1-t z)^{-a},  \label{eq:hfIntReprF21}\\
  {}_pF_q\left( (a_p);(b_q);z \right) & = \frac{\Gamma(b_q)}{\Gamma(a_p)\Gamma(b_q-a_p)}
                                        \int\limits_0^1 \dm t
                                        \frac{t^{a_p-1}}{(1-t)^{1+a_p-b_q}} %\nonumber\\
  {}_{p-1}F_{q-1}\left( (a_{p-1});(b_{q-1}); t z\right),  \label{eq:hfIntReprFPQ}\\
  F_1\left( a; b,b^\prime;c;w, z \right) & = \frac{\Gamma(c)}{\Gamma(a)\Gamma(c-a)} \int\limits_0^1 \dm t \frac{t^{a-1} (1-t)^{c-a-1}}{(1-w t)^{b} (1 - z t)^{b^\prime}}.  \label{eq:hfIntReprF1}
\end{align}

\bibliographystyle{JHEP}
\bibliography{sfRRR}
\end{document}